\documentclass[11pt]{article}
\pdfoutput=1

\usepackage{jheppub} 

\usepackage{amsmath,amssymb,epsfig,amsfonts}
\usepackage{graphicx}
\usepackage{epsf}
\usepackage{makeidx}
\usepackage{multirow}
\usepackage[all]{xy}
\usepackage{hyperref}
\usepackage{verbatim} 
\usepackage{rotating}
\usepackage{xcolor}
\usepackage{multirow}
\usepackage{bm}
\usepackage{cleveref}
\usepackage{slashed}
\usepackage[normalem]{ulem}

\usepackage{graphicx}
\usepackage{latexsym,amsmath,amsfonts,amssymb}
\usepackage{mathrsfs}
\usepackage{bbm}
\usepackage{bm}
\usepackage{subfigure}
\usepackage{paralist}
\usepackage[section]{placeins}

\usepackage{tikz}
\usepackage{diagbox}
\usetikzlibrary{positioning}
\usetikzlibrary{calc}
\usetikzlibrary{decorations.pathreplacing,calligraphy}

\usepackage{xstring}
\usetikzlibrary{decorations.pathmorphing} 
\usetikzlibrary{decorations.markings} 
\usetikzlibrary{arrows} 
\usetikzlibrary{shapes} 
\usetikzlibrary{matrix} 
\usetikzlibrary{positioning} 
\usepackage[english]{babel} 
\usepackage[autostyle]{csquotes}

\tikzstyle{brane}=[draw]
\tikzset{D7/.style={circle, draw=black, inner sep=0pt, fill=white, minimum size=3mm}}
\tikzset{hasse/.style={circle, fill,inner sep=2pt}}
\tikzset{flavor/.style={regular polygon,fill=white,regular polygon sides=4,inner sep=2.5pt, draw}}
\tikzset{gauge/.style={circle, draw,inner sep=2.5pt}}
\tikzset{gaugeb/.style={circle, draw,fill=black,inner sep=2.5pt}}
\tikzset{gauger/.style={circle, draw,fill=cyan,inner sep=2.5pt}}
\tikzset{gaugeg/.style={circle, draw,fill=red,inner sep=2.5pt}}
\tikzset{SUd/.style={circle, draw=black, inner sep=0pt, fill=yellow, minimum size=2mm}}
\tikzset{bd/.style={circle, draw=black, inner sep=0pt, fill=black, minimum size=2mm}}
\tikzset{wd/.style={circle, draw=black, inner sep=0pt, fill=white, minimum size=2mm}}
\tikzset{Dynkin/.style={circle, draw=black, inner sep=0pt, fill=white, minimum size=2mm}}
\tikzstyle{ligne}=[draw, thick] 
\tikzset{doublearrow/.style={ draw=black!75, color=black!75, thick, double distance=3pt, }}

\numberwithin{equation}{section}  

\newtheorem{conjecture}{Conjecture}[section]

\newcommand{\be}{\begin{equation}}
\newcommand{\ee}{\end{equation}}
\newcommand{\ba}{\begin{aligned}}
\newcommand{\ea}{\end{aligned}}

\def\half{{\frac{1}{2}}}

\def\unit{{1\kern-.65ex {\rm l}}}
\def\1{{1\kern-.65ex {\rm l}}}

\def\CA{{\cal A}}
\def\CB{{\cal B}}

\def\CE{{\cal E}}

\def\CJ{{\cal J}}

\def\CM{{\cal M}}
\def\CN{{\cal N}}

\def\CS{{\cal S}}

\newcount\hour \newcount\minute
\hour=\time \divide \hour by 60
\minute=\time
\def\now{%
\ifnum \hour<13
  \ifnum \hour=0 \advance \hour by 12 \number\hour:\else \number\hour:\fi%
     \ifnum \minute<10 0\fi%
     \number\minute%
\ A.M.%
\else \advance \hour by -12 \number\hour:%
  \ifnum \minute<10 0\fi%
  \number\minute%
  \ P.M.%
\fi%
}

\makeatother

\def\mb{\mathbb}

\def\mc{\mathcal}
\def\bp{\begin{pmatrix}}
\def\ep{\end{pmatrix}}

\newcommand{\nn}{\nonumber}
\newcommand{\matarray}[1]{\begin{matrix} #1 \end{matrix}}

\newcommand{\bea}{\begin{equation} \begin{aligned}}
 \newcommand{\eea}{\end{aligned} \end{equation}}
\newcommand{\bit}{\begin{itemize}} 
\newcommand{\eit}{\end{itemize}} 

\newcommand{\Z}{\mathbb{Z}}
\newcommand{\C}{\mathbb{C}}
\newcommand{\R}{\mathbb{R}}
\newcommand{\Q}{\mathbb{Q}}

\renewcommand{\t}{\widetilde }

\renewcommand{\d}{\partial }

\newcommand{\m}{\mathfrak{m}}

\newcommand{\ov}{\over}

\newcommand{\h}{\widehat}

\newcommand{\MG}{{\mathbf X}} 

\newcommand{\FT}{{\mathcal{T}_\MG^{\rm 5d}}} 

\newcommand{\FTfour}{{\mathscr{T}_\MG^{\rm 4d}}} 

\newcommand{\EQfour}{{\text{EQ}^{(4)}}} 
\newcommand{\MQfour}{{\text{MQ}^{(4)}}} 
\newcommand{\MQfive}{{\text{MQ}^{(5)}}} 

\begin{document}

\baselineskip=18pt  
\numberwithin{equation}{section}  
\allowdisplaybreaks  


%
%


\thispagestyle{empty}

\vspace*{0.8cm} 
\begin{center}
{{\Huge  Coulomb and Higgs Branches from \\ 
\smallskip 
Canonical Singularities, Part 1: }\\

\bigskip {\LARGE
Hypersurfaces with Smooth Calabi-Yau Resolutions
}}

 \vspace*{1.2cm}
Cyril Closset$\,^\flat$,  Sakura Sch\"afer-Nameki$\,^\sharp$, Yi-Nan Wang$\,^{\natural}$\\

\vspace*{1.0cm}
{\it $^\flat$ School of Mathematics, University of Birmingham,\\
Watson Building, Edgbaston, Birmingham B15 2TT, UK}

\medskip
{\it $^{\sharp, \natural}$ Mathematical Institute, University of Oxford, \\
Andrew-Wiles Building,  Woodstock Road, Oxford, OX2 6GG, UK}\\

\medskip
{\it $^\natural$ School of Physics and State Key Laboratory of Nuclear Physics and Technology,\\
Peking University, Beijing 100871, China\\ }

\medskip
{\it  $^\natural$ Center for High Energy Physics, \\
Peking University, Beijing 100871, China}

\vspace*{0.8cm}
\end{center}
\vspace*{.2cm}

\noindent
Compactification of M-theory and of IIB string theory on threefold canonical singularities gives rise to superconformal field theories (SCFTs) in 5d and 4d, respectively. 
The resolutions and deformations of the singularities encode salient features of the  SCFTs and of their moduli spaces. In this paper, we build on Part 0 of this series \cite{Closset:2020scj} and further explore the physics of SCFTs arising from isolated hypersurface singularities.  We study in detail these canonical isolated hypersurface singularities that admit a smooth Calabi-Yau (crepant)  resolution. Their 5d and 4d physics is discussed and their 3d reduction and mirrors (the magnetic quivers) are determined in many cases.
As an explorative tool, we provide a {\tt Mathematica} code which computes key quantities for any canonical  isolated hypersurface singularity, including the 5d rank, the 4d Coulomb branch spectrum and central charges, higher-form symmetries in 4d and 5d, and crepant resolutions. 

\newpage

\tableofcontents


\section{Introduction}
\label{sec:introduction}
Supersymmetric quantum field theories (SQFT) can often be embedded within string theory by using either branes, geometry, or a mixture of both (distinct approaches are often related by string dualities). Over the years, string-theory techniques have led to a deeper understanding of the rich landscape of superconformal field theories (SCFT) in  space-time dimension $d >2$, and especially for $d>4$. These SCFTs can also be analysed using SQFT methods. A typical strategy is to analyse the low-energy effective field theories obtained after breaking conformal invariance (explicitly or spontaneously). Here, we are interested in the geometric engineering approach, wherein conformal field theories correspond to certain geometric singularities, and the massive phases correspond to smoothing the singularity. 

This paper is part of a series started in \cite{Closset:2020scj}, wherein we explore the geometric engineering of 4d and 5d SCFTs at threefold isolated hypersurface singularities (IHS)
\be
\MG = \{F(x)=0\; \;|\; \; x\in \C^4\} \,.
\ee
  We assume that $\MG$ is a canonical threefold singularity, which is essentially the Calabi-Yau condition. Consequently, Type II string theory or M-theory on this geometric background preserves 8 supercharges.%
 \footnote{This is a subtle point, since any discussion of supersymmetry implicitly assumes the existence of a Ricci-flat metric. Our discussion of the singularities will remain entirely algebraic, thus bypassing (and postponing) any deeper questions about the metric.} 
 Then, we expect the existence of two distinct maps from geometries to theories in 4d and 5d, respectively
 \bea
&& \text{IIB}  \;\; :&\quad \MG \mapsto \FTfour~,\;\qquad\qquad&
&&\text{M-theory} \;\; :& \quad \MG \mapsto \FT~.
 \eea
On the one hand, the infrared limit of Type IIB string theory on $\MG$ is expected to give us a 4d $\CN=2$ SCFT \cite{Shapere:1999xr,  Xie:2015rpa}. On the other hand, the infrared limit of M-theory on $\MG$ should similarly give rise to a 5d SCFT \cite{Seiberg:1996bd, Morrison:1996xf}. The 4d SCFT $\FTfour$ and the 5d SCFT $\FT$ are closely related upon compactification to 3d, as discussed in \cite{Closset:2020scj, Closset:2020afy, CSNWII}.

There are two types of smoothing of $\MG$ we may consider: deformations or crepant resolutions ({\it i.e.} resolutions which preserve the Calabi-Yau condition). This corresponds to turning on some complex structure moduli or some K\"ahler moduli, respectively. In Type IIB, families of deformation are `classical', in that they are not subject to worldsheet nor D-brane instanton corrections. The general family of deformed singularities, denoted by $\h \MG$, gives us the Seiberg-Witten geometry on the extended Coulomb branch of the 4d SCFT $\FTfour$. Similarly, families of crepant resolutions $\pi : \t\MG \rightarrow \MG$, denoted by $\t\MG$,  are often `classical' in M-theory, where they correspond to the real extended Coulomb branch of the 5d SCFT $\FT$. By contrast, deformations in M-theory and crepant resolutions in IIB are quantum-corrected, due to M2-brane instantons wrapping 3-cycles and to worldsheet, D1- and D3-brane instantons wrapping exceptional curves and divisors, respectively. These `quantum' geometries in M-theory and Type IIB correspond to going onto the Higgs branch  of $\FT$ or $\FTfour$, respectively. Note that this class of geometric engineering of SCFTs reverses the old QFT slogan about the Higgs branch being protected from quantum corrections (which is true for Lagrangian theories, but misleading otherwise): from the viewpoint of string theory, the Coulomb branch is `classical' and the Higgs branch is `quantum'.

Topologically, the crepant resolutions of canonical IHS are rather more subtle than their deformations. While the latter always have the homotopy type of a bouquet of 3-spheres, crepant resolutions of canonical IHS can have a rich topology, including compact 2-, 3- and 4-cycles. Moreover, crepant resolutions, even if they exist, are not always smooth.  We might have residual terminal singularities in the resolved threefold $\t \MG$, or it might happen that $\t\MG$ is smooth but the exceptional divisors themselves are not. In this paper, we will focus on isolated hypersurface singularities whose crepant resolutions are smooth. They form a small but important subset of all canonical IHS, to which we can apply a number of techniques to study the corresponding SCFT moduli spaces.

Part of the motivation for this paper is to explain some simple tools to study the key properties of deformed and resolved canonical hypersurface singularities (details which were ommited in~\cite{Closset:2020scj}). All the necessary computations can be easily implemented on a computer. We attached an ancillary {\tt Mathematica} notebook {\tt Basic-Data-IHS.nb} to the arXiv submission of this paper. The code takes as an input any  canonical isolated hypersurface singularity, whose classification \cite{yau2005classification, Xie:2015rpa, Davenport:2016ggc}
will be streamlined and updated in section \ref{sec:Classi} below. (See also \cite{Buican:2021xhs} for a related discussion.) We then provide two functions: 
\begin{enumerate}
\item {\tt BasicIHSdata}[$F(x_1, x_2, x_3, x_4), \{x_1, x_2, x_3, x_4\}$]: this outputs all the basic characteristics of the 4d and 5d SCFTs such as ranks, flavor ranks, dimensions of the Higgs branches, 4d Coulomb branch spectrum and central charges $a$ and $c$, and the higher form symmetries.
\item {\tt SmoothOperator}[$F(x_1, x_2, x_3, x_4), \{x_1, x_2, x_3, x_4\}$]: determines a resolution of the hypersurface singularity. \end{enumerate}
This computational tool will hopefully be useful to the practitioner.

We will then apply these methods to explore hypersurface singularities that admit a crepant resolution that is smooth (without residual terminal singularities). It is interesting to organise the discussion in terms of the 5d rank, $r$, which is the number of exceptional divisors in the crepant resolution. We scanned through a large database of canonical IHS, which included a large number theories of low $r$. While it appears that there exists an infinite number of canonical IHS at any $r$ (certainly, this is true at small $r$), we would like to put forward the following:

\paragraph{\bf Conjecture (1):}   For any $r>0$, there exists a finite number of canonical IHS whose crepant resolution is smooth.

\medskip
\noindent
We call such a singularity with a smooth resolution a `smoothable model'. In this paper, we list all smoothable models for $r\leq 4$ -- this list is complete to the best of our current knowledge.
 For $r\geq 5 $, the scan through our dataset becomes impractical,  but we can focus on a subset of `fully smooth models', which are the smoothable models with exceptional divisors which are themselves smooth. We can scan for  such models using the following:
 
 \paragraph{\bf Conjecture (2):}   For any smoothable model with smooth exceptional divisors, the associated 4d SCFT $\FTfour$ has an integral CB spectrum ({\it i.e.} the conformal dimensions of the CB operators satisfy $\Delta\in \Z$).

\medskip
\noindent
 This conjecture is based on observation in many examples. If true, this is a remarkable mathematical relation between the smoothness of the exceptional divisors and the spectrum of the deformed singularity. Assuming Conjecture (2), we can efficiently explore the space of all `fully smooth' models up to $r=10$, by first restricting to models with an integral 4d CB spectrum.

For all these smoothable models, we performed a detailed analysis of their 4d and 5d properties, including a determination of their magnetic quivers or quiverines whenever possible. The use of magnetic quivers (MQ)  for 5d SCFTs was initiated in \cite{Bourget:2019aer}, where they were motivated by brane-web constructions. The MQ associated to a 5d SCFT is a 3d $\mathcal{N}=4$ theory whose Coulomb branch is isomorphic to the Higgs branch of the 5d SCFT.  Magnetic quivers have been a very powerful tool to explore the moduli spaces of superconformal theories with 8 supercharges, with most approaches based on brane-webs \cite{Bourget:2020gzi, Bourget:2020asf, Bourget:2020xdz, vanBeest:2020kou, VanBeest:2020kxw, Akhond:2020vhc, Bourget:2021siw, Bourget:2021csg, Bourget:2021jwo,Akhond:2021knl, vanBeest:2021xyt, Nawata:2021nse, Sperling:2021fcf}. From a geometric perspective, the magnetic quivers were studied {\it e.g.}  in  \cite{Closset:2020scj, Closset:2020afy, Collinucci:2020kdm, Collinucci:2021ofd, DeMarco:2021try}.

Further studies of more generic canonical singularities and of their 4d and 5d interpretation, as well as a more in-depth discussion of the 3d electric and magnetic quiver(ines), will be presented in part 2 of this series  \cite{CSNWII}.

\medskip
\noindent
This paper is organised as follows. In section~\ref{sec:Summary}, we start with a lightning review of the geometry of isolated hypersurface singularities and of their relation to 5d and 4d SCFTs.  
 In section~\ref{sec:Link}, we discuss the topology of the link of the singularity, and its interpretation in terms of higher-form symmetries of the 5d and 4d SCFTs. 
  In section~\ref{sec:Class}, we review the classification of canonical IHS and explain how to use the ancillary {\tt Mathematica} code. 
 In section \ref{sec:Smooth}, the (fully) smoothable models of rank  $r\leq 10$ are discussed in detail, highlighting some examples with noteworthy properties, such as the models with 4d central charges satisfying $a>c$ and $a=c$, and a fully resolvable model that corresponds to the AD$[E_7, E_7]$ SCFT in 4d, for which we propose a 5d IR quiver gauge theory. An appendix contains details on other smoothable models, including the infinite series of type AD$[D_{2n}, D_{2n}]$.




\section{Hypersurface Singularities and SCFTs: A Practical Summary}
\label{sec:Summary}

In this section, we review key aspects of the geometry of canonical threefold hypersurface singularities, $\MG$, and of its relation to the SCFTs $\FTfour$ and $\FT$. 
By definition, a canonical singularity  \cite{reidCanonical3Folds}  admits a resolution $\pi: \t\MG \rightarrow \MG$ such that
\be\label{def can through res}
K_{\t\MG} = \pi^* K_{\MG} + \sum_k a_k S_k~,\qquad \qquad S_k \subset \pi^\ast(0)~,
\ee
with $a_k \in \R$ such that $a_k\geq 0$, $\forall k$. Here, $K_X$ denotes the canonical divisor of $X$, and $S_k$ denote the exceptional divisors. A canonical singularity is called terminal if $a_k>0$, $\forall k$. 

In this paper, we focus on canonical singularities that admit a (complete) crepant resolution -- that is, $a_k=0$, $\forall k$.%
\footnote{In previous works \protect\cite{Closset:2020scj, Closset:2020afy}, we studied canonical singularities without crepant resolution, which are either terminal singularities or singularities that admit a {\it partial} crepant resolution ($a_k=0$ for some but not all $k$).} 
 We  also focus exclusively on quasi-homogeneous isolated hypersurface singularities (IHS) in $\C^4$,
\be
\MG \, \cong \, \big\{ F(x)=0\; \big|\; x\in \C^4 \big\}~,
\ee
which are defined by a single quasi-homogeneous  polynomial with a unique critical point at the origin -- $F \in \C[x_1, x_2, x_3, x_4]$ such that $d F(x) = F(x)=0$ if and only if $x=0$.

\subsection{Deformation and 4d Coulomb Branch Spectrum}
 Consider the quasi-homogeneous IHS $\MG$, with the $\C^\ast$-action:
\be\label{def F weights}
F(\lambda^q x) = \lambda \, F(x)~, \qquad \text{with} \qquad x_i \rightarrow \lambda^{q_i} x_i~, \qquad\qquad i=1, \cdots, 4~,
\ee
for some  scaling weights $q_i \in \Q_{>0}$.  Let us recall key properties of the deformation of such singularities \cite{Arnold1975}, and its interpretation in terms of the Coulomb branch of the 4d SCFT $\FTfour$.

\paragraph{Canonical singularity condition.} The quasi-homogeneous IHS $\MG$ is canonical if and only if the weights $q_i$ are such that:
\be
\sum_{i=1}^4  q_i  >1~.
\ee
If this inequality is not satisfied, $\MG$ is worse than canonical.
This condition can also be written in terms of the reduced central charge \cite{Katz:1997eq}:
\be
\h c\equiv \sum_{i=1}^4 (1-2q_i) <2~.
\ee
Here, $\h c$ is the reduced central charge of the 2d $\CN=(2,2)$ LG model with four fields $x_i$ and superpotential $W=F(x)$.

\paragraph{Milnor ring and multiplicity.} A deformation of an hypersurface singularity $\MG\cong \{F=0\}$ is a (smooth) hypersurface of the generic form:
\be
\h F(x)= F(x)+\sum_{l=1}^\mu t_l x^{\frak{m}_l}=0~,
\ee
where $\{x^{\frak{m}_l}\}_{l=1}^\mu$ is a basis for the so-called Milnor ring of the singularity:
\be
\CM(F) ={\C[x_1, x_2, x_3, x_4]/\CJ}~, \qquad\quad \CJ\equiv \left(\d_{x_1} F, \cdots, \d_{x_4} F \right)~.
\ee
The integer
\be
\mu \equiv {\rm dim}\;  \CM(F)~,
\ee
is the multiplicity of the singularity (also known as the Milnor number). For quasi-homogeneous singularities, $\CM(F)$ has a basis in terms of $\mu$ distinct monomials:
\be\label{xm monomials}
x^{\m_l} \equiv x_1^{\m_{l,1}} \cdots x_4^{\m_{l,4}}~, \qquad l=1, \cdots, \mu~,
\ee
with the integers $\m_{l, i}\geq 0$. These monomials can be determined for any $F$ by standard Gr\"obner basis algorithms implemented {\it e.g.} in \textsc{Singular}. Two hypersurface singularities $\{F_1=0\}$ and $\{ F_2=0\}$ are biholomorphically equivalent if and only if their Milnor rings are isomorphic \cite{Yau1982}.
 The multiplicity of $\MG$ is related to the scaling weights according to
\be\label{formula for mu quasihom}
\mu = \prod_{i=1}^4 \left({1\ov q_i}-1\right)~.
\ee

\paragraph{The spectrum of $\MG$.} The monomials \eqref{xm monomials} are assigned weights $Q_l \in \Q$ according to
\be
Q_l =q(\m_l) \equiv  \sum_{i=1}^4    q_i \m_{l, i}~.
\ee
Let us (partially) order the monomials by their weights, with the lowest weight being $Q_1= 0$ for $x^{\m_1}=1$.  Let us also introduce the shifted weights
\be\label{spectral nbr def}
\ell_l =  Q_l + \sum_{i=1}^4 q_i -1~,
\ee
called the spectral numbers.
These rational numbers are valued in  the open interval $(0,2)$.
 The so-called {\it spectrum} of the quasi-homogeneous singularity is then given by the ordered set:  
\be\label{spectrum def}
\CS_\MG= \{ \ell_l\}_{l=1}^\mu~, \qquad \ell_1 \leq \ell_{2} \leq \cdots \leq \ell_\mu~.
\ee
The ordered monomials $x^{\m_l}$ and $x^{\m_{\mu-l+1}}$ of $\CM(F)$ form pairs such that
\be
\ell_l + \ell_{\mu-l+1} = 2  \qquad \text{if} \qquad  \ell_l  < 1~\,.
\ee
There can also be unpaired monomials such that $\ell_l=1$.
The spectral numbers define a grading of the Milnor ring
\be
\CM(F) = \bigoplus_{\ell} \CM(F)_\ell~,.
\ee
There is also a simple formula for the Poincar\'e series of the graded ring
\be
\label{PMF}
P_{\CM(F)}(q) \equiv \sum_\ell {\rm dim} \CM(F)_\ell \; q^\ell = {1\ov q} \prod_{i=1}^4 {q^{q_i} -q \ov 1- q^{q_i}}~,
\ee
which reduces to \eqref{formula for mu quasihom} in the limit $q\rightarrow 1$. The Poincar\'e series gives an efficient way of computing the spectrum for quasi-homogeneous hypersurface singularities, bypassing the need for finding an explicit basis of $\CM(F)$. We denote by $\h\MG$ a (generic) deformation of $\MG$.

Another important number that characterizes the singularity $\MG$ is its modality, $m(\MG)$, which is defined as the number of deformations of $\MG$ which preserve its multiplicity. For quasi-homogeneous singularities, this is given by
\be\label{modality def}
m(\MG) = \#\big\{ x^{\m_l} \; \big| \; \ell_l \geq  \sum_{i=1}^4 q_i \big\}~.
\ee
The corresponding deformation parameters $t_l$ are the `moduli' of the singularity. 

\paragraph{The Coulomb branch spectrum of $\FTfour$.} In the Type IIB engineering, the family of deformations $\{\h\MG_t\}$ is interpreted as the Seiberg-Witten geometry of the 4d SCFT $\FTfour$~\cite{Shapere:1999xr, Xie:2015rpa}. The spectrum of $\MG$ is equivalent to the spectrum of (extended) Coulomb branch (CB) dimensions of $\FTfour$, as follows. To each monomial $x^{\m_l}$, one assigns the CB dimension
\be
\Delta_l \equiv \Delta(t_l)=\frac{1-Q_l}{\sum_{i=1}^4 q_i-1}=\frac{\sum_{i=1}^4 q_i- \ell_l}{\sum_{i=1}^4 q_i-1}~,
\ee
which are the scaling dimensions  of the deformation parameters $t_l$ divided by $\sum_i q_i-1$.
There are three classes of deformations, depending on $\Delta_l$
\begin{itemize}
\item $\Delta_l>1$: these geometric deformations are interpreted as VEVs of CB operators of conformal dimensions $\Delta_l$ and  $U(1)_{\rm r}$ charge ${\rm r}= 2 \Delta_l$
\be\label{VEV CB 4d}
t_l = \left\langle (\CE_{{\rm r}, (0,0)})_l\right\rangle~.
\ee
We denote by $\h r$ the number of such operators.

\item $\Delta_l=1$:  these geometric deformations  are interpreted as the mass deformations of $\FTfour$
\be\label{def f from spec}
\delta S =   t_l \int d^4 x\int d^4\theta  (\h B_1)_l~,\qquad l=1, \cdots, f~.
\ee
They correspond to conserved currents,  {\it i.e.} flavor symmetries. We may equivalently view the deformation parameters as VEVs for the scalars $\phi_l$ in the  $U(1)$ background vectors multiplets spanning a maximal torus of the flavour symmetry group
\be\label{mass VEVs}
t_l = \langle \phi_l\rangle \equiv m_l~.
\ee
In particular, $f$ is the rank of the flavor symmetry group of $\FTfour$.

\item $\Delta_l<1$: these geometric deformations correspond to the $\h r$ monomials  which are paired with the ones with $\Delta_{\mu-l}>1$, such that $\Delta_l+ \Delta_{\mu-l}=2$. They are interpreted as chiral deformations of the SCFT by the CB operators themselves
\be
\delta S = t_l \int d^4 x\int d^4\theta  (\CE_{\mathrm{r}, (0,0)})_{\mu-l}~.
\ee
In particular, if $\Delta_l=0$ ($\Delta_{\mu-l}=2$), this is a marginal deformation of the SCFT -- indeed, this corresponds to deforming the IHS equation by a monomial with $Q_l=1$, thus preserving the scaling symmetry, which is necessary for conformal invariance. Note that we can also have deformations with $0<\Delta_l<1$, corresponding to  relevant deformations of the SCFT ($1<\Delta_{\mu-l} <2$), or else deformations with 
$\Delta_l<0$, corresponding to irrelevant deformations ($\Delta_{\mu-l}>2$). The relevant deformations only exists if there are generators of the Milnor ring with scaling dimensions in the small range
\be
{\h c\ov 2}<Q_{\mu-l} < 1~.
\ee
Note also that the modality \eqref{modality def} is the total number of marginal and irrelevant operators of the 4d SCFT; thus, the `moduli' of the singularity are precisely those deformations that do not trigger an RG flow away from the fixed point $\FTfour$.

\end{itemize}

\noindent
Let us denote by $\CB_{\h r}$ the Coulomb branch spanned by \eqref{VEV CB 4d}. In particular, $\h r$ is the rank of the SCFT $\FTfour$. We define the extended Coulomb branch $\CM_C[\FTfour]$ as the total space of the Coulomb branch fibered over the mass parameters \eqref{mass VEVs}, namely
\be
\CB_{\h r} \rightarrow \CM_C[\FTfour] \rightarrow \{m\}~,
\ee
The extended CB  obviously has complex dimension $\h r +f$. Note also the relation
\be
\mu= 2 \h r+f~.
\ee
The multiplicity $\mu$ gives the dimension of the charge lattice $\Gamma$ of BPS particles on the CB, which arise as D3-branes wrapping the compact 3-cycles of the deformed hypersurface
\be
\Gamma \cong H_3(\h\MG, \Z)\cong \Z^{\mu}~.
\ee
Let us also denote by $\CM_{3,3}$ the intersection form on the third homology
\be
(\CM_{3,3})_{\alpha\beta}= S^3_\alpha \cdot S^3_\beta~,
\ee
where the 3-cycles $\{S^3_\alpha\}_{\alpha=1}^\mu$ give some basis of $H_3(\h\MG, \Z)$.

\paragraph{Central charges.} From the Coulomb branch spectrum, we can compute the conformal anomaly coefficients $a$ and $c$ according to  \cite{Shapere:2008zf, Xie:2015rpa}
\be
a= {R(A)\ov 4} + {R(B)\ov 6} + {5 \h r \ov 24}~, \qquad\qquad
c= {R(B)\ov 3} + {\h r\ov 6}~.
\ee
Here, we defined
\be
R(A) = \sum_{\Delta_l >1}( \Delta_l -1)~, \qquad R(B)={\mu \ov 4 (\sum_{i=1}^4 q_i -1)}= {1\ov 4} \mu \Delta_{\rm max}~,
\ee
with $ \Delta_{\rm max}\equiv \Delta_{1}$ the largest CB operator dimension. Here, we also used the fact that there is no hypermultiplets on the Coulomb branch of $\FTfour$ because $H_2(\h \MG,\Z)=0$ \cite{Xie:2015rpa}.
From $a$ and $c$, we may also compute the `effective number' of hypermultiplets and of vector multiplets of the SCFT
\be
n_h=-16a+20c~, \qquad n_v=8a-4c\,.
\ee
For Lagrangian SCFTs, this gives the actual number of matter fields and gauge fields, respectively. For further discussions of the spectrum of 4d SCFTs in related contexts, see {\it e.g.}  \cite{Wang:2016yha, Caorsi:2018zsq, Argyres:2020wmq, Cecotti:2021ouq}.

\subsection{Crepant Resolution and 5d Coulomb Branch}
\label{sec:resolution}

For any canonical threefold singularity, there exists a crepant resolution $\pi:\t \MG\rightarrow \MG$, such that the resolved singularity $\t \MG$ has at worst terminal singularities~\cite{reid1983minimal,reid1985young}. 
In M-theory, the resolution $\t\MG$ corresponds to probing the extended Coulomb branch of the 5d SCFT $\FT$ \cite{Intriligator:1997pq}. (See also {\it e.g.} \cite{Jefferson:2017ahm, Closset:2018bjz, Closset:2019juk, Saxena:2020ltf, Closset:2021lhd} for some further discussions.)

\paragraph{Topology of the crepant resolution $\t\MG$.} 
We will call `a crepant resolution' any (partial) resolution $\pi : \t\MG\rightarrow \MG$ that introduces $r$ exceptional divisors with $a_k=0$ in \eqref{def can through res}. This is excepted to preserves 5d $\CN=1$ (or 4d $\CN=2$) supersymmetry. A smooth (or `complete') crepant resolution, corresponding to $a_k=0$, $\forall k$, in  \eqref{def can through res}, is a crepant resolution that is smooth. 

In general, there may be many distinct crepant resolutions of $\MG$, but some of their key topological properties are invariant.  The set of residual terminal singularities is such an invariant, and so are the Betti numbers
\bea
\label{betti Res}
&b_1(\t \MG)=0~, \qquad\quad
&&b_2(\t \MG)= r+f~, \cr
&b_3(\t \MG)= b_3~, \qquad\quad
&&b_4(\t \MG) =r~.
\eea
Here, $r$ is the number of exceptional divisors in the crepant resolution, and $f= \rho(\MG)$ is the rank of the divisor class group of the singularity $\MG$. We can also understand $f$ as the number of 2-cycles of $\t\MG$ that are dual to non-compact divisors. They correspond to $U(1)$ flavor symmetries of the 5d SCFT $\FT$. Note also that the crepant resolution $\t\MG$ generally has three-cycles, as we will explain momentarily. 

There is a remarkable relation between $f$ as defined here and the integer $f$ defined in \eqref{def f from spec} from the spectrum of the singularity. It is a non-trivial mathematical fact that the two quantities agree \cite{caibar1999minimal}
\be\label{feqrho}
f=\rho(\MG)=\#\{\ell_l \,|\, \ell_l=1\}~.
\ee
Physically, this can be interpreted in terms of generalised geometric transitions, wherein $f$ 2-cycles in the resolution $\t\MG$ become $f$ 3-cycles in the deformation $\h\MG$ \cite{Closset:2020scj}.%
\footnote{The proper mathematical understanding of this relation involves discussing the mixed Hodge structure of the singularity, which we will review in more detail in \protect\cite{CSNWII}.} A third way to compute $f$ is from the five-dimensional link of the singularity, as we will review in section~\ref{sec:Link}.

\paragraph{Newton polytope and weightings.} For any quasi-homogeneous IHS, the Betti numbers \eqref{betti Res} can be computed in the following way~\cite{caibar1999minimal}. Let us define the Newton polytope $\mc{N}(F)$ associated to $\{F(x)=0\}$ as the polytope in the lattice of monomials $M\cong\mb{Z}^4$
\be
\mc{N}(F)=\{(\m_{n,1},\m_{n,2},\m_{n,3},\m_{n,4})\in M=\mb{Z}^4\, |\, \m_n\in F\}\,.
\ee
Here,  $\m_n\in F$ means that the monomial $x^{\m_n}$ enters the polynomial $F(x)$ with non-zero coefficient.  
Let us also define the weightings $\alpha$ as the maps $\alpha: M\rightarrow \mb{Z}$, which can be viewed as vectors $(\alpha_1,\alpha_2,\alpha_3,\alpha_4)$ in the dual lattice $N\cong {\rm Hom}(M, \Z)$. The evaluation map on a monomial $\m$ reads
\be
\alpha(\m)=\sum_{i=1}^4 \alpha_i\m_i~,
\ee
and the evaluation map on a polynomial $F$ is defined as
\be
\alpha(F)=\mathrm{min}_{\m\in F}\,  \alpha(\m)\,.
\ee
We then define the sets
\be
W_+(F)=\left\{\alpha\in N^+\; |\;\,\forall \m\in\mc{N}(F)\ ,\ \alpha(\m-(1,1,1,1))\geq -1\right\}\,.
\ee
and
\be
W(F)=W_+(F)\cup\{(1,0,0,0)\ ,(0,1,0,0)\ ,(0,0,1,0)\ ,(0,0,0,1)\}\,.
\ee
$N^+$ is the open cone of $N$ satisfying $\alpha_i>0$ $(i=1,\dots,4)$. We can then define a non-compact toric fourfold $T_\Sigma$ with toric fan $\Sigma$. The elements of $W(F)$ are in one-to-one correspondence with the 1d cone of $\Sigma$,  and one must choose a set of 2d, 3d and 4d cones of $\Sigma$, such that $\Sigma$ is maximally triangulated. The resolved threefold $\t \MG$ is then exactly the anti-canonical hypersurface $-K(T_\Sigma)$ of $T_\Sigma$,  and each triangulation corresponds to a different (toric) resolution of $\MG$. Finally, each element of $W_+(F)$ corresponds to a compact exceptional divisor $S_i\subset \t\MG$, which may be reducible. 

\paragraph{Computation of $r$ and $b_3$.} The topological numbers $r$ and $b_3$ are independent of the choice of resolution, and they can computed as follows \cite{caibar1999minimal, Caibarb3} (analogously to the Batyrev formula~\cite{Batyrev:1994hm}). Consider the dual polytope $\Gamma$ of $W(F)$, defined as
\be
\Gamma=\left\{\m\in M^+\; |\;\,\forall \alpha\in W(F)\ ,\ \alpha(\m-(1,1,1,1))\geq -1\right\}\,.
\ee
The dual face $\Gamma_\alpha\subset \Gamma$ of $\alpha\in W(F)$, defined as
\be
\Gamma_\alpha=\{\m\in\Gamma|\alpha(\m-(1,1,1,1))=-1\}
\ee
can have different dimensions, denoted by $\mathrm{dim}(\Gamma_\alpha)$. The explicit formula for $r$ is then
\be
r=\sum_{\alpha\in W_+(F)}\begin{cases}
0 & \text{if}\; \mathrm{dim}(\Gamma_\alpha)=0~,\\
 l'(\Gamma_\alpha)+1\qquad \qquad&  \text{if}\; \mathrm{dim}(\Gamma_\alpha)=1~,\\
1\quad & \text{if}\;  \mathrm{dim}(\Gamma_\alpha)> 1~,
\end{cases}
\ee
where $l'(\Gamma_\alpha)$ is number of interior lattice points in $\Gamma_\alpha$. When $\mathrm{dim}(\Gamma_\alpha)=1$ and $l'(\Gamma_\alpha)>0$, the compact divisor corresponding to $\alpha$ has $l'(\Gamma_\alpha)+1$ irreducible components.

To compute $b_3$, the number of 3-cycles in the crepant resolution, we first note that each compact divisor associated to the weighting $\alpha\in W_+(F)$ has the structure of $\mb{P}^1$ fibration over a complex curve with genus $g_C(\alpha)$. The numbers $g_C(\alpha)$ are computed as
\be
g_C(\alpha)= \begin{cases}
l'(\Gamma_\alpha)\qquad &\text{if}\; \mathrm{dim}(\Gamma_\alpha)=2~,\\
0\quad &\text{if}\; \mathrm{dim}(\Gamma_\alpha)\neq 2~.
\end{cases}.
\ee
The formula for $b_3$ is then
\be
b_3=2\sum_{\alpha\in W_+(F)} g_C(\alpha)\,.
\ee
In other words, there is a one-to-one correspondence between the 3-cycles and the 1-cycles at the base of the ruled exceptional divisors. Note that the resolved threefold $\t\MG$ is itself simply-connected.

\paragraph{Resolution sequence.} Finally, let us discus the resolution sequence itself. A toric resolution sequence is defined as follows. For any IHS, we start with the non-compact ambient space $\mb{C}^4$ with coordinates $(x_1,x_2,x_3,x_4)$, with the set of weightings $\{(1,0,0,0)\ ,$ $(0,1,0,0)\ ,$ $(0,0,1,0)\ ,$ $(0,0,0,1)\}$. Each step in the resolution sequence is a weighted blow-up of the ambient space at the locus $y_1=\dots=y_k=0$, where $y_i=0$ is the toric divisor $S_i^\Sigma$ in the ambient space corresponding to the weighting $\alpha_i$. It takes one of the following forms \cite{caibar1999minimal,Lawrie:2012gg,Apruzzi:2019opn}:

\begin{enumerate}
\item{$(y_1^{(a_1)},y_2^{(a_2)},y_3^{(a_3)},y_4^{(a_4)};\delta)$: weighted blow-up of the locus $y_1=y_2=y_3=y_4=0$ with weight $(a_1,a_2,a_3,a_4)$. Here, the only possible weights are $(3,2,1,1)$, $(2,1,1,1)$ and $(1,1,1,1)$. The exceptional divisor $\{\delta=0\}$ in the ambient space is associated to the weighting
\be
\alpha(\delta)=\sum_{i=1}^4 a_i \alpha(y_i)\,.
\ee
In the hypersurface equation, one replaces
\be
y_1\rightarrow y_1\delta^{a_1}~, \;\;  \;
 y_2\rightarrow y_2\delta^{a_2}~, \;\; \;
 \ y_3\rightarrow y_3\delta^{a_3}~, \;\;\;  y_4\rightarrow y_4\delta^{a_4}~,
\ee
and then divide the equation by $\delta^{a_1+a_2+a_3+a_4-1}$. There is a new SR ideal generator $y_1 y_2 y_3 y_4$ after the resolution.
}
\item{$(y_1,y_2,y_3;\delta)$: blow up the locus $y_1=y_2=y_3=0$. The exceptional divisor $\delta=0$ in the ambient space is associated to the weighting
\be
\alpha(\delta)=\alpha(y_1)+\alpha(y_2)+\alpha(y_3)\,.
\ee
In the hypersurface equation, one replaces
\be
y_1\rightarrow y_1\delta~, \;\;  \;
 y_2\rightarrow y_2\delta~, \;\;  \;
  y_3\rightarrow y_3\delta~,
\ee
and then divide the equation by $\delta^2$. There is a new SR ideal generator $y_1 y_2 y_3$ after the resolution.
}
\item{$(y_1,y_2;\delta)$: blow up the locus $y_1=y_2=0$. The exceptional divisor $\delta=0$ in the ambient space is associated to the weighting
\be
\alpha(\delta)=\alpha(y_1)+\alpha(y_2)\,.
\ee
In the hypersurface equation, one replaces
\be
y_1\rightarrow y_1\delta~, \;\;  \;  y_2\rightarrow y_2\delta~,
\ee
and then divide the equation by $\delta$. There is a new SR ideal generator $y_1 y_2$ after the resolution.
}
\end{enumerate}
\noindent
After the resolution sequence, one can also compute the topological quantities associated to each exceptional divisor. In particular, we can compute the intersection numbers of the exceptional divisors $S_i$ inside $\t\MG$, according to
\be\label{SSSdata}
S_i\cdot S_j\cdot S_k|_{\t\MG}=(-K_\Sigma)\cdot S_i^\Sigma\cdot S_j^\Sigma\cdot S_k^\Sigma\,.
\ee
This determines the prepotential on the 5d CB of $\FT$~\cite{Intriligator:1997pq}.


\section{Link Topology and Higher-form Symmetries}
\label{sec:Link}

The five-manifold $L_5$ at the boundary of a non-compact Calabi-Yau three-fold encodes global symmetries of the field theories, both in 4d and 5d.  In particular, the torsion part of $H_\ast(L_5, \Z)$ encodes the higher-form symmetries \cite{Gaiotto:2014kfa} of both the 5d SCFT $\FT$ and the 4d SCFT $\FTfour$ \cite{Morrison:2020ool,Albertini:2020mdx, 
DelZotto:2020esg, Bhardwaj:2020phs, Closset:2020scj}. For isolated hypersurface singularities, $L_5$ is simply connected, and the 5d field theory $\FT$ does not have any 1-form symmetries. On the other hand, we often have  non-zero torsion in $H_2(L_5, \Z)$, which can be computed from the weights $q_i$ of the quasi-homogeneous polynomial $F(x)$  \cite{RANDELL1975347, 10.1007/BFb0070047}.  In M-theory, this translates into 3-form symmetries of $\FT$ that act on M5-branes wrapping relative 3-cycles (and/or its dual 0-form symmetry acting on wrapped M2-branes). In Type IIB string theory, this torsion group is related to the 1-form symmetry of the 4d SCFT $\FTfour$, with the charged line operators arising from wrapped D3-branes. 

In the rest of this section, we first review an explicit formula for the link homology, and we then briefly discuss its 5d and 4d interpretation. 

\subsection{Topology of the Link}
\label{subsec:LinkTopo}

The link of the singularity $\MG$ is the $5$-dimensional compact manifold $L_5(\MG)$ obtained by intersecting the hypersurface with a seven-sphere $S^7_\varepsilon$ centered at the origin of $\C^4$
 \be
L_{5}(\MG)= \Big\{ x \in \C^4 \; \Bigg|\;  F(x)=0~, \quad \sum_{i=1}^4 |x_i|^2 = \varepsilon \Big\}~,
\ee 
for a small radius $\varepsilon>0$. For quasi-homogeneous singularities, we may pick any value of $\varepsilon$ due to the scaling symmetry. Therefore, by sending $\varepsilon\rightarrow \infty$, we see that the link is also the {\it boundary manifold} of $\MG$ and of any local desingularization of the singularity; in particular, for any smooth deformation $\h \MG$ of $\MG$, we   have $\d \h\MG \cong L_{5}(\MG)$. The five-manifold $L_5$ is connected and simply-connected, and its homology groups are
\bea\label{link hom}
H_0(L_5, \Z)& \,\cong& H_5(L_5, \Z)\cong \Z~, \qquad
H_2(L_5, \Z) &\,\cong &\Z^f &\oplus  \mathfrak{t}_2 \cr
H_1(L_5, \Z) &\,\cong &H_4(L_5, \Z)\cong  0~,  \qquad 
H_3(L_5, \Z)&\,\cong &\Z^f &\,,
\eea
where 
\be 
\mathfrak{t}_2 = \text{Tor}\, H_2(L_5, \Z)
\ee
denotes the torsion part of the second homology. 
This leads to our third definition of the `flavor rank' $f$, as the Betti number of the link
\be
f= b_2(L_5) = b_3(L_5) \,.
\ee
This gives us an intuitive way of understanding the `generalised geometric transition' mentioned below \eqref{feqrho}, generalising the conifold transition~\cite{Candelas:1989js}: as we resolve or deform the singularity $\MG$, we have $f$ exceptional 2-cycles that survive `at the tip' inside $\t\MG$, or $f$ 3-cycles  that survive inside $\h\MG$, respectively.

\paragraph{The  homology of $L_5(\MG)$: explicit formulas.} The non-trivial quantities $f$ and $ \mathfrak{t}_2$ in \eqref{link hom} can be computed from the scaling weights $q_i$ in \eqref{def F weights} as follows. Let us define the quantities
\be
v_i\equiv  {1\ov q_i}~,\quad \qquad a_i \equiv {\rm Numerator}(v_i)~.
\ee
Let us consider the set $I=\{1, 2, 3, 4\}$, and let $I_s$ denote any particular subset $I_s \subset I$ of $s$ elements. 
Then, we have the explicit formula \cite{RANDELL1975347, 10.1007/BFb0070047}
\be
f= b_2(L_5)=\sum_{s=0}^4 \sum_{I_s \subset I} (-1)^{4-s} {\prod_{j\in I_s} v_j \ov {\rm LCM}_{j\in I_s}(a_j)}~.
\ee
The sum includes all subsets $I_s$ of $I$, including the trivial subset $I_0 =\{\}$, which contributes $+1$. More generally, we have the following formula for the torsion part, $ \mathfrak{t}_2$ \cite{10.1007/BFb0070047}.  For every $I_s \subset I$,  we define
\be
\kappa(I_s)=\sum_{t=0}^s \sum_{I_t \subset I_s} (-1)^{s-t}{ \prod_{i\in I_t} v_i \ov {\rm LCM}_{j\in I_t}(a_j)}~,
\ee
and
\be
k(I_s)=\begin{cases}
  \kappa(I_s)\quad & \text{if}\;  4-s  \in 2\Z+1~, \\ 0 & \text{if}\;  4-s  \in 2\Z~. \end{cases}
\ee
We then define a set of integers, $C(I_s)$, inductively, starting with
\be
C(I_0)= C(\{\})= \gcd_{i\in I}(a_i)= \gcd(a_1, \cdots, a_4)~,
\ee
and
\be
C(I_s)= {\gcd_{i\in I-I_s}(a_i)\ov  \prod_{I_t \subsetneq I_s} C(I_t)}~.
\ee
Then, for the integers $j>1$, we define
\be
d_j= \prod_{I_s | k(I_s)\geq j} C(I_s)~, \qquad j=1, \cdots, J~, \qquad \qquad J \equiv \max_{I_s\subset I} k(I_s)~.
\ee
We then have:
\be\label{f and t2 general}
\mathfrak{t}_2 \cong  \Z_{d_1} \oplus \cdots \oplus \Z_{d_J} \,,\qquad
f=\kappa(I)\,.
\ee

\paragraph{The torsion group $\mathfrak{t}_2$ from the deformation $\h\MG$.} Using a long-exact sequence for the (relative) homology groups of the deformed threefold $\h\MG$, as reviewed in the next subsection, one can show that the torsion group $ \mathfrak{t}_2$ is also isomorphic to the cokernel of the map $\CM_{3,3}: \mb{Z}^\mu \rightarrow \mb{Z}^\mu$ given by the intersection form on $H_3(\h\MG, \Z)$, namely
\be
 \mathfrak{t}_2\cong \Z^{\mu}/ \CM_{3,3}\Z^\mu~.
\ee
 We computed this cokernel in many examples, allowing us to check the general formula \eqref{f and t2 general} for $\mathfrak{t}_2$, which appeared as a conjecture in \cite{10.1007/BFb0070047}, on a case-by-case basis.%
\footnote{A proof of \eqref{f and t2 general} is given in \protect\cite{RANDELL1975347} for Type I singularities and in \protect\cite{10.2307/1998598} for Type XIX singularities, in the notation of section~\protect\ref{sec:Classi} below. As far as we are aware, the general case remains a conjecture.}

\subsection{Defects and Higher-Form Symmetries in 5d from M-theory}
\label{sec:HFS}

Consider the canonical singularity $\MG$, any desingularisation $\MG_6$ of $\MG$ (which will be either the deformation $\h\MG$ or the crepant resolution $\t\MG$), and the boundary five-manifold $L_5\cong \d\MG_6$. M-theory on $\MG_6$ provides us with a natural set of defect operators of the 5d low-energy theory, from membranes wrapping  non-compact $k$-cycles. The wrapped M2- and M5-branes give rise to defect operators of dimension $q=k-3$ and $q=6-k$, respectively, which can be charged under a non-trivial $q$-form symmetry. The sets of such charged `electric' and `magnetic' defect operators of dimension $q$ are denoted by
\be
\Gamma^{(q)}_{\rm M2}=\mathfrak{h}_{(k=3-q)}~,\qquad\quad
\Gamma^{(q)}_{\rm M5}=\mathfrak{h}_{(k=6-q)}~,
\ee
respectively. They are computed as
\be\label{hdef}
\mathfrak{h}_{(k)}\equiv {\rm Tor}\big( H_{k}(\MG_6 , L_5,\mathbb{Z}) /\text{im}(f_{k})\big)~,
\ee
where $f_{k}$ is the map of $f_{k} : H_k(\MG_6,\Z)\rightarrow H_k(\MG_6,L_5,\Z)$ that sends compact $k$-cycles to relative $k$-cycles. It sits inside a long exact sequence for the relative homology (with integral coefficients)
\be\label{longexactseq}
\cdots \rightarrow\ 
H_{k} (L_5) \stackrel{h_{k}}{\longrightarrow} 
H_{k} (\MG_6) \stackrel{f_{k}}{\longrightarrow} 
H_{k} (\MG_6, L_5) \stackrel{g_{k}}{\longrightarrow} 
H_{k-1} (L_5) \stackrel{h_{k-1}}{\longrightarrow}
H_{k-1} (\MG_6) \rightarrow \cdots \,.
\ee
The objects counted by $\mathfrak{h}_{(k)}$ are the defects whose charge cannot be screened by the dynamical particles that arise from membranes wrapping compact $k$-cycles of $\MG_6$. For $q \leq 1$, for definiteness, let us define the `defect groups' of the 5d theory, in its massive phase determined by $\MG_6$
\be\label{def Dq}
{\bf D}_\FT^{(q)} \cong  \Gamma^{(q)}_{\rm M2}\oplus \Gamma^{(3-q)}_{\rm M5}~.
\ee
The defect operators in $\Gamma^{(q)}_{\rm M2}$ and $\Gamma^{(3-q)}_{\rm M5}$ can have a non-trivial Dirac pairing given by the intersection of the relative $(3-k)$- and $(3+k)$-cycles inside $\MG_6$. The higher-form symmetry of the 5d theory is determined after one chooses a polarisation of the defect group \eqref{def Dq}, 
\be
\Lambda^{(q)}_\FT \subset {\bf D}_\FT^{(q)}~,
\ee
by fixing a maximal set of mutually commuting defects  \cite{Aharony:2013hda, DelZotto:2015isa, Garcia-Etxebarria:2019cnb, Morrison:2020ool, DelZotto:2020esg, Bhardwaj:2021pfz, Bhardwaj:2021zrt, Bhardwaj:2020phs}. The higher-form symmetry group itself is given by the Pontryagin dual group
\be
\h \Lambda^{(q)}_\FT = \text{Hom}\big(\Lambda^{(q)}_\FT, U(1)\big)~.
\ee
Note that it is generally a mixture of $q$- and $(3-q)$-form symmetries. We may also choose, for simplicity, the `purely electric' or the `purely magnetic' polarisation, consisting in picking one of the two summands in \eqref{def Dq}.

Given a general local Calabi-Yau threefold in M-theory, there are two defects groups of interest
\be
{\bf D}_\FT^{(0)} \cong  \mathfrak{h}_{(3)}\oplus  \mathfrak{h}_{(3)}~, \qquad\qquad
{\bf D}_\FT^{(1)} \cong  \mathfrak{h}_{(2)}\oplus  \mathfrak{h}_{(4)}~,
\ee
corresponding to the `0-form symmetry' and the `1-form symmetry', respectively, allowing for a slight abuse of language. Using the exact sequence \eqref{longexactseq}, we can write the groups $\mathfrak{h}_{(k)}$ in terms of the homology of the boundary:
\be\label{hk expand}
\mathfrak{h}_{(k)}={\rm Tor}\big( \text{im} (g_{k}) \big)={\rm Tor}\big( \text{ker} (h_{k-1})\big) \subset  {\rm Tor}\, H_{k-1}(L_5)~. 
\ee
The charged defect operators wrap relative $k$-cycles $\Sigma_k$ with non-trivial boundary on $L_5$, $[\d\Sigma_k]\in {\rm Tor}\, H_{k-1}(L_5)$. For $L_5$ any connected, orientable closed manifold, there are two independent torsion groups in its homology
\be
\mathfrak{t}_1\equiv {\rm Tor}\, H_2 (L_5, \mathbb{Z}) \cong {\rm Tor}\, H_3 (L_5, \mathbb{Z})~, \qquad \qquad
\mathfrak{t}_2\equiv {\rm Tor}\, H_2 (L_5, \mathbb{Z})~.
\ee
Let us now see how these torsion groups are related to higher-form symmetries in 5d.

\paragraph{1-form symmetry and Coulomb-branch defects.} For the sake of generality, let us first discuss the possibility of a non-trivial $1$-form symmetry in 5d \cite{Morrison:2020ool, DelZotto:2020esg, Bhardwaj:2021zrt}.  We will assume that the smooth threefold $\MG_6$ is simply-connected and that it has no torsion in homology, which is the case for all examples we have studied (such as toric threefolds or hypersurface singularities). It directly follows from \eqref{longexactseq} and \eqref{hk expand} that the `electric' part of the defect group is isomorphic to the the torsion part of the first homology group of the boundary. 
\be
\mathfrak{h}_{(2)} \cong  \mathfrak{t}_1~.
\ee
Alternatively, we can use the definition \eqref{hdef} of $\mathfrak{h}_{(2)}$ as the cokernel of the map $f_2$, which gives
\be
\mathfrak{h}_{(2)}   \cong {\rm Tor}\big( \Z^{b_4}/\CM_{4,2}\Z^{b_2}\big)~,
\ee
where $b_j$ denote the Betti numbers of $\MG_6$. Similarly, we have
\be
\mathfrak{h}_{(4)}   \cong {\rm Tor}\big( \Z^{b_2}/\CM_{2,4}\Z^{b_4}\big) \cong \mathfrak{h}_{(2)} ~,
\ee
The 1-form (and/or 2-form) symmetry is then determined by the intersection pairing between the compact 2-cycles and 4-cycles of $\MG_6$. The charged defect operators naturally live on the Coulomb branch of the 5d SCFT, in which case $\MG_6=\t \MG$, $b_2=r+f$ and $b_4=f$. For the 5d SCFTs $\FT$ of interest in this paper, which arise at hypersurface singularities, the link manifold $L_5$ is simply connected, and therefore we find that $\mathfrak{h}_{(2)} =\mathfrak{h}_{(4)} =0$.

\paragraph{0- and 3-form symmetry and Higgs-branch defects.}  The 0-form/3-form symmetries of $\FT$ act on defect operators that are more naturally defined on the Higgs branch of $\FT$. Let us then take $\MG_6=\h\MG$ a deformed singularity, and assume that $H_2(\h\MG)=0$ for simplicity, which is the case, in particular, for the generic deformation of an hypersurface singularity. We then find that
\be
\mathfrak{h}_{(3)}    \cong  \mathfrak{t}_2 \cong {\rm Tor}\big( \Z^{b_3}/\CM_{3,3}\Z^{b_3}\big)~,
\ee
with $b_3=\mu=2\h r+f$ for an IHS. It would be interesting to have a more detailed understanding of these charged HB defect operators from the 5d field-theory point of view. This is left as a challenge for future work.

\subsection{Defects and Higher-Form Symmetries in 4d from Type IIB}
In the Type IIB compactification on $\MG_6$, the natural set of charged defect operators of the 4d field theory arise from wrapped D-branes. We may define the defect group
\be\label{defect D1D3D5}
{\bf D}_\FTfour^{(0)} \cong
\Gamma^{(0)}_{\rm D1}\oplus  \Gamma^{(2)}_{\rm D5}\oplus  
\Gamma^{(2)}_{\rm D3}\oplus  \Gamma^{(0)}_{\rm D3}
\cong  \mathfrak{h}_{(2)} \oplus \mathfrak{h}_{(4)} \oplus \mathfrak{h}_{(2)} \oplus \mathfrak{h}_{(4)}~,
\ee
corresponding to D1-, D5- and D3-branes wrapping 2- and 4-cycles, as well as the defect group
\be\label{D31form}
{\bf D}_\FTfour^{(1)} \cong  \Gamma^{(1)}_{\rm D3}\cong  \mathfrak{h}_{(3)}\cong  \mathfrak{t}_2~,
\ee
from D3-brane wrapping relative 3-cycles inside $\MG_6$. Since our focus is on hypersurface singularities, we will not consider further the interesting possibilities suggested by \eqref{defect D1D3D5}. In other words, our 4d theories $\FTfour$ will have no defects charged under non-trivial 0-form or 2-form symmetries.  On the other hand, the defect group~\eqref{D31form} is generally non-trivial, and it encodes all the (`electric' and `magnetic') line operators of $\FTfour$ \cite{Closset:2020scj, DelZotto:2020esg} -- see also \cite{Bhardwaj:2021pfz, Hosseini:2021ged, Bhardwaj:2021zrt, Buican:2021xhs} for related discussions. It is interesting to note that, for 4d SCFTs $\FTfour$ engineered at IHS, non-trivial 1-form symmetries can only arise if the SCFT is non-isolated, as shown in \cite{Buican:2021xhs}.%
\footnote{This claim was checked systematically in \protect\cite{Buican:2021xhs}, and it is easily checked with the `experimental' methods of this paper. In particular, in our finite database of models discussed in section~\ref{sec:Smooth}, 56$\%$ of the models are isolated (no marginal deformations) and all those isolated models have $\mathfrak{t}_2=0$. Only 15$\%$ of all models (amounting to 35$\%$ of the models with marginal deformations) have a non-trivial one-form symmetry. 
}

The actual one-form symmetry $\h \Lambda^{(1)}_\FTfour$ of $\FTfour$ is obtained upon choosing a polarisation
\be
\Lambda^{(1)}_\FTfour \subset {\bf D}_\FTfour^{(1)}~,\qquad \quad
\h \Lambda^{(1)}_\FTfour  = \text{Hom}\big(\Lambda^{(1)}_\FTfour, U(1)\big)~.
\ee
For our hypersurface singularities, the torsion in the second homology of the link always takes the form:
\be
\mathfrak{t}_2\cong \mathfrak{f}\oplus  \mathfrak{f}~,
\ee
so that there exists a `purely electric' choice of polarisation for which the one-form symmetry group is isomorphic to $\mathfrak{f}$ \cite{Closset:2020scj}.

\section{Classification of Canonical IHS and {\tt Mathematica} Code}
\label{sec:Class} 

In this section, we review the complete classification of canonical isolated singularities, which is equivalent to the Kreuzer-Skarke classification of 2d $\CN=(2,2)$ Landau-Ginzburg (LG) superpotentials  \cite{Kreuzer:1992bi}. This 2d perspective clarifies some aspects of the discussion in  \cite{yau2005classification, Xie:2015rpa}.

We then present an ancillary {\tt Mathematica}~code \cite{Mathematica} (available with the arXiv version of this paper) which implements the computations explained in the previous sections for any such singularity.  The code computes the spectrum of $\MG$, the basic topological data of its crepant resolution $\t\MG$, as well as the homology of the link $L_5$. This directly gives all the `basic data' associated to the 5d SCFT $\FT$ and the 4d SCFT $\FTfour$. The code also implements the crepant resolution procedure explained in section~\ref{sec:resolution}.

\subsection{Classification of Canonical Isolated Hypersurface Singularities  in $\C^4$}
\label{sec:Classi}

Consider a quasi-homogeneous IHS in $\C^4$, defined by the polynomial
\be\label{poly F}
F(x_1, x_2, x_3, x_4)\in \C[x_1, x_2, x_3, x_4]~,
\ee
with the scaling weights $q_i$ as in \eqref{def F weights}. 
Such quasi-homogeneous polynomials $F(x)$ were classified by Kreuzer and Skarke (KS) many years ago \cite{Kreuzer:1992bi}.  A classification of isolated hypersurface threefold singularities was also given by Yau and Yu (YY)  \cite{yau2005classification}, which was used by Xie and Yau in their studies of geometrically-engineered 4d $\CN=2$ SCFTs \cite{Xie:2015rpa}. The YY classification is slightly incomplete, however  \cite{Davenport:2016ggc}. 

It is very useful to think of the polynomial \eqref{poly F} as defining a 2d Landau-Ginzburg (LG) model with four chiral superfields $x_i$  and a superpotential $W= F(x)$ \cite{Shapere:1999xr}. Any quasi-homogeneous IHS then defines a 2d $\CN=(2,2)$ SCFT as well as the 4d SCFT $\FTfour$, with the two perspectives related by a 4d/2d correspondence  \cite{Cecotti:2010fi}, provided that the reduced central charge of the 2d theory satisfies \cite{Lerche:1989uy, Gukov:1999ya, Shapere:1999xr}:
\be\label{hat c can cond}
\h c \equiv \sum_{i=1}^4 (1-2 q_i) < 2~.
\ee
In fact, the IHS is canonical if and only if \eqref{hat c can cond} holds \cite{reidCanonical3Folds, Markushevich1986}. 
Such LG models were recently classified by Davenport and Melnikov \cite{Davenport:2016ggc}.%
\footnote{Note that~\protect\cite{Davenport:2016ggc} also considered LG models with five fields such that $\h c<2$. We only consider four fields.} 
Let us review their result, while keeping the YY notation for ease of comparison. One finds that the YY classification needs to be refined, by introducing `subtypes' for some of the 19 `types'. Since this classification is essentially equivalent the KS classification of quasi-homogeneous polynomials, we will call it the KS-YY classification.

\begin{table}[htbp]
$$
\begin{array}{|c|c|c|c|c|c|c||c|c|c|c|c|c|}
\hline
\multicolumn{2}{|c|}{\text{Type}} & F(x_1, x_2, x_3, x_4)   & \text{Skeleton and links}  &  \text{Skeleton} \cr \hline \hline
 \text{I} & \{1,1\} &\; { \footnotesize x_1^a +x_2^b+x_3^c+ x_4^d} \; &
 \begin{tikzpicture}[baseline=1mm]
\node[] (1) [label = above: {\scriptsize$a$}]{$\circ$};
\node[] (2) [right = of 1, label = above: {\scriptsize$b$}]{$\circ$};
\node[] (3) [right = of 2, label = above: {\scriptsize$c$}]{$\circ$};
\node[] (4) [right = of 3, label = above: {\scriptsize$d$}]{$\circ$};
\end{tikzpicture}
 & (S_{1,1})^{\oplus 4}
\\ \hline\hline
 \text{II} & \{2,1\} &\;   x_1^a +x_2^b+x_3^c+ x_3 x_4^d \; &
 \begin{tikzpicture}[baseline=1mm]
\node[] (1) [label = above: {\scriptsize$a$}]{$\circ$};
\node[] (2) [right = of 1, label = above: {\scriptsize$b$}]{$\circ$};
\node[] (3) [right = of 2, label = above: {\scriptsize$c$}]{$\circ$};
\node[] (4) [right = of 3, label = above: {\scriptsize$d$}]{$\circ$};
\draw[->] (4) -- (3);
\end{tikzpicture}
  &    (S_{1,1})^{\oplus 2} \oplus  S_{2,1} 
\\ \hline\hline
  \text{III} & \{3,1\} &\;   x_1^a +x_2^b+x_3^c x_4+ x_3 x_4^d \; &
 \begin{tikzpicture}[baseline=1mm]
\node[] (1) [label = above: {\scriptsize$a$}]{$\circ$};
\node[] (2) [right = of 1, label = above: {\scriptsize$b$}]{$\circ$};
\node[] (3) [right = of 2, label = above: {\scriptsize$c$}]{$\circ$};
\node[] (4) [right = of 3, label = above: {\scriptsize$d$}]{$\circ$};
\draw[->] (4) to [out = 145, in = 35]  (3);\draw[->] (3)  to [out = -35, in = 215]  (4);
\end{tikzpicture}
  & (S_{1,1})^{\oplus 2}\oplus  S_{2,2} 
\\ \hline\hline 
 \text{IV} & \{4,1\} &\;   x_1^a +x_1 x_2^b+x_3^c+ x_3 x_4^d \; &
 \begin{tikzpicture}[baseline=1mm]
\node[] (1) [label = above: {\scriptsize$a$}]{$\circ$};
\node[] (2) [right = of 1, label = above: {\scriptsize$b$}]{$\circ$};
\node[] (3) [right = of 2, label = above: {\scriptsize$c$}]{$\circ$};
\node[] (4) [right = of 3, label = above: {\scriptsize$d$}]{$\circ$};
\draw[->] (2) -- (1);
\draw[->] (4) -- (3);
\end{tikzpicture}
  & S_{2,1}\oplus  S_{2,1}
\\ \hline\hline
  \text{V} & \{5,1\} &\;   x_1^a x_2 +x_1 x_2^b+x_3^c + x_3 x_4^d \; &
 \begin{tikzpicture}[baseline=1mm]
\node[] (1) [label = above: {\scriptsize$a$}]{$\circ$};
\node[] (2) [right = of 1, label = above: {\scriptsize$b$}]{$\circ$};
\node[] (3) [right = of 2, label = above: {\scriptsize$c$}]{$\circ$};
\node[] (4) [right = of 3, label = above: {\scriptsize$d$}]{$\circ$};
\draw[->] (2) to [out = 145, in = 35]  (1);\draw[->] (1)  to [out = -35, in = 215]  (2);
\draw[->] (4) -- (3);
\end{tikzpicture}
  & S_{2,2} \oplus  S_{2,1} 
\\ \hline\hline 
  \text{VI}&  \{6,1\} &\;   x_1^a x_2 +x_1 x_2^b+x_3^c x_4+ x_3 x_4^d \; &
 \begin{tikzpicture}[baseline=1mm]
\node[] (1) [label = above: {\scriptsize$a$}]{$\circ$};
\node[] (2) [right = of 1, label = above: {\scriptsize$b$}]{$\circ$};
\node[] (3) [right = of 2, label = above: {\scriptsize$c$}]{$\circ$};
\node[] (4) [right = of 3, label = above: {\scriptsize$d$}]{$\circ$};
\draw[->] (2) to [out = 145, in = 35]  (1);\draw[->] (1)  to [out = -35, in = 215]  (2);
\draw[->] (4) to [out = 145, in = 35]  (3);\draw[->] (3)  to [out = -35, in = 215]  (4);
\end{tikzpicture}
  & S_{2,2} \oplus  S_{2,2} 
\\ \hline\hline 
 \text{VII}& \{7,1\} &\;   x_1^a + x_2^b+x_2 x_3^c+ x_3 x_4^d \; &
 \begin{tikzpicture}[baseline=1mm]
\node[] (1) [label = above: {\scriptsize$a$}]{$\circ$};
\node[] (2) [right = of 1, label = above: {\scriptsize$b$}]{$\circ$};
\node[] (3) [right = of 2, label = above: {\scriptsize$c$}]{$\circ$};
\node[] (4) [right = of 3, label = above: {\scriptsize$d$}]{$\circ$};
\draw[->] (3) -- (2);
\draw[->] (4) -- (3);
\end{tikzpicture}
  & S_{1,1}\oplus  S_{3,1}
\\ \hline\hline
 \text{VIII}_1& \{8,1\}&\; \matarray{  x_1^a + x_2^b+x_2 x_3^c+ x_2 x_4^d \\+ x_3^p x_4^q }&
 \begin{tikzpicture}[baseline=1mm]
\node[] (1) [label = above: {\scriptsize$a$}]{$\circ$};
\node[] (2) [right = of 1, label = above: {\scriptsize$b$}]{$\circ$};
\node[] (3) [right = of 2, label = above: {\scriptsize$c$}]{$\circ$};
\node[] (4) [right = of 3, label = above: {\scriptsize$d$}]{$\circ$};
\draw[->] (3) -- (2);
\draw[dashed] (3)  to [out = -25, in = 205]  (4);
\draw[->] (4) to [out = 145, in = 35] (2);
\end{tikzpicture}
  & S_{1,1}\oplus  S_{3,2}
\\ \hline
 \text{VIII}_2& \{8,2\}&\;   \matarray{x_1^a + x_2^b+x_2 x_3^c+ x_2 x_4^d \\+x_1 x_3^p x_4^q } &
 \begin{tikzpicture}[baseline=1mm]
\node[] (1) [label = above: {\scriptsize$a$}]{$\circ$};
\node[] (2) [right = of 1, label = above: {\scriptsize$b$}]{$\circ$};
\node[] (3) [right = of 2, label = above: {\scriptsize$c$}]{$\circ$};
\node[] (4) [right = of 3, label = above: {\scriptsize$d$}]{$\circ$};
\draw[->] (3) -- (2);
\draw[dashed] (3)  to [out = -25, in = 205] node[] (L1) {$\bullet$}  (4);
\draw[->] (4) to [out = 145, in = 35] (2);
\draw[dashed] (1)  to [out = -25, in = 195]   (L1);
\end{tikzpicture}
  & S_{1,1}\oplus  S_{3,2}
\\ \hline\hline
 \text{IX}_1 &\{9,1\}&\;   \matarray{x_1^a + x_2^b x_4 +x_3^c x_4+ x_2 x_4^d \\+x_2^p x_3^q } &
 \begin{tikzpicture}[baseline=1mm]
\node[] (1) [label = above: {\scriptsize$a$}]{$\circ$};
\node[] (2) [right = of 1, label = above: {\scriptsize$b$}]{$\circ$};
\node[] (3) [right = of 2, label = above: {\scriptsize$c$}]{$\circ$};
\node[] (4) [right = of 3, label = above: {\scriptsize$d$}]{$\circ$};
\draw[->] (3) -- (4);
\draw[->] (2)  to [out = -25, in = 205]  (4);
\draw[->] (4) to [out = 145, in = 35] (2);
\draw[dashed] (2)  to (3);
\end{tikzpicture}
  & S_{1,1}\oplus  S_{3,4}
\\ \hline 
 \text{IX}_2& \{9,2\}&\;  \matarray{ x_1^a + x_2^b x_4 +x_3^c x_4+ x_2 x_4^d \\+x_1 x_2^p x_3^q} &
 \begin{tikzpicture}[baseline=1mm]
\node[] (1) [label = above: {\scriptsize$a$}]{$\circ$};
\node[] (2) [right = of 1, label = above: {\scriptsize$b$}]{$\circ$};
\node[] (3) [right = of 2, label = above: {\scriptsize$c$}]{$\circ$};
\node[] (4) [right = of 3, label = above: {\scriptsize$d$}]{$\circ$};
\draw[->] (3) -- (4);
\draw[->] (2)  to [out = -25, in = 205]  (4);
\draw[->] (4) to [out = 145, in = 35] (2);
\draw[dashed] (2)  to node[] (L1) {$\bullet$}  (3);
\draw[dashed] (1)  to [out = -25, in = 205]   (L1);
\end{tikzpicture}
  & S_{1,1}\oplus  S_{3,4}
\\ \hline\hline
 \text{X}& \{10,1\}&\;   x_1^a + x_2^b x_3 +x_3^c x_4+ x_2 x_4^d  \; &
 \begin{tikzpicture}[baseline=1mm]
\node[] (1) [label = above: {\scriptsize$a$}]{$\circ$};
\node[] (2) [right = of 1, label = above: {\scriptsize$b$}]{$\circ$};
\node[] (3) [right = of 2, label = above: {\scriptsize$c$}]{$\circ$};
\node[] (4) [right = of 3, label = above: {\scriptsize$d$}]{$\circ$};
\draw[->] (2) -- (3);
\draw[->] (3) -- (4);
\draw[->] (4) to [out = 145, in = 35] (2);
\draw[dashed] (2)  to (3);
\end{tikzpicture}
  & S_{1,1}\oplus  S_{3,3}
\\ \hline \hline
 \text{XI}& \{11,1\}&\;   x_1^a + x_1 x_2^b +x_2 x_3^c + x_3 x_4^d  \; &
 \begin{tikzpicture}[baseline=1mm]
\node[] (1) [label = above: {\scriptsize$a$}]{$\circ$};
\node[] (2) [right = of 1, label = above: {\scriptsize$b$}]{$\circ$};
\node[] (3) [right = of 2, label = above: {\scriptsize$c$}]{$\circ$};
\node[] (4) [right = of 3, label = above: {\scriptsize$d$}]{$\circ$};
\draw[->] (4) -- (3);
\draw[->] (3) -- (2);
\draw[->] (2) -- (1);
\end{tikzpicture}
  & S_{4,1}
\\ \hline \hline
 \text{XII}_1& \{12,1\}&\;   \matarray{x_1^a + x_1 x_2^b +x_1 x_3^c + x_2 x_4^d \\+ x_2^p x_3^q  } &
 \begin{tikzpicture}[baseline=1mm]
\node[] (1) [label = above: {\scriptsize$a$}]{$\circ$};
\node[] (2) [right = of 1, label = above: {\scriptsize$b$}]{$\circ$};
\node[] (3) [right = of 2, label = above: {\scriptsize$c$}]{$\circ$};
\node[] (4) [right = of 3, label = above: {\scriptsize$d$}]{$\circ$};
\draw[->] (2) -- (1);
\draw[->] (3) to [out = 145, in = 35] (1);
\draw[->] (4) to [out = 145, in = 35] (2);
\draw[dashed] (2)  to (3);
\end{tikzpicture}
  & S_{4,2}
\\ \hline 
 \text{XII}_2& \{12,2\}&\;   \matarray{x_1^a + x_1 x_2^b +x_1 x_3^c + x_2 x_4^d\\ + x_2^p x_3^q x_4  } &
 \begin{tikzpicture}[baseline=1mm]
\node[] (1) [label = above: {\scriptsize$a$}]{$\circ$};
\node[] (2) [right = of 1, label = above: {\scriptsize$b$}]{$\circ$};
\node[] (3) [right = of 2, label = above: {\scriptsize$c$}]{$\circ$};
\node[] (4) [right = of 3, label = above: {\scriptsize$d$}]{$\circ$};
\draw[->] (2) -- (1);
\draw[->] (3) to [out = 145, in = 35] (1);
\draw[->] (4) to [out = 145, in = 35] (2);
\draw[dashed] (2)  to  node[] (L1) {$\bullet$}  (3);
\draw[dashed] (L1)  to [out = -25, in = 205]   (4);
\end{tikzpicture}
  & S_{4,2}
\\ \hline 
 \end{array}
$$
\caption{The KS-YY classification of isolated hypersurface singularities in $\C^4$: Types I to XIII.
 Note that types I--VI exhaust the possibilities using skeleton with up to $n=2$ fields, while the cases I--X exhaust the possibilities using skeletons with up to $n=3$ fields.
 \label{tab:YYClassification 1/3}}
\end{table}

\begin{table}
$$
\begin{array}{|c|c|c|c|c|c|c||c|c|c|c|c|c|}
\hline
\multicolumn{2}{|c|}{\text{Type}} & F(x_1, x_2, x_3, x_4)  & \text{Skeleton and links}  &  \text{Skeleton} \cr \hline \hline
 \text{XIII}_1& \{13,1\}&\; \matarray{ x_1^a + x_1 x_2^b + x_2 x_3^c + x_2 x_4^d\\ + x_3^p x_4^q   } &
 \begin{tikzpicture}[baseline=1mm]
\node[] (1) [label = above: {\scriptsize$a$}]{$\circ$};
\node[] (2) [right = of 1, label = above: {\scriptsize$b$}]{$\circ$};
\node[] (3) [right = of 2, label = above: {\scriptsize$c$}]{$\circ$};
\node[] (4) [right = of 3, label = above: {\scriptsize$d$}]{$\circ$};
\draw[->] (2) -- (1);
\draw[->] (3) -- (2);
\draw[->] (4) to [out = 145, in = 35] (2);
\draw[->] (4) to [out = 145, in = 35] (2);
\draw[dashed] (3)  to   (4);
\end{tikzpicture}
  & S_{4,3}
\\ \hline 
 \text{XIII}_2& \{13,2\}&\; x_1^a + x_1 x_2^b + x_2 x_3^c + x_2 x_4^d+x_1  x_3^p x_4^q    \; &
 \begin{tikzpicture}[baseline=1mm]
\node[] (1) [label = above: {\scriptsize$a$}]{$\circ$};
\node[] (2) [right = of 1, label = above: {\scriptsize$b$}]{$\circ$};
\node[] (3) [right = of 2, label = above: {\scriptsize$c$}]{$\circ$};
\node[] (4) [right = of 3, label = above: {\scriptsize$d$}]{$\circ$};
\draw[->] (2) -- (1);
\draw[->] (3) -- (2);
\draw[->] (4) to [out = 145, in = 35] (2);
\draw[->] (4) to [out = 145, in = 35] (2);
\draw[dashed] (3)  to   node[] (L1) {$\bullet$} (4);
\draw[dashed] (L1)  to [out = 205, in = -25]   (1);
\end{tikzpicture}
  & S_{4,3}
\\ \hline \hline
 \text{XIV}_1& \{14,1\}&\; \matarray{x_1^a + x_1 x_2^b + x_1 x_3^c + x_1 x_4^d\\
 + x_2^p x_3^q + x_3^r x_4^s + x_2^u x_4^v  }  \; &
 \begin{tikzpicture}[baseline=1mm]
\node[] (1) [label = above: {\scriptsize$a$}]{$\circ$};
\node[] (2) [right = of 1, label = above: {\scriptsize$b$}]{$\circ$};
\node[] (3) [right = of 2, label = above: {\scriptsize$c$}]{$\circ$};
\node[] (4) [right = of 3, label = above: {\scriptsize$d$}]{$\circ$};
\draw[->] (2) -- (1);
\draw[->] (3) to [out = 145, in = 35] (1);
\draw[->] (4) to [out = 145, in = 35] (1);
\draw[dashed] (2)  to (3);
\draw[dashed] (3)  to (4);
\draw[dashed] (4)  to [out = 205, in = -25]   (2);
\end{tikzpicture}
  & S_{4,4}
\\ \hline
 \text{XIV}_2 & \{14,2\} &\; \matarray{x_1^a + x_1 x_2^b + x_1 x_3^c + x_1 x_4^d\\
 + x_2^p x_3^q x_4 + x_3^r x_4^s + x_2^u x_4^v  }    \; &
 \begin{tikzpicture}[baseline=1mm]
\node[] (1) [label = above: {\scriptsize$a$}]{$\circ$};
\node[] (2) [right = of 1, label = above: {\scriptsize$b$}]{$\circ$};
\node[] (3) [right = of 2, label = above: {\scriptsize$c$}]{$\circ$};
\node[] (4) [right = of 3, label = above: {\scriptsize$d$}]{$\circ$};
\draw[->] (2) -- (1);
\draw[->] (3) to [out = 145, in = 35] (1);
\draw[->] (4) to [out = 145, in = 35] (1);
\draw[dashed] (2)  to  node[] (L1) {$\bullet$}  (3);
\draw[dashed] (L1)  to [out = 25, in = 155]   (4);
\draw[dashed] (3)  to (4);
\draw[dashed] (4)  to [out = 205, in = -25]   (2);
\end{tikzpicture}
  & S_{4,4}
\\ \hline
 \text{XIV}_3 & \{14,3\} &\; \matarray{x_1^a + x_1 x_2^b + x_1 x_3^c + x_1 x_4^d\\
 + x_2^p x_3^q x_4 +x_2  x_3^r x_4^s + x_2^u x_4^v  }    \; &
 \begin{tikzpicture}[baseline=1mm]
\node[] (1) [label = above: {\scriptsize$a$}]{$\circ$};
\node[] (2) [right = of 1, label = above: {\scriptsize$b$}]{$\circ$};
\node[] (3) [right = of 2, label = above: {\scriptsize$c$}]{$\circ$};
\node[] (4) [right = of 3, label = above: {\scriptsize$d$}]{$\circ$};
\draw[->] (2) -- (1);
\draw[->] (3) to [out = 145, in = 35] (1);
\draw[->] (4) to [out = 145, in = 35] (1);
\draw[dashed] (2)  to  node[] (L1) {$\bullet$}  (3);
\draw[dashed] (L1)  to [out = 25, in = 155]   (4);
\draw[dashed] (3)  to node[] (L2) {$\bullet$}  (4);
\draw[dashed] (2)  to [out = 25, in = 155]   (L2);
\draw[dashed] (4)  to [out = 205, in = -25]   (2);
\end{tikzpicture}
  & S_{4,4}
\\ \hline 
 \text{XIV}_4 & \{14,4\} &\; \matarray{x_1^a + x_1 x_2^b + x_1 x_3^c + x_1 x_4^d\\
 + x_2^p x_3^q x_4 +x_2  x_3^r x_4^s + x_2^u x_3 x_4^v  }    \; &
 \begin{tikzpicture}[baseline=1mm]
\node[] (1) [label = above: {\scriptsize$a$}]{$\circ$};
\node[] (2) [right = of 1, label = above: {\scriptsize$b$}]{$\circ$};
\node[] (3) [right = of 2, label = above: {\scriptsize$c$}]{$\circ$};
\node[] (4) [right = of 3, label = above: {\scriptsize$d$}]{$\circ$};
\draw[->] (2) -- (1);
\draw[->] (3) to [out = 145, in = 35] (1);
\draw[->] (4) to [out = 145, in = 35] (1);
\draw[dashed] (2)  to  node[] (L1) {$\bullet$}  (3);
\draw[dashed] (L1)  to [out = 25, in = 155]   (4);
\draw[dashed] (3)  to node[] (L2) {$\bullet$}  (4);
\draw[dashed] (2)  to [out = 25, in = 155]   (L2);
\draw[dashed] (4)  to [out = 210, in = -30] node[] (L3) {$\bullet$}  (2);
\draw[dashed] (3)  to   (L3);
\end{tikzpicture}
  & S_{4,4}
\\ \hline \hline
 \text{XV}_1 & \{15,1\} &\; x_1^a x_2 + x_1 x_2^b + x_1 x_3^c + x_3 x_4^d
 + x_2^p x_3^q    \; &
 \begin{tikzpicture}[baseline=1mm]
\node[] (1) [label = above: {\scriptsize$a$}]{$\circ$};
\node[] (2) [right = of 1, label = above: {\scriptsize$b$}]{$\circ$};
\node[] (3) [right = of 2, label = above: {\scriptsize$c$}]{$\circ$};
\node[] (4) [right = of 3, label = above: {\scriptsize$d$}]{$\circ$};
\draw[->] (2) to [out = 145, in = 35]  (1);\draw[->] (1)  to [out = -35, in = 215]  (2);
\draw[->] (4) -- (3);
\draw[->] (3) to [out = 145, in = 35] (1);
\draw[dashed] (2)  to   (3);
\end{tikzpicture}
  & S_{4,6}
\\ \hline  
 \text{XV}_2 & \{15,2\} &\; x_1^a x_2 + x_1 x_2^b + x_1 x_3^c + x_3 x_4^d
 + x_2^p x_3^q x_4    \; &
 \begin{tikzpicture}[baseline=1mm]
\node[] (1) [label = above: {\scriptsize$a$}]{$\circ$};
\node[] (2) [right = of 1, label = above: {\scriptsize$b$}]{$\circ$};
\node[] (3) [right = of 2, label = above: {\scriptsize$c$}]{$\circ$};
\node[] (4) [right = of 3, label = above: {\scriptsize$d$}]{$\circ$};
\draw[->] (2) to [out = 145, in = 35]  (1);\draw[->] (1)  to [out = -35, in = 215]  (2);
\draw[->] (4) -- (3);
\draw[->] (3) to [out = 145, in = 35] (1);
\draw[dashed] (2)  to  node[] (L1) {$\bullet$}  (3);
\draw[dashed] (4)  to [out = 205, in = -25]   (L1);
\end{tikzpicture}
  & S_{4,6}
\\ \hline
 \end{array}
$$
\caption{The KS-YY classification of isolated hypersurface singularities in $\C^4$: Types XIII to XVI. \label{tab:YYClassification 2/3}}
\end{table}

\begin{table}
$$
\begin{array}{|c|c|c|c|c|c|c||c|c|c|c|c|c|}
\hline
\multicolumn{2}{|c|}{\text{Type}} & F(x_1, x_2, x_3, x_4)  & \text{Skeleton and links}  &  \text{Skeleton} \cr \hline \hline
 \text{XVI}_1 & \{16,1\} &\; \matarray{x_1^a x_2 + x_1 x_2^b + x_1 x_3^c + x_1 x_4^d\\
 +x_2^p x_3^q+ x_3^r x_4^s+ x_2^u x_4^v }     \; &
 \begin{tikzpicture}[baseline=1mm]
\node[] (1) [label = above: {\scriptsize$a$}]{$\circ$};
\node[] (2) [right = of 1, label = above: {\scriptsize$b$}]{$\circ$};
\node[] (3) [right = of 2, label = above: {\scriptsize$c$}]{$\circ$};
\node[] (4) [right = of 3, label = above: {\scriptsize$d$}]{$\circ$};
\draw[->] (2) to [out = 145, in = 35]  (1);\draw[->] (1)  to [out = -35, in = 215]  (2);
\draw[->] (3) to [out = 145, in = 35] (1);
\draw[->] (4) to [out = 145, in = 35] (1);
\draw[dashed] (2)  to (3);
\draw[dashed] (3)  to (4);
\draw[dashed] (4)  to [out = 205, in = -25]   (2);
\end{tikzpicture}
  & S_{4,9}
\\ \hline 
 \text{XVI}_2 & \{16,2\} &\; \matarray{x_1^a x_2 + x_1 x_2^b + x_1 x_3^c + x_1 x_4^d\\
 +x_2^p x_3^q x_4+ x_3^r x_4^s + x_2^u x_4^v }     \; &
 \begin{tikzpicture}[baseline=1mm]
\node[] (1) [label = above: {\scriptsize$a$}]{$\circ$};
\node[] (2) [right = of 1, label = above: {\scriptsize$b$}]{$\circ$};
\node[] (3) [right = of 2, label = above: {\scriptsize$c$}]{$\circ$};
\node[] (4) [right = of 3, label = above: {\scriptsize$d$}]{$\circ$};
\draw[->] (2) to [out = 145, in = 35]  (1);\draw[->] (1)  to [out = -35, in = 215]  (2);
\draw[->] (3) to [out = 145, in = 35] (1);
\draw[->] (4) to [out = 145, in = 35] (1);
\draw[dashed] (2)  to node[] (L1) {$\bullet$}  (3);
\draw[dashed] (3)  to (4);
\draw[dashed] (4)  to [out = 205, in = -25]   (2);
\draw[dashed] (L1)  to [out = 25, in = 155]   (4);
\end{tikzpicture}
  & S_{4,9}
\\ \hline
 \text{XVI}_3 & \{16,3\} &\; \matarray{x_1^a x_2 + x_1 x_2^b + x_1 x_3^c + x_1 x_4^d\\
 +x_2^p x_3^q x_4+ x_3^r x_4^s + x_2^u x_3 x_4^v }     \; &
 \begin{tikzpicture}[baseline=1mm]
\node[] (1) [label = above: {\scriptsize$a$}]{$\circ$};
\node[] (2) [right = of 1, label = above: {\scriptsize$b$}]{$\circ$};
\node[] (3) [right = of 2, label = above: {\scriptsize$c$}]{$\circ$};
\node[] (4) [right = of 3, label = above: {\scriptsize$d$}]{$\circ$};
\draw[->] (2) to [out = 145, in = 35]  (1);\draw[->] (1)  to [out = -35, in = 215]  (2);
\draw[->] (3) to [out = 145, in = 35] (1);
\draw[->] (4) to [out = 145, in = 35] (1);
\draw[dashed] (2)  to node[] (L1) {$\bullet$}  (3);
\draw[dashed] (3)  to (4);
\draw[dashed] (L1)  to [out = 25, in = 155]   (4);
\draw[dashed] (4)  to [out = 210, in = -30] node[] (L3) {$\bullet$}  (2);
\draw[dashed] (3)  to   (L3);
\end{tikzpicture}
  & S_{4,9}
\\ \hline
 \text{XVII}_1 & \{17,1\} &\; \matarray{x_1^a x_2 + x_1 x_2^b + x_2 x_3^c + x_1 x_4^d\\
 + x_1^p x_3^q+ x_2^r x_4^s }     \; &
 \begin{tikzpicture}[baseline=1mm]
\node[] (1) [label = above: {\scriptsize$a$}]{$\circ$};
\node[] (2) [right = of 1, label = above: {\scriptsize$b$}]{$\circ$};
\node[] (3) [right = of 2, label = above: {\scriptsize$c$}]{$\circ$};
\node[] (4) [right = of 3, label = above: {\scriptsize$d$}]{$\circ$};
\draw[->] (2) to [out = 145, in = 35]  (1);\draw[->] (1)  to [out = -35, in = 215]  (2);
\draw[->] (3) to (2);
\draw[->] (4) to [out = 155, in = 30] (1);
\draw[dashed] (3)  to [out = 215, in = -25]   (1);
\draw[dashed] (4)  to [out = 210, in = -30]   (2);
\end{tikzpicture}
  & S_{4,8}
\\ \hline 
 \text{XVII}_2 & \{17,2\} &\; \matarray{x_1^a x_2 + x_1 x_2^b + x_2 x_3^c + x_1 x_4^d\\
 + x_1^p x_3^q+ x_2^r x_3 x_4^s }     \; &
\begin{tikzpicture}[baseline=1mm]
\node[] (1) [label = above: {\scriptsize$a$}]{$\circ$};
\node[] (2) [right = of 1, label = above: {\scriptsize$b$}]{$\circ$};
\node[] (3) [right = of 2, label = above: {\scriptsize$c$}]{$\circ$};
\node[] (4) [right = of 3, label = above: {\scriptsize$d$}]{$\circ$};
\draw[->] (2) to [out = 145, in = 35]  (1);\draw[->] (1)  to [out = -35, in = 215]  (2);
\draw[->] (3) to (2);
\draw[->] (4) to [out = 155, in = 30] (1);
\draw[dashed] (3)  to [out = 210, in = -30]   (1);
\draw[dashed] (4)  to [out = 210, in = -30] node[] (L1) {$\bullet$}   (2);
\draw[dashed] (L1)  to    (3);
\end{tikzpicture}
  & S_{4,8}
\\ \hline 
 \text{XVII}_3 & \{17,3\} &\; \matarray{x_1^a x_2 + x_1 x_2^b + x_2 x_3^c + x_1 x_4^d\\
 + x_1^p x_3^q+ x_2^r   x_4^s }     \; &
 \begin{tikzpicture}[baseline=1mm]
\node[] (1) [label = above: {\scriptsize$a$}]{$\circ$};
\node[] (2) [right = of 1, label = above: {\scriptsize$b$}]{$\circ$};
\node[] (3) [right = of 2, label = above: {\scriptsize$c$}]{$\circ$};
\node[] (4) [right = of 3, label = above: {\scriptsize$d$}]{$\circ$};
\draw[->] (2) to [out = 145, in = 35]  (1);\draw[->] (1)  to [out = -35, in = 215]  (2);
\draw[->] (3) to (2);
\draw[->] (4) to [out = 155, in = 30] (1);
\draw[dashed] (3)  to [out = 210, in = -30] node[] (L2) {$\bullet$}   (1);
\draw[dashed] (4)  to [out = 210, in = -30] node[] (L1) {$\bullet$}   (2);
\draw[dashed] (L1)  to    (3);
\draw[dashed] (L2)  to [out = -30, in = -140]   (4);
\end{tikzpicture}
  & S_{4,8}
\\ \hline \hline
 \text{XVIII}_1 & \{18,1\} &\;  x_1^a x_3 + x_1 x_2^b + x_2 x_3^c + x_2 x_4^d+ x_3^p x_4^q
     \; &
 \begin{tikzpicture}[baseline=1mm]
\node[] (1) [label = above: {\scriptsize$a$}]{$\circ$};
\node[] (2) [right = of 1, label = above: {\scriptsize$b$}]{$\circ$};
\node[] (3) [right = of 2, label = above: {\scriptsize$c$}]{$\circ$};
\node[] (4) [right = of 3, label = above: {\scriptsize$d$}]{$\circ$};
\draw[->] (2) to (1);
\draw[->] (3) to (2);
\draw[->] (4) to [out = 145, in = 35] (2);
\draw[->] (1)  to [out = -25, in = 205]  (3);
\draw[dashed] (3)  to    (4);
\end{tikzpicture}
  & S_{4,7}
\\ \hline\hline
 \text{XIX} & \{19,1\} &\;  x_1^a x_3 + x_1 x_2^b + x_3^c x_4 + x_2 x_4^d
     \; &
 \begin{tikzpicture}[baseline=1mm]
\node[] (1) [label = above: {\scriptsize$a$}]{$\circ$};
\node[] (2) [right = of 1, label = above: {\scriptsize$b$}]{$\circ$};
\node[] (3) [right = of 2, label = above: {\scriptsize$c$}]{$\circ$};
\node[] (4) [right = of 3, label = above: {\scriptsize$d$}]{$\circ$};
\draw[->] (2) to (1);
\draw[->] (3) to (4);
\draw[->] (4) to [out = 145, in = 35] (2);
\draw[->] (1)  to [out = -25, in = 205]  (3);
\end{tikzpicture}
  & S_{4,5}
\\ \hline
 \end{array}
$$
\caption{The KS-YY classification of isolated hypersurface singularities in $\C^4$: Types XVII to XIX. \label{tab:YYClassification 3/3}}
\end{table}

\paragraph{The full KS-YY classification.}  The YY classification distinguishes between 19 types of polynomials $F(x)$, denoted by the roman numerals I to XIX, and we denote the subtypes necessary for the full classification by subscripts.%
\footnote{We also use the notation $\{\text{type},\text{subtype}\}$ with $\text{type}\in \{1,2, \cdots, 19\}$, as in the {\tt Mathematica}  notebook.} 
 Then, for a given type and subtype, the positive integers $(a,b,c,d)$ fully determine the singularity (up to a choice of solution to some linear Diophantine equation, in some cases, which we discuss momentarily). For instance, the singularity IX$_2(2,3,4,2)$ is given by $F=  x_1^2 + x_2^3 x_4 +x_3^4 x_4+ x_2 x_4^2 +x_1 x_2 x_3^2$. The complete classification is given in tables~\ref{tab:YYClassification 1/3}, \ref{tab:YYClassification 2/3} and  \ref{tab:YYClassification 3/3}.

It is very important to note that this classification includes many redundacies: several singularities of distinct types can be physically equivalent -- two quasi-homogeneous singularities are physically equivalent if and only they have same scaling weights, up to ordering. From the 2d LG model perspective, these scaling dimensions $(q_i)$ are the fundamental quantities, and any two superpotentials $W=F_1(x)$ and $W=F_2(x)$ that realises the same scaling dimensions must flow to the same 2d SCFT. From the 4d SCFT point of view, two singularities with the same scaling weights obviously give rise to the same 4d CB spectrum, as per \eqref{PMF}.

In order to explain the KS-YY classification, let us introduce some useful terminology and notation \cite{Kreuzer:1992bi, Davenport:2016ggc}. First of all, an (LG) field $x_i$ that appears in $F(x)$ is called a `root' if it appears as $x_i^{m_i}$, and it is called a `pointer' if it appears as $x_i^{m_i} x_j$, for some positive integer $m_i$. One can show that every $x_i$ in $F(x)$ must be either a root or a pointer. We say that a pointer $x_i$ `points at' the field $x_j$.  One can then introduce a convenient graphical notation for the polynomial. Each $x_i$, root or pointer, is denoted by a dot accompanied with the integer $m_i$, and for each pointer we draw an arrow from $x_i$ to $x_j$. For instance, the polynomial in two variables $F=x_1^{m_1} + x_1 x_2^{m_2}$ is represented by
\bea
 \begin{tikzpicture}[baseline=1mm]
\node[] (1) [label = above: {\scriptsize$m_1$}]{$\circ$};
\node[] (2) [right = of 1, label = above: {\scriptsize$m_2$}]{$\circ$};
\draw[->] (2) -- (1);
\end{tikzpicture} \,.
\eea
With four fields $x_1, \cdots, x_4$, there are 19 distinct possible `skeletons', corresponding to the 19 classes in the YY classification. As we see from the tables, some of the skeletons are reducible. The basic building blocks are the skeletons denoted by $S_{n, \alpha}$ in \cite{Davenport:2016ggc}, where $n$ denotes the number of fields connected by arrows. For $n=1, 2, 3, 4$, there are $1$, $2$, $4$ and $9$ distinct skeletons, respectively \cite{Kreuzer:1992bi, Davenport:2016ggc}.

Whenever a skeleton contains a pair of pointers $x_i$ and $x_j$ that point to the same node $x_k$ (with $i\neq j \neq k$), the corresponding polynomial is not an isolated singularity. One should introduce a `link' or a `pointing link', which is an additional monomial involving $x_i$ and $x_j$ such that the singularity is isolated. The link is denoted by the dashed line connecting the two nodes
\be\label{link expl}
x_i^{m_i} x_k + x_j^{m_j} x_k + x_i^p x_j^q \; : \;  
 \begin{tikzpicture}[baseline=1mm]
\node[] (3) [label = above: {\scriptsize$m_k$}]{$\circ$};
\node[] (2) [right = of 3, label = above: {\scriptsize$m_j$}]{$\circ$};
\node[] (1) [left = of 3, label = above: {\scriptsize$m_i$}]{$\circ$};
\draw[->] (1) -- (3);\draw[->] (2) -- (3);
\draw[dashed] (1)  to [out = -25, in = 205]   (2);
\end{tikzpicture} \,.
\ee
The link, and therefore the singularity, exists only if there exists positive integers $p$ and $q$ such that the polynomial is quasi-homogeneous. We must then solve a Diophantine equation determined by the scaling weight $q_i$ of the four fields. For instance, in \eqref{link expl}, if $q_k$ is the scaling weight of $x_j$, we must have $p {1-q_k\ov m_i}+q {1-q_k\ov m_j}=1$. 

A pointing link is a link that is itself a pointer to another field $y$, of the form $x_i^p x_j^q y$. A link associated to a field $x_k$ can only point to a field $y$ that is not $x_k$ nor any of the fields that are pointed at by the fields $x_i$ and $x_j$ involved in the link. For instance, we can have
\be
x_i^{m_i} x_k + x_j^{m_j} x_k + x_i^p x_j^q y \; : \;  
 \begin{tikzpicture}[baseline=1mm]
\node[] (3) [label = above: {\scriptsize$m_k$}]{$\circ$};
\node[] (2) [right = of 3, label = above: {\scriptsize$m_j$}]{$\circ$};
\node[] (1) [left = of 3, label = above: {\scriptsize$m_i$}]{$\circ$};
\draw[->] (1) -- (3);\draw[->] (2) -- (3);
\node[] (4) [right = of 2, label = above: {\scriptsize$n$}]{$\circ$};
\draw[dashed] (1)  to [out = -25, in = 205] node[] (L1) {$\bullet$}    (2);
\draw[dashed] (L1)  to [out = -15, in = 215]    (4);
\end{tikzpicture}
\ee
We denote the pointed link as shown here. As for simple links, pointed links can only exists if they preserve the quasi-homogeneity of the polynomial. 

\subsection{Computing the Basic Data of Any Canonical IHS: {\tt Mathematica} Code}

Given any $F(x)$ defining a canonical threefold singularity $\MG$, we can easily compute all the mathematical and physical quantities introduced in the previous sections, which we call `the basic data' associated to $\MG$. We implemented these computations on a computer using {\tt Mathematica}~\cite{Mathematica}, and we provided the {\tt Mathematica} notebook {\tt Basic-Data-IHS.nb} as an ancillary file to the arXiv version of this paper. Here, we describe the main routines in that notebook. 

\paragraph{Basic Characteristics of IHSs.}  The code implements many computations which are collated through the main routine:
\be
\mathtt{BasicIHSdata}[F(x_1,x_2,x_3,x_4),\{x_1,x_2,x_3,x_4\}]~.
\ee
The code also generate any allowed polynomial $F(x)$ given a type, subtype and the positive integers $(a,b,c,d)$:
\be
\mathtt{FfromABCD}[\{t,s\}, \{a,b,c,d\}]~,
\ee
with $\{t,s\}=\{\text{type},\text{subtype}\}$. The `basic data' routine can be called directly as:
\be
\mathtt{BasicIHSdataFromABCD}[\{t,s\}, \{a,b,c,d\}]~.
\ee
The output is given as a list with the following information:
\begin{itemize}
\item $F(x)$ itself;
\item the scaling weights $(q_i)$;
\item $r$: the rank of the 5d SCFT $\FT$  (that is, the number of exceptional divisors in $\t\MG$);
\item $f$: the flavor rank preserved on the Coulomb branch of $\FTfour$ and $\FT$;
\item $d_H=\h r +f$: the quaternionic Higgs branch dimension of $\FT$;
\item $\h{r}$: the rank of the 4d SCFT $\FTfour$;
\item $\h{d}_H= r+f$: the quaternionic  Higgs branch dimension of $\FTfour$;
\item $\mu= 2\h r+f$: the multiplicity (Milnor number) of $\MG$;
\item $\text{modality}$: the modality $m(\MG)$ of $\MG$;
\item $n_h$: the effective number of hypermultiplets of $\FTfour$;
\item $n_v$: the effective number of vector multiplets of $\FTfour$;
\item $b_3$: the number of 3-cycles in the resolved threefold $\t \MG$;
\item $\Delta A$: the difference between the `virtual dimension' $n_h-n_v$ of the HB of $\FTfour$ and the actual HB dimension, $\Delta A=n_h-n_v-\h{d}_H$;
\item $g_C$: the list $\{g_1, _2,  \cdots\}$ of genera for the higher-genus ($g>0$) complex curves at the base of the exceptional divisors inside $\t \MG$;
\item specL: the spectrum $\{\ell_l\}$ of the Milnor ring $\CM(F)$;
\item CB spectrum: the Coulomb branch spectrum of $\FTfour$;
\item $a$ and $c$: the conformal anomaly coefficients of $\FTfour$;
\item marginal Defs: the number of marginal deformations of $\FTfour$;
\item ${\rm Tor} H_2$: the torsional part of the second homology of the link $L_5$, ${\rm Tor}\,H_2(L_5,\mb{Z})=  \mathfrak{t}_2$.
\end{itemize}

\paragraph{Smoothing Operator.}
The crepant resolution procedure described in secton~\ref{sec:resolution} is encoded in an additional routine, which constructs the (partially) resolved Calabi-Yau threefold $\MG$ of any IHS $\{F(x_1,x_2,x_3,x_4)=0\}$. Using the function:
\be
 {\tt SmoothOperator}[F(x_1,x_2,x_3,x_4), \{x_1,x_2,x_3,x_4\}]~,
 \ee
  one gets an output in form of $\{$Resolution sequence, Resolved equation, Patches$\}$. Note that there can be several distinct resolution sequences for a given IHS, and the code only generates one of them.

The set ``Resolution sequence'' contains the crepant resolution sequence of the IHS, in the form  $\{\mathrm{Blp}_1,\mathrm{Blp}_2,\dots,\mathrm{Blp}_n\}$. Each Blp$_i$ is a blow-up of the toric ambient space, given in the form $\{\{\{y_1,y_2,\dots,y_k\},\{w_1,w_2,\dots,w_k\}\}, \delta\}$, which corresponds to the blow-up of the locus $y_1=y_2=\dots=y_k=0$ with weights $(w_1,w_2,\dots,w_k)$, with the exceptional divisor $\delta=0$. Using the notations of section~\ref{sec:resolution}, Blp$_i$ is also written as $(y_1^{(w_1)},y_2^{(w_2)},\dots,y_k^{(w_k)};\delta)$, and the superscripts $w_i$ are omitted if they all equal to one.

The set ``Patches'' contains the coordinate patches of the toric ambient space $T_\Sigma$ after the blow-ups, which are in one-to-one correspondence with the 4d cones of the toric fan $\Sigma$. This coordinate patch structure also extends to the anti-canonical hypersurface $\t\MG=-K(T_\Sigma)$. On each coordinate patch $\{y_1,y_2,y_3,y_4\}$, where $y_i$ are the  local coordinates of $\t\MG$, one can set $y_1=y_2=y_3=y_4=0$ unless the intersection locus is not contained in $\t\MG$.

We always use the notation $\delta_i=0$ ($\delta[i]=0$ in the code) for the exceptional divisors inside the blow-up of $T_\Sigma$. They may not intersect $\t\MG=-K(T_\Sigma)$ eventually, and they may be reducible in $\t\MG$. Hence, the total number of $\delta_i$ does not equal to the 5d rank $r$ in general.

Note that this code also works in the generic case with residual terminal singularities. In these cases, the resolution sequence is a partial crepant resolution that gives rise to $\t\MG$ with at most terminal singularities.

As an example, take $F=x_1^5+x_2^2 x_1+x_3^5+x_4^2 x_3$. We input {\tt SmoothOperator[$x_1^5 + x_2^2 x_1 + x_3^5 + x_4^2 x_3, \{x_1, x_2, x_3, x_4\}$]}, and get the output $\{$Resolution sequence, Resolved equation, Patches$\}$ where
 \be
 \ba
& \mathrm{Resolution\ sequence}=\\
 &\qquad\quad \{\{\{\{x_1, x_2, x_3, x_4\}, \{1, 1, 1, 1\}\}, \delta[
   1]\}, \{\{\{x_2, x_4, \delta[1]\}, \{1, 1, 1\}\}, \delta[2]\}\}\,,
  \ea
  \ee
and after applying this to the singular IHS, we obtain the resolved equation (including the proper transforms):  
\be
\mathrm{Resolved\ equation}=x_1 x_2^2+x_3 x_4^2+(x_1^5+x_3^5)\delta[1]^2\,.
\ee
The patches are then given by 
\be
\ba
\mathrm{Patches}&=\{\{x_1, x_2, x_3, \delta[1]\}, \{x_2, x_4, x_1, \delta[2]\}, \{x_2, \delta[
  1], x_1, \delta[2]\}, \{x_4, \delta[1], x_1, \delta[
  2]\}, \cr
  &\{x_1, x_3, x_4, \delta[1]\}, \{x_2, x_4, x_3, \delta[
  2]\}, \{x_2, \delta[1], x_3, \delta[2]\}, \{x_4, \delta[
  1], x_3, \delta[2]\}\}\,.
  \ea
 \ee
The resolution sequence and the resolved equation match \eqref{D2nD2n-res}-\eqref{D2nD2n-resEq}.

\section{Smoothable Models}  
\label{sec:Smooth}

In the previous section, we spelled out a general approach to determine the  basic properties of the SCFTs $\FT$ and $\FTfour$ for any IHS $\MG$. We also explained how to explicitly construct the crepant resolution $\pi: \t\MG \rightarrow \MG$.  This allows us to determine the smoothness properties of the resolved model $\t\MG$ (whether it contains residual singularities or not), as well as possible 5d infrared gauge theory descriptions.  In earlier works \cite{Closset:2020scj, Closset:2020afy}, we studied interesting models with terminal singularities (whether $\MG$ was a terminal singularity with $r=0$, or a singularity with $r>0$ that contains remnant terminal singularities after a crepant resolution).  In this section, we would like to focus on {\it smoothable models} -- that is, singularities $\MG$ whose crepant resolution $\t\MG$ is a smooth CY threefold (below, we also call such a CY threefold `a smooth model').

\subsection{Tabulating the Smoothable Models}

We found it useful to organise the smoothable singularities $\MG$  into three categories:
\begin{enumerate}
\item  Smooth models  with smooth exceptional divisors  and $b_3(\t \MG)=0$
\item  Smooth models  with smooth exceptional divisors and $b_3(\t \MG)>0$
\item  Smooth models with singular divisors (for any $b_3(\t\MG)$).
\end{enumerate}
It is also useful to organise the models by their 5d rank, $r$, {\it i.e.} by the number of exceptional divisors. For low rank, we gathered (conjecturally) all such models by brute force, by generating a large dataset of 15,142 
physically distinct models with $r>0$.%
\footnote{We considered all canonical IHS of type I to XIX with $(a,b,c,d)$ `small enough', and then identified the many models with identical scaling weights. For definiteness, we chose the cap $a,b,c,d\leq K$ with $K=60$ for the 19 types (and for each subtype) in the KS-YY classification. This gives us a total of 39,094 physically distinct models. 61$\%$ of these models have $r=0$, and we do not consider them further in this paper.} The number of models for $r\leq 10$ in that dataset is given in Table~\ref{tab:dataset}. 
\begin{table}[t]
\renewcommand{\arraystretch}{1.2}
\begin{center}
\begin{tabular}{ |c||c|c|c|c|c|c|c|c|c|c|} 
 \hline
$r$                       &    \; $1$  \; & \;  $2$ \; &   \;$3$\; &   $4$  & $5$  &  $6$  &  $7$  &  $8$  &  $9$  &  $10$     \\
\hline
\hline
$\#$ of models & $1167$ & $649$            & $693$     & $666$       & $506$     & $549$     & $486$     & $495$     & $491$     & $461$\\
\hline
$\#$ of $\Delta \CA\in \Z$ models & $161$ & $128$            & $143$     & $131$       & $113$     & $154$     & $98$     & $117$     & $137$     & $117$\\
\hline
$\#$ of smoothable models & $12$ &  $24$ & $32$ & $23$ & $32$ &$39$ &$23$ & $41$& $?$& $?$\\
    \hline
$\#$ of $\Z$CB models & $3$ & $6$            & $10$     & $9$       & $12$     & $18$     & $7$     & $9$     & $12$     & $12$\\
\hline
    $\#$ of `fully smooth' models & $3$ &$5$ & $10$ & $7$ & $9$ & $14$ & $5$ &$8$ &$7$ &$7$ \\
    \hline
\end{tabular}
\end{center} \caption{Number of low-rank models in our dataset. Here, $\Z$CB refers to the models with integral CB dimensions and `fully smooth' refers to the smoothable models with smooth exceptional divisors. 
}
\label{tab:dataset}
\end{table}

For $r=1,2,3$, we resolved all the models explicitly and found all the smoothable models in our dataset. For larger $r$, this becomes impractical, but we can reduce strongly the number of cases to consider by realising that a necessary condition for having a smoothable model is that $\Delta\CA \in \Z$,  because $\Delta\CA\notin \Z$ can only arise from residual terminal singularities \cite{Closset:2020scj}. We are also particularly interested in the smoothable models with smooth exceptional divisors (the `fully smooth models', for short). We then make the following conjecture, based on explicit analysis of our dataset:

{\conjecture
For any smoothable model $\MG$ with smooth exceptional divisors, the associated 4d SCFT $\FTfour$ has an integral CB spectrum.
\label{conj1}}

\medskip
\noindent
Note that the converse is not true. On the other hand, we propose the following:

{\conjecture
Any model $\MG$ with an integral CB spectrum is smoothable (not necessarily with smooth divisors).
}

\medskip
\noindent
In summary, we have the following inclusions of sets, at any fixed rank $r>0$:
\be
\{\text{`fully smooth' }\}\subset  \{\text{integral CB}\}\subset \{\text{smoothable}\}
\subset \{\text{$\Delta\CA\in \Z$}\}\subset \{\text{rank-$r$ models}\}~,
\ee
where the first and second inclusions are conjectures. The number of models at fixed rank appears to be infinite, but we conjecture that the number of smoothable models of a fixed rank is finite. For small $r$, we also believe that all these models are included in our dataset.%
\footnote{We do not have a proof that our lists of `smoothable' and `fully smooth' models up to $r=10$ is complete within the infinite class of models defined by canonical IHSs. This is largely due to the fact that the correlation between rank $r$ and the hypersurface equation types is not one we understand systematically at this point. It would be very interesting to determine more direct relations between the $(a,b,c,d)$'s defining the equation $F(x)=0$ and the rank $r$, if possible.}
The numbers of models with various properties (in our dataset) are shown in Table~\ref{tab:dataset}.  For $r\leq 10$, we considered all the models with integral CB spectrum and extracted all the `fully smooth' models from that set. They are discussed in detail in the following, and in appendix.

\paragraph{Models with $b_3=0$.}
We first discuss the smoothable models with smooth exceptional divisors with $b_3(\MG)=0$. We studied all such models with rank $r\leq 10$ in our dataset (assuming Conjecture \ref{conj1} is true). We also considered the infinite series  AD$[D_{2n},D_{2n}]$, which engineers the 4d SCFT of type $[G,G']=[D_{2n},D_{2n}]$  \cite{Cecotti:2010fi}.  These $b_3=0$ models are summarized in tables \ref{tab:b3zero1} and \ref{tab:b3zero2}. 

This class of models have the simplest 5d interpretation, as they correspond to 5d SCFTs $\FT$ whose Coulomb-branch low-energy physics is described entirely by $r$ 5d $U(1)$ vector multiples, with a prepotential determined by the intersection numbers  \eqref{SSSdata} in the smooth resolved threefold. Similarly, the 4d SCFT $\FTfour$ is such that the Higgs-branch low-energy physics consists of only $\h d_H= r+f$ hypermultiplets \cite{Closset:2020scj}.

\paragraph{Models with $b_3>0$.}
Similarly, the smoothable models with smooth exceptional divisors but $b_3>0$ are listed until rank $r=10$ in tables \ref{tab:SmoothSmoothb31}, \ref{tab:SmoothSmoothb32}, \ref{tab:SmoothSmoothb33}, \ref{tab:SmoothSmoothb34} and  \ref{tab:SmoothSmoothb35}. 

Note that, in all these tables, we color the labels $\{t, s\}\{a,b,c,d\}$ in $(i)$ black for the cases with a Lagrangian $\FTfour$;  $(ii)$ blue for the cases with a $D_p^b(G)$ trinion description studied in \cite{Closset:2020afy}; 
$(iii)$ red  for the cases with a generalised $D_p^b(G)$ description of $\FTfour$ (to be discussed below); $(iv)$ orange  for the cases without a known description of $\FTfour$ yet. 
For the case $(i)$, the Lagrangian quiver description of $\FTfour$ is written down in the table. For the other cases, we  simply list the CB spectrum. Many of theses models were already discussed in \cite{Closset:2020scj,Closset:2020afy}.

The physical interpretation of the 3-cycles in $\t\MG$ was already discussed in  \cite{Closset:2020scj}. The 5d SCFT $\FT$ has an `enhanced Coulomb branch' with $\half b_3$ free hypermultiplets at any generic point on the CB. The 4d SCFT $\FTfour$ has an Higgs branch whose low-energy physics includes $\half b_3$ free abelian vector multiplets in addition to the $\h d_H$ hypermultiplets. 

Once we have a resolved threefold with 3-cycles, it is generally possible to perform a sort of geometric transition that results in a distinct local geometry \cite{Jefferson:2018irk}, as we will see in some examples below. Such transitions change the asymptotic of the threefold. Physically, they correspond to mass deformations that lift the Higgs branch component of the enhanced CB.  To precisely characterise the physics of enhanced Coulomb branches in 5d is an interesting question, which is left for future work (for a discussion of 4d enhanced Coulomb branches, see \cite{Argyres:2016xmc}).

\paragraph{Smoothable models with singular divisors.} Finally, one can consider the smoothable models whose exceptional divisors are not all smooth. The interpretation of such singularities in the divisors (while the threefold itself is smooth) appears to be more subtle, and we hope to come back to this in future work. Here, we simply list all the smoothable models up to rank $r=4$ in the tables~\ref{smoothableModelsI},~\ref{smoothableModelsII},~\ref{smoothableModelsIII},~\ref{smoothableModelsIV},~\ref{smoothableModelsV}, ~\ref{smoothableModelsVI} and ~\ref{smoothableModelsVII}.  


\subsection{The AD$[E_7,E_7]$ model}
\label{sec:ADE7E7}

We now illustrate one example of `fully smooth' model with $b_3=0$, which turns out to have $r=5$. We consider the geometry that is associated to the Argyres-Douglas-type theory $[G,G']=[E_7,E_7]$ \cite{Cecotti:2010fi}. This is the Type$\{4,1\}\{3,3,3,3\}$ singularity, with basic data:
 \be
\small
\begin{tabular}{ | c ||c |c | c|c|| c|c| c|| c| c|c|}\hline
$F$& $r$ &$f$ & $G_F^{\rm 5d}$  &$d_H$ &$\h r$& $\h d_H$  & $\Delta \CA_r$  & $b_3$ & $\mathfrak{t}_2$  \\ [0.5ex] 
\hline 
$x_1^3+x_2^3 x_1+x_3^3+x_4^3 x_3$ & $5$ & $7$ & $SU(2)^6\times U(1)$ & $28$ & $21$ & 12 & 0 & 0 & --\\\hline
\end{tabular}
\ee

\paragraph{Deformations and 4d SCFT.} In this case, the CB spectrum of $\FTfour$ contains operators with $\Delta=7$ and $\Delta=9$, but there is no operator with $\Delta=8$. Although the CB spectrum is fully integral,
\be
\Delta= \{2^4,3^5,4^4,5^3,6^2,7^2,9\}\,,
\ee
it cannot match any set of Casimir operators of Lie groups. Hence there is no Lagrangian description of $\FTfour$. There is, however, a generalised quiver using $D_p^b(G)$ building blocks. The conjectured 4d generalised $D_p(G)$ description is:%
\footnote{We thank Simone Giacomelli for discussions of this model and for suggesting this particular quiver. }
\be\label{QuiverE7E7}
[1]- SU(3) - SU(5) - SU(7) -   D_2(SO(20)) -  SO(8) -  D_2(SO(8)) \,,
\ee
where the $D_2(SO(2N))$ factor contributes the CB operators of dimensions $\Delta = 3, 5, 7, \cdots, N-1$, and the $SO(2N)$ gauge groups contribute $\Delta = 2, 4, \cdots, 2N-2, N$. The flavor symmetry of $D_2 (SO(20))$ is  $SO(22)\times U(1)$, of which we gauge an $SU(7)\times SO(8)$ subgroup.  Note that $D_2(SO(8))$ is the $E_6$ MN theory \cite{Minahan:1996fg, Cecotti:2013lda}.

\paragraph{Resolutions and 5d SCFT.}

\begin{figure}
\centering
\includegraphics[height=5cm]{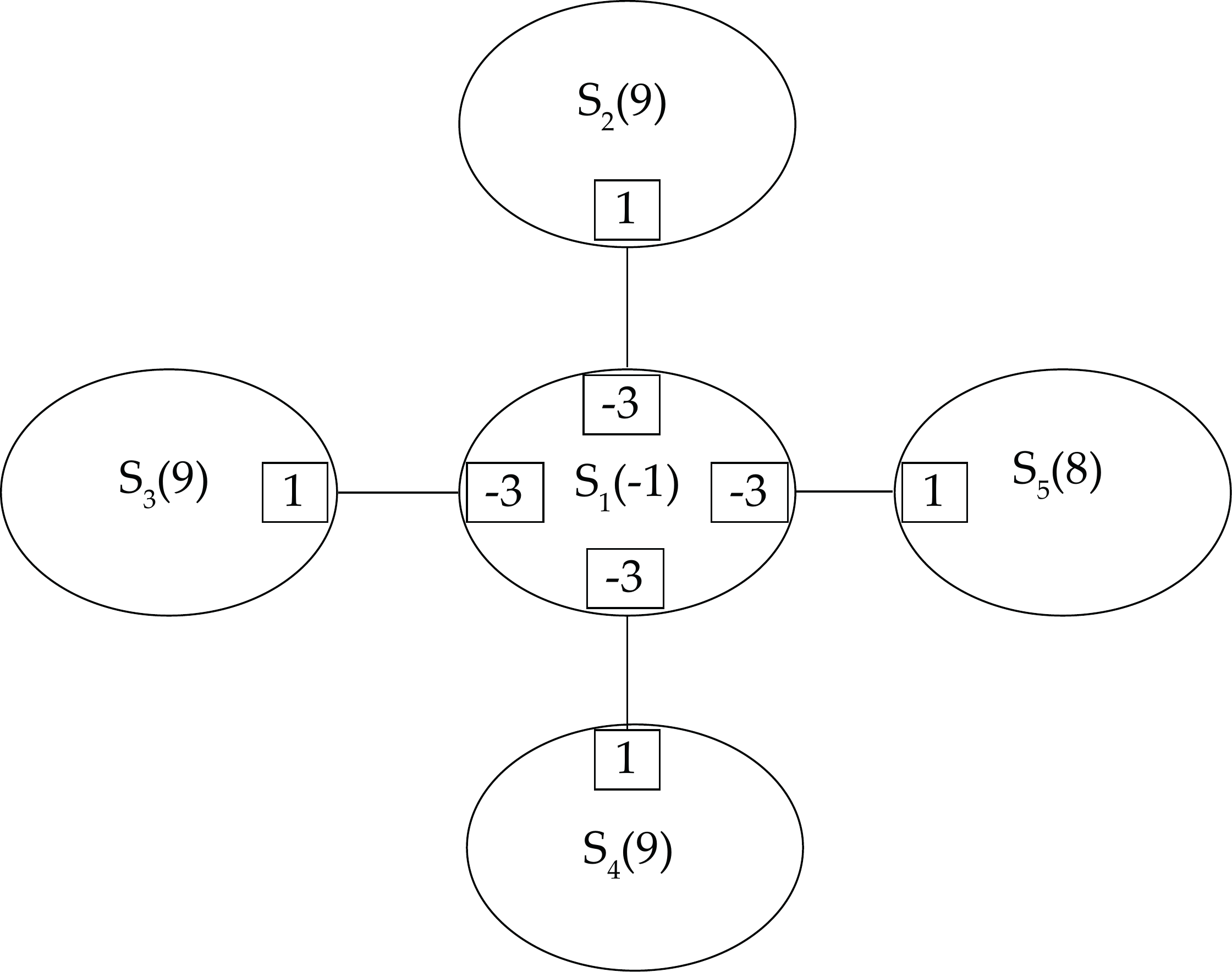}\qquad 
\includegraphics[height=5cm]{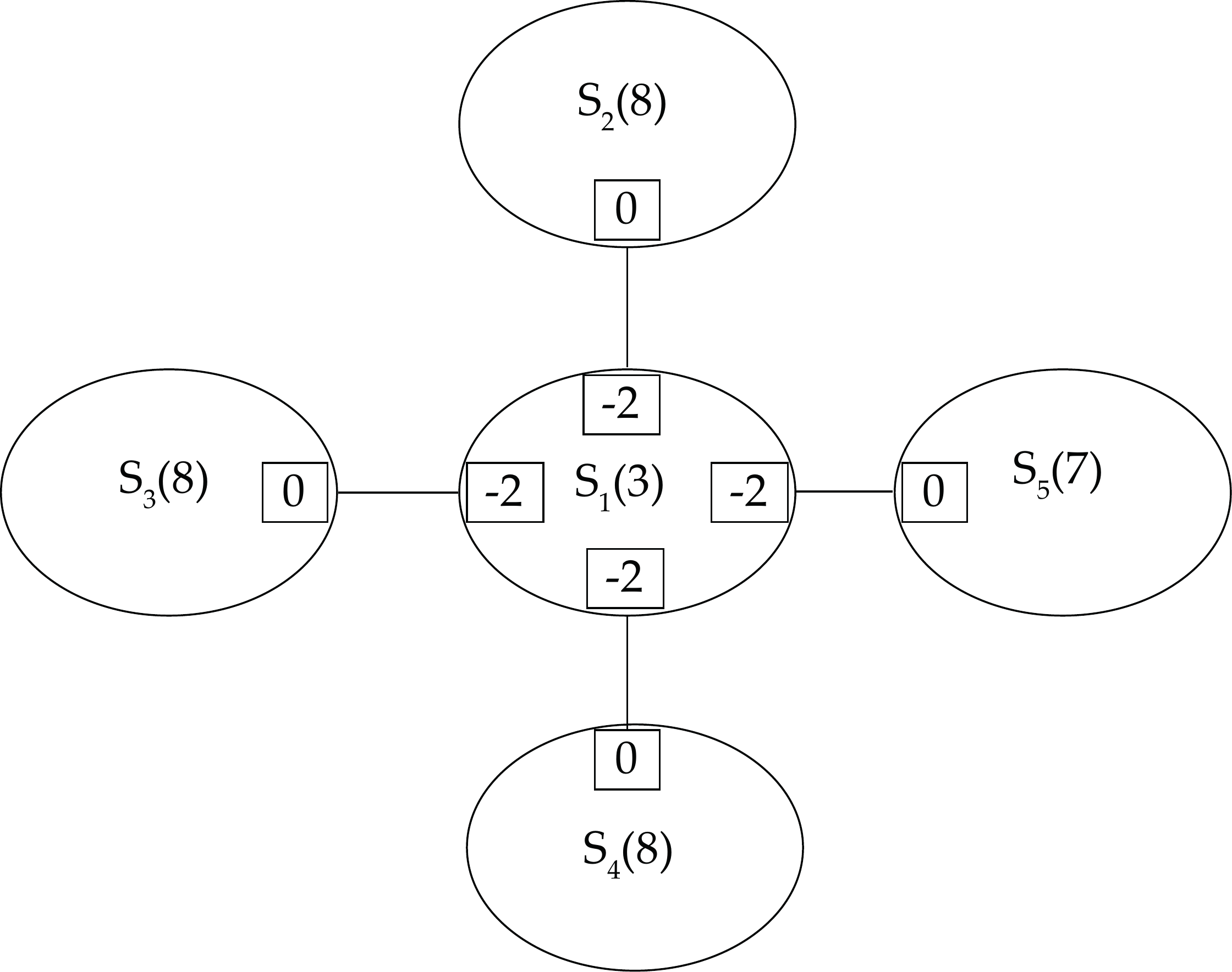}
\caption{Resolved AD$[E_7, E_7]$ geometry: compact surfaces are $S_i$ and gluing curves are links between these, with self-intersection numbers shown in the boxes. 
The LHS is the model after resolution of the IHS, and the RHS after the flop. }\label{fig:E7E7}
\end{figure}

This is a fully resolvable IHS, and a resolution sequence is given by:
\be
\ba
&(x_1,x_2,x_3,x_4;\delta_1)\ ,\ (x_1,x_3,\delta_1;\delta_2)\ ,\ (\delta_1,\delta_2;\delta_3)\,.
\ea
\ee
After the resolution, the equation
\be
x_1^3\delta_2+x_2^3 x_1\delta_1+x_3^3\delta_2+x_4^3 x_3\delta_1=0
\ee
is smooth. The exceptional divisor $\delta_1=0$ has three components:
\be
\delta_1=0:\quad x_1^3+x_4^3=0\,.
\ee
The exceptional divisors $\delta_2=0$ and $\delta_3=0$ are irreducible, thus resulting in total in a rank $r=5$ theory. 
The intersection numbers of the compact surfaces are shown on the LHS of figure \ref{fig:E7E7}. The three components of $\delta_1=0$ are $S_2$, $S_3$ and $S_4$, which are all $\mb{P}^2$s. $S_1$ corresponds to $\delta_3=0$ and $S_5$ corresponds to $\delta_2=0$. 

\medskip
\noindent The conventions for the intersection diagrams are as follows (similarly to the ones in \cite{Bhardwaj:2018yhy,Tian:2021cif}):

\begin{itemize}
\item[(1)] Each oval denotes an irreducible compact divisor $S_i$. The number in the bracket after $S_i$ is the self-triple intersection number $S_i^3$.

\item[(2)]  The line segment between ovals $S_i$ and $S_j$ corresponds to the complete intersection curve $S_i\cdot S_j$. The numbers in the boxes at the end of the line segment describe the self-intersection numbers of the curves inside their respective divisors. The number on $S_i$ corresponds to $S_i\cdot S_j^2$ while the number of $S_j$ corresponds to $S_j\cdot S_i^2$.

\item[(3)]   For each triangle among three ovals $S_i$, $S_j$ and $S_k$, we use a boxed number to denote the triple intersection number $S_i\cdot S_j\cdot S_k$.
\end{itemize}

\noindent
To see the structure of the curves more clearly, we may do four flops by blowing up $S_2$, $S_3$, $S_4$, $S_5$ once and blowing down $S_1$ four times, which gives the local model shown on the right-hand-side of figure \ref{fig:E7E7}. 
Here, $S_1$ is a gdP$_6$ of type $4\mathbf{A}_1$, which is also known as the (resolution of) Cayley's cubic surface \cite{heath2002density}. We plot the Mori cone generators of $S_1$ and their representations under the Picard group generators $h,e_i$ ($h^2=1\ ,\ h\cdot e_i=0\ ,\ e_i\cdot e_j=-\delta_{ij}$), showing the four $(-2)$-curves $S_1\cdot S_2$, $S_1\cdot S_3$, $S_1\cdot S_4$ and $S_1\cdot S_5$:
\be
\includegraphics[height=6cm]{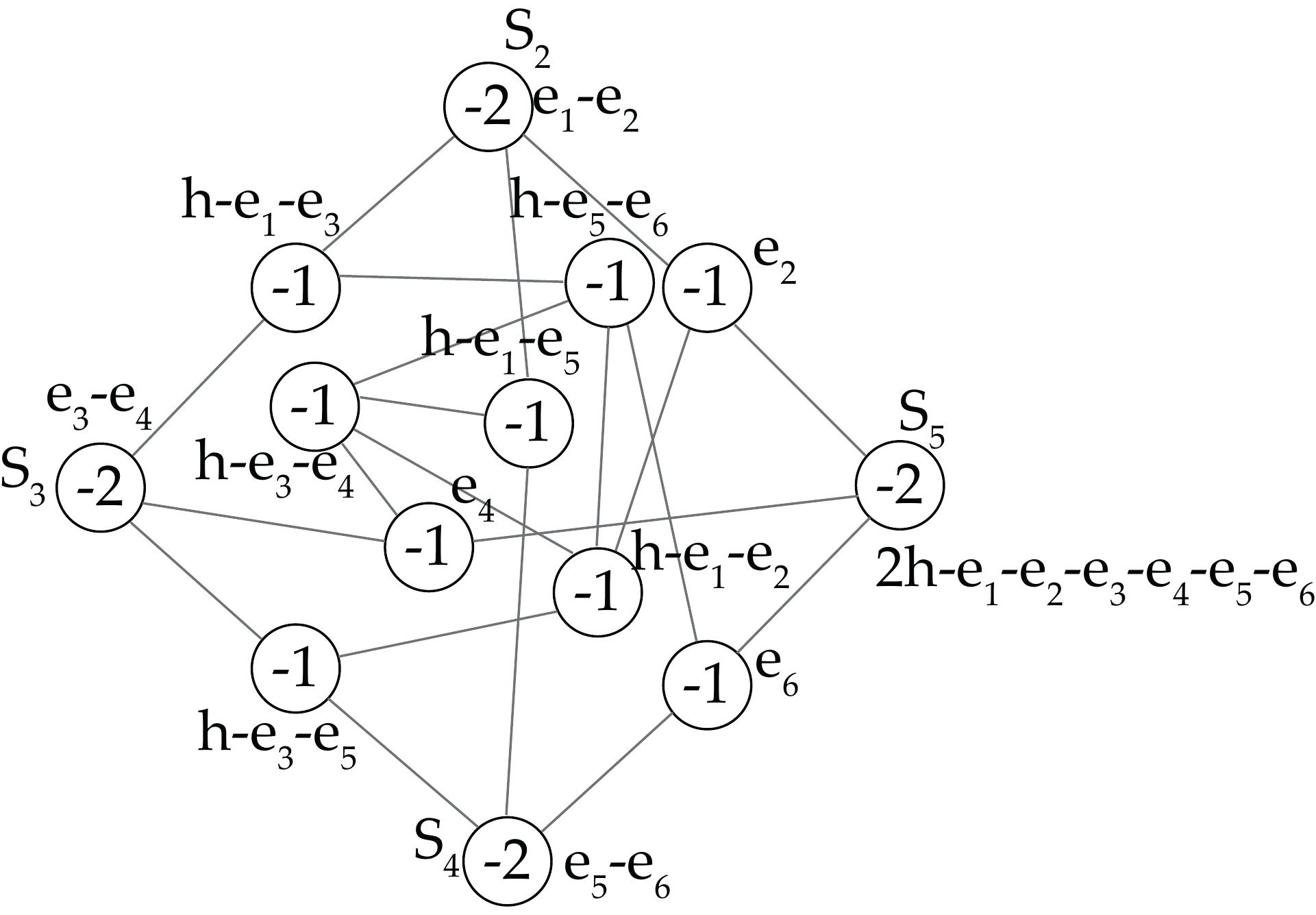}
\ee
We also plot the curves on $S_2$, $S_3$, $S_4$ and $S_5$:
\be
\includegraphics[height=3cm]{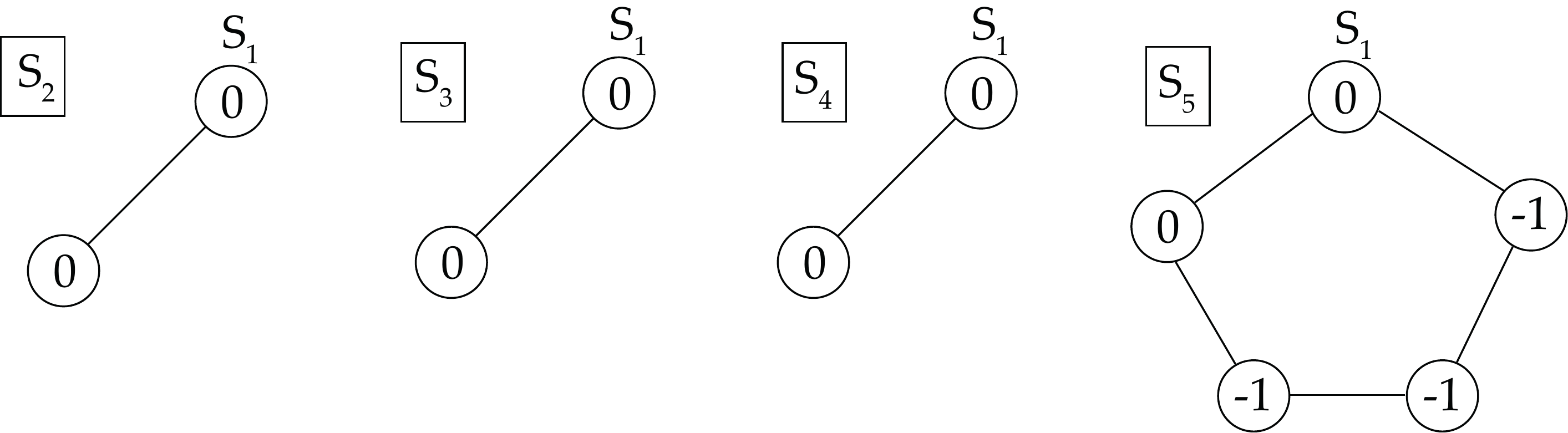}
\ee
Now we can combine the curves on each $S_i$ and get the combined fiber diagram (CFD) introduced in \cite{Apruzzi:2019vpe, Apruzzi:2019opn, Apruzzi:2019enx, Apruzzi:2019kgb} of $\FT$ by applying the procedure in \cite{Apruzzi:2019kgb}. The CFDs are -- in brief -- a collection of (usually rational) curves, which encode the flavor symmetries (in terms of the subgraph of green-colored nodes) and the $(-1)$-self-intersection curves that can be flopped and the associated hypermultiplet decoupled. 

The CFD is constructed by connected to the same $S_i\cdot S_j$ are combined, with the proper multiplicities. For example, on $S_1$ the six $(-1)$-curves between the four $(-2)$-curves have multiplicities two, which correspond to $(-2)$ nodes in the CFD. Hence the CFD can be read off as:
\be
\includegraphics[height=4cm]{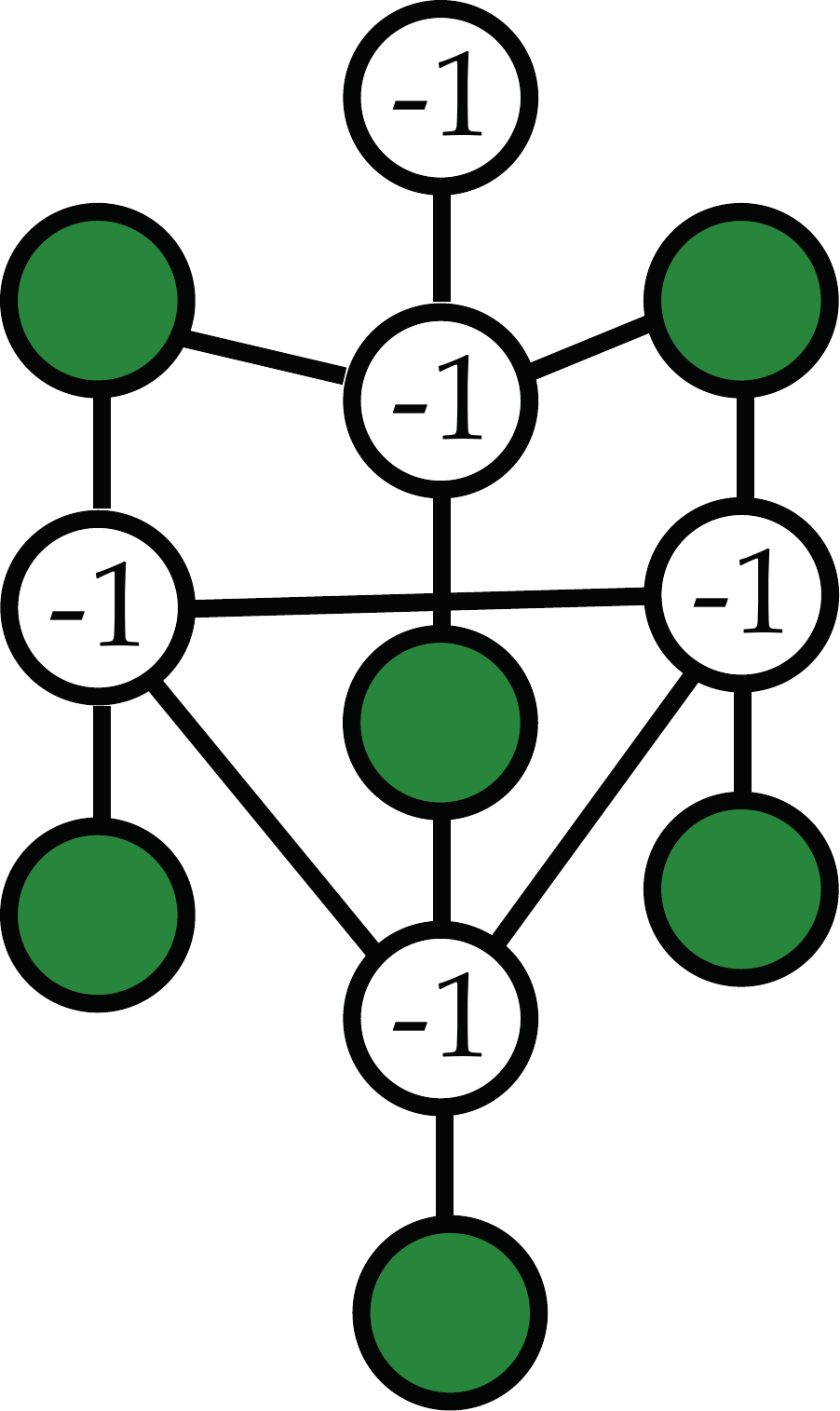}
\ee
The UV flavor symmetry algebra is therefore $\mathfrak{g}_F=\mathfrak{su}(2)^6\oplus \mathfrak{u}(1)$.

On the geometry, we can also identify a ruling structure by taking $S_1\cdot S_2$ and $S_1\cdot S_5$ to be the section curves, $S_1\cdot S_3$ and $S_1\cdot S_4$ to be the fiber. From the spectrum of massless hypermultiplets in the IR gauge theory limit, we can identify the following 5d gauge theory description:
\be
 \begin{tikzpicture}[x=.6cm,y=.6cm]
\node at (-3.3,0) {$SU(2)_0$};
\node at (0,0) {$SU(4)_{-\frac{1}{2}}$};
\node at (3.3,0) {$SU(2)_0$};
\node at (0,1.7) {$1\mathbf{AS}+1\mathbf{F}$};
\draw[ligne, black](-2,0)--(-1.3,0);
\draw[ligne, black](1.3,0)--(2,0);
\draw[ligne, black](0,0.5)--(0,1.3);
\end{tikzpicture}
\ee
If we decouple the two $SU(2)$ gauge factors by sending the volume of $S_3$ and $S_4$ to infinity, the surfaces $S_1, S_2, S_5$ give rise to IR gauge theory $SU(4)_{-\frac{1}{2}}+1\mathbf{AS}+5\mathbf{F}$.
It would be interesting to see whether this IR gauge theory description can inform the computation of the magnetic quiver for this theory, and to then match this with the conjectured 
4d quiver (\ref{QuiverE7E7}).


\begin{table}[ht]
$$
\begin{array}{|c|c|c|c|}\hline
\hbox{Type $(a,b,c,d)$} & F &  \hbox{Trinion Representation} & \hbox{CB Spectrum} \cr \hline\hline 
\{1,1\}\{3,4,4,4\} & x_2^4 + x_3^4 + x_4^4 + x_1^3 &  
\begin{tikzpicture}[x=.6cm,y=.6cm]
\node at (-2.3,0) {$D_4^{12}$};
\node at (0,0) {$E_6$};
\node at (2.3,0) {$D_{4}^{12}$};
\node at (0,2) {$D_{2}^{12}$};
\draw[ligne, black](-1.6,0)--(-0.6,0);
\draw[ligne, black](0.6,0)--(1.6,0);
\draw[ligne, black](0,0.5)--(0,1.3);
\end{tikzpicture} & 2^6,3^7,5^3,6^6,8,9^3,12 \cr \hline 
\multirow{2}{*}{\{1,1\}\{2,3,10,10\} } &\multirow{2}{*}{ $x_3^{10}+x_4^{10}+x_2^3+x_1^2$} & 
\begin{tikzpicture}[x=.6cm,y=.6cm]
\node at (-2.3,0) {$D_{10}^{30}$};
\node at (0,0) {$E_8$};
\node at (2.3,0) {$D_{10}^{30}$};
\node at (0,2) {$D_{1}^{24}$};
\draw[ligne, black](-1.6,0)--(-0.6,0);
\draw[ligne, black](0.6,0)--(1.6,0);
\draw[ligne, black](0,0.5)--(0,1.3);
\end{tikzpicture} 

& 2^7,3^8,5^6,6^9,8^5,9^8,11^4,12^7,14^3\\
 & & &  15^6,17^2,18^5,20,21^4,24^3,27^2,30 \\\hline
\multirow{2}{*}{ \{1,1\}\{2,5,6,6\} } & \multirow{2}{*}{$x_3^6+x_4^6+x_2^5+x_1^2$} &

 \begin{tikzpicture}[x=.6cm,y=.6cm]
\node at (-2.3,0) {$D_{6}^{30}$};
\node at (0,0) {$E_8$};
\node at (2.3,0) {$D_{6}^{30}$};
\node at (0,2) {$D_{1}^{20}$};
\draw[ligne, black](-1.6,0)--(-0.6,0);
\draw[ligne, black](0.6,0)--(1.6,0);
\draw[ligne, black](0,0.5)--(0,1.3);
\end{tikzpicture} 

 &
2^3,3^4,4^5,5^4,7^2,8^3,9^4,10^5,12\\
 & & & 13^2,14^3,15^4,18,19^2,20^3,24,25^2,30\\ \hline
\multirow{2}{*}{\{2,1\}\{3,6,9,2\} } & \multirow{2}{*}{$x_3^9+x_4^2 x_3+x_2^6+x_1^3$} &

 \begin{tikzpicture}[x=.6cm,y=.6cm]
\node at (-2.3,0) {$D_{3}^{18}$};
\node at (0,0) {$SO(20)$};
\node at (2.3,0) {$D_{6}^{18}$};
\node at (0,2) {$D_{2}^{18}$};
\draw[ligne, black](-1.6,0)--(-1,0);
\draw[ligne, black](1,0)--(1.6,0);
\draw[ligne, black](0,0.5)--(0,1.3);
\end{tikzpicture} 

&
2^5,3^4,4^7,5^3,6^5,7^4,8^3,9^3\\
& & & 10^4,11,12^3,13,14,15,16,18 \\ \hline
\{2,1\} \{4,4,8,2\} & x_3^8+x_4^2 x_3+x_1^4+x_2^4 
 &

  \begin{tikzpicture}[x=.6cm,y=.6cm]
\node at (-2.3,0) {$D_{4}^{16}$};
\node at (0,0) {$SO(18)$};
\node at (2.3,0) {$D_{4}^{16}$};
\node at (0,2) {$D_{2}^{16}$};
\draw[ligne, black](-1.6,0)--(-1,0);
\draw[ligne, black](1,0)--(1.6,0);
\draw[ligne, black](0,0.5)--(0,1.3);
\end{tikzpicture} 
  &
2^8,4^8,5^2,6^6,8^6,9,10^3,12^3,14,16\\\hline
\multirow{2}{*}{ \{2,1\}\{2,5,5,6\} } &\multirow{2}{*}{$x_3^5+x_4^6 x_3+x_1^2+x_2^5$}
 &

  \begin{tikzpicture}[x=.6cm,y=.6cm]
\node at (-2.3,0) {$D_{6}^{24}$};
\node at (0,0) {$E_8$};
\node at (2.3,0) {$D_{5}^{30}$};
\node at (0,2) {$D_{1}^{20}$};
\draw[ligne, black](-1.6,0)--(-1,0);
\draw[ligne, black](1,0)--(1.6,0);
\draw[ligne, black](0,0.5)--(0,1.3);
\end{tikzpicture} 

  &
2^7,4^5,6^7,8^6,10^4,12^6,14^4,\\
 & & & 16^2,18^4,20^2,22,24^2,26,30\\\hline
\multirow{2}{*}{ \{2,1\}\{3,6,12,2\} } & \multirow{2}{*}{$x_3^{12}+x_4^2 x_3+x_1^3+x_2^6$ }
&
  \begin{tikzpicture}[x=.6cm,y=.6cm]
\node at (-2.3,0) {$D_{6}^{24}$};
\node at (0,0) {$SO(26)$};
\node at (2.3,0) {$D_{3}^{24}$};
\node at (0,2) {$D_{2}^{24}$};
\draw[ligne, black](-1.6,0)--(-1,0);
\draw[ligne, black](1,0)--(1.6,0);
\draw[ligne, black](0,0.5)--(0,1.3);
\end{tikzpicture} 
  &
2^9,4^9,5^2,6^8,8^8,9,10^6,12^6,\\
& & & 13,14^4,16^4,18^2,20^2,22,24\\\hline
\end{array}
$$
\caption{Models with integral spectrum and a 4d SCFT description as a $D_p^b(G)$ trinion.\label{tab:Trinions}}
\end{table}

\subsection{4d SCFTs: $D_p^b(G)$-Trinions}
\label{sec:b3-1-form}

We now turn our attention `fully smooth' models that have $b_3>0$. The 5d interpretation becomes more difficult in general, due to existence of an enhanced Coulomb branch. 
As before, the 4d SCFTs $\FT$ may not have a Lagrangian description, as can be seen from tables~\ref{tab:SmoothSmoothb31}, \ref{tab:SmoothSmoothb32}, \ref{tab:SmoothSmoothb33}, \ref{tab:SmoothSmoothb34} and  \ref{tab:SmoothSmoothb35}. Some of these models have an interpretation in terms of the $D_p^b(G)$-trinions introduced in \cite{Closset:2020afy}. 
Using the notation for the trinions introduced therein, namely:
\be
\begin{tikzpicture}[x=.6cm,y=.6cm]
\node at (-3,0) {$D_{p_2}^{b_2}(G)$};
\node at (0,0) {$G$};
\node at (3,0) {$D_{p_1}^{b_1}(G)$};
\node at (0,2) {$D_{p_3}^{b_3}(G)$};
\draw[ligne, black](-1.6,0)--(-0.6,0);
\draw[ligne, black](0.6,0)--(1.6,0);
\draw[ligne, black](0,0.5)--(0,1.3);
\end{tikzpicture}
\ee
the models with 4d trinion description are summarized in table \ref{tab:Trinions}.

\subsection{4d SCFTs with $c-a<0$}
Another interesting class of models are those for which the conformal anomalies of $\FTfour$ satisfy the inequality $c<a$. Such 4d SCFTs are slightly peculiar, and appear to be more `rare'; from the point of view of the superconformal index, $c-a<0$ is analogous to having $c<0$ for a 2d CFT -- see {\it e.g.} \cite{DiPietro:2014bca, Buican:2015ina, ArabiArdehali:2015ybk, Beem:2017ooy}. It turns out that there are only 8 canonical IHS for which $c<a$ in our database, and they are all smoothable models with $b_3>0$.
Their basic data is shown in table \ref{tab:abigggerc}.
\begin{table}
$$
\begin{array}{|c|c|c|c|c|c|c|c|}\hline
\text{Type}&  F(x,y,z,t)& r=\h d_H &f &d_H=\h r &   24(c-a) & b_3 & \mathfrak{t}_2\cr \hline\hline
\{1,1\}\{2, 5,5, 5\}\; &\; x^2 + y^5 + z^5+t^5&\; 2   \; &\;  0\;&\; 32 \; & \;  -4\; &\; 12\; & \; \Z_2^{12}\;\\
\{1,1\}\{2,3,8,8\}\; &\; x^2 + y^3 + z^8+ t^8   \; &\; 2&\;  0\;&  \;49 & \; -1\; & \;  6\; & \; \Z_3^6\\
\{1,1\}\{2, 3, 9, 9\}\; &\;  x^2 + y^3 + z^9+t^9   \;&\; 3\; &\;  0\;&\;  64\; &\; -4\; & \; 14 & \; \Z_2^{14}\;\\
\{1,1\}\{3,4,4,4\}\; &\; x^3 + y^4 + z^4+t^4   \;&\; 3 &\;  0\;&\;  27\; &\; -3\; & \; 12 &\; \Z_3^{6}\;\\
\{2,1\}\{2,7,7,3\}\; &\; x^2 + y^7 + z^7+ t^3 z   \; &\; 3 &\;  0\;&\;  45\; &\; -3\; & \;12  &\;\Z_2^{12}\;\\
\{1,1\}\{2, 3,10,10\}\; &\; x^2 + y^3 + z^{10}+t^{10}   \; &\; 5&\;  0\;&\;  81\; &\; -3\; & \; 16 & \;\Z_3^8 \\
\{2,1\}\{2, 3, 15, 7\}\; &\; x^2 + y^3  + z^{15}+ t^7 z  \;&\; 5 &\;  0\;&\;  91\; &\; -1\; & \; 12 &\; \Z_2^{12}\;\\
\{1,1\}\{2,5,6,6\}\; &\; x^2 + y^5 + z^6+ t^6   \; &\; 6&\;  0\;&\;  50\; &\; -2\; & \; 16 &\; \Z_5^{4}\;\\\hline
\end{array}
$$
\caption{Table of 4d SCFTs with $c<a$ engineered from IHS in IIB. \label{tab:abigggerc}}
\end{table}

The 4d Coulomb branch spectrum of all these singularities except for the type \{1,1\}\{2,3,8,8\} is integer-valued, $\Delta_\alpha\in \Z$. (And these 7 models are actually `fully smooth'.) Note that we have:
\be
24(c-a)=n_h - n_v= r+f -\half b_3~,
\ee
in agreement with the physical interpretation of the 3-cycles as vector multiplets in IIB. 
We will now provide the corresponding 4d SCFTs  $\FTfour$ (whose description is simply related to the electric and magnetic quiverines in 3d, since $f=0$ \cite{Closset:2020scj, Closset:2020afy}). We also study the resolution of these singularities and identify the 5d gauge theory phase if it exists. 

\subsubsection{Rank 2: \{1,1\}\{2,5,5,5\}}
\label{sec:2555}

This case was also discussed in \cite{Closset:2020scj}. We have the Lagrangian description of $\FTfour$ as the quiver:
\be
  \begin{tikzpicture}[x=.7cm,y=.7cm]
\draw[ligne, black](0,0)--(10,0);
\node[bd] at (0,0) [label=below:{{\scriptsize Spin(5)}}] {};
\node[bd] at (1,0) [label=above:{{\scriptsize$Sp(3)$}}] {};
\node[bd] at (2,0) [label=below:{{\scriptsize Spin(11)}}] {};
\draw[ligne, black](2,0)--(2,2);
\node[bd] at (2,1) [label=above:{{\qquad\;\scriptsize$Sp(2)$}}] {};
\node[bd] at (2,2)[label=left:{{\scriptsize Spin(1)}}] {};
\node[bd] at (3,0) [label=above:{{\scriptsize$Sp(4)$}}] {};
\node[bd] at (4,0) [label=below:{{\scriptsize Spin(9)}}] {};
\node[bd] at (5,0) [label=above:{{\scriptsize $Sp(3)$}}] {};
\node[bd] at (6,0) [label=below:{{\scriptsize Spin(7)}}] {};
\node[bd] at (7,0) [label=above:{{\scriptsize$Sp(2)$}}] {};
\node[bd] at (8,0) [label=below:{{\scriptsize Spin(5)}}] {};
\node[bd] at (9,0) [label=above:{{\scriptsize$Sp(1)$}}] {};
\node[bd] at (10,0) [label=below:{{\scriptsize Spin(3)}}] {};
\end{tikzpicture}
\ee
Note that the six ${\rm Spin}(2n+1)$ ($n>0$) gauge groups give rise to a $\mathfrak{f}=\mb{Z}_2^6$ 1-form symmetry (the vector of ${\rm Spin}(2n+1)$ does not break the $\Z_2$ center), in agreement with the geometric determination of $\mathfrak{t}_2$.

Let us also discuss the resolution of this model. Starting with the hypersurface equation 
$x_1^2+x_2^5+x_3^5+x_4^5=0$, the resolution sequence is
\be
\ba
&(x_1^{(2)},x_2^{(1)},x_3^{(1)},x_4^{(1)};\delta_1)\ ,\ (x_1,\delta_1;\delta_2)\ ,\ (\delta_1,\delta_2;\delta_3)
\ea
\ee
The resolved equation is
\be
x_1^2\delta_2+(x_2^5+x_3^5+x_4^5)\delta_1=0\,.
\ee
The set $\delta_1=0$ is empty, so there are only two exceptional divisors $S_1:\ \delta_2=0$ and $S_2:\ \delta_3=0$. The triple intersection numbers are
\be
\includegraphics[height=2cm]{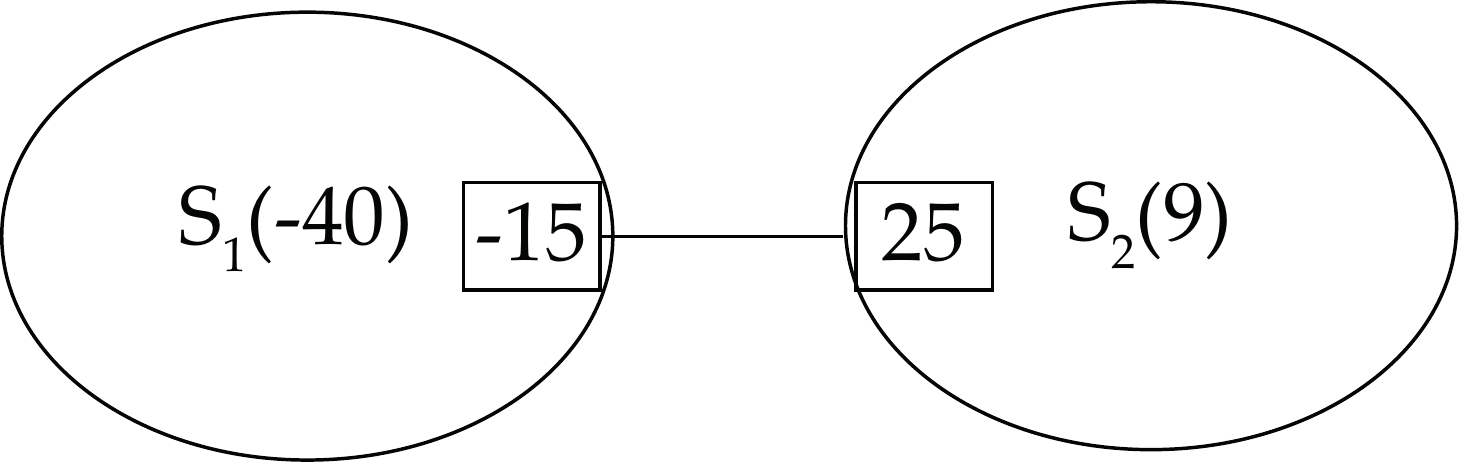}
\ee
Hence $S_1\cdot S_2$ is a genus-6 curve, and $S_1$ is a ruled surface over that curve. Note that  $S_2$ is a $\mb{P}^2$ and $S_1\cdot S_2$ has equation $\delta_2=\delta_3=x_2^5+x_3^5+x_4^5=0$, which is a degree-5 curve on $\mb{P}^2$ with genus $g=6$. Since $S_1$ has 12 3-cycles, the total $b_3$ of the resolved CY3 is $b_3(\t\MG)=12$.

The presence of the 3-cycles in $\t\MG$ makes the M-theory interpretation more subtle, as mentioned above. We can obtain a distinct and more `conventional' model by performing a geometric transition, as described in \cite{Jefferson:2018irk}, to obtain a new  Calabi-Yau threefold with distinct asymptotics and $b_3=0$. In the present example, one performs a complex structure deformation of the geometry, such that the intersection curve $S_1\cdot S_2$ has six double point singularities. Then we perform flop operations six times by blowing up six double points on $S_2$. After these flops, the triple intersection numbers of the new surfaces $S_1'$ and $S_2'$ are
\be
\includegraphics[height=2cm]{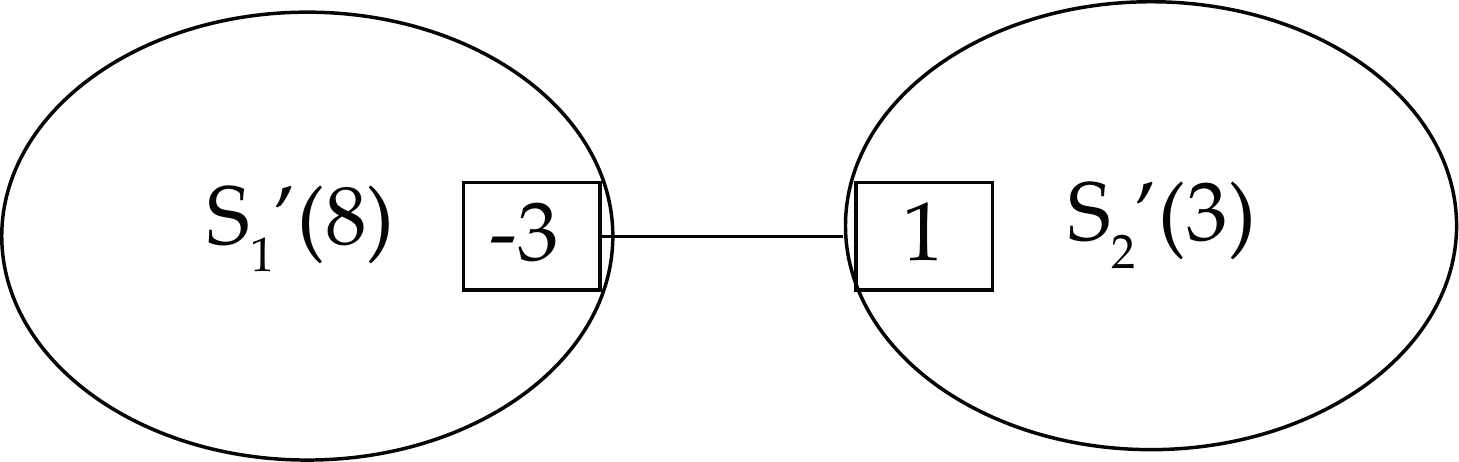}
\ee
Hence $S'_1$ is a $\mb{F}_3$ and $S'_2$ is a $dP_6$. These triple intersection numbers exactly give rise to the 5d SCFT with IR description $SU(3)_{9/2}+5\bm{F}$, $Sp(2)+3\bm{F}+2\bm{AS}$ or $G_2+5\bm{F}$, and UV flavor symmetry $G_F^{\rm 5d}=Sp(6)$.
Before the flops, the geometry gives $f=0$. After the flops, there are six new 2-cycles in the flopped geometry, which give rise to the flavor rank $f' =6$ that matches $G_F^{\rm 5d}$.

\subsubsection{Rank 2: \{1,1\}\{2,3,8,8\}}
\label{sec:2388}

The 4d CB spectrum of this model is non-integral. In \cite{Closset:2020afy}, a non-Lagrangian description of $\FTfour$ is found, where $D_p^b$ denotes the $D_p^b(G)$ theory~\cite{Cecotti:2013lda,Giacomelli:2017ckh, Giacomelli:2020ryy}, here with $G=E_6$ for the central gauge group:
\be
 \begin{tikzpicture}[x=.6cm,y=.6cm]
\node at (-4,0) {$\FTfour=$};
\node at (-2.3,0) {$D_8^{12}$};
\node at (0,0) {$E_6$};
\node at (2.3,0) {$D_{1}^{9}$};
\node at (0,2) {$D_{8}^{12}$};
\draw[ligne, black](-1.6,0)--(-0.6,0);
\draw[ligne, black](0.6,0)--(1.6,0);
\draw[ligne, black](0,0.5)--(0,1.3);
\end{tikzpicture} 
\ee

The singularity $x_1^2+x_2^3+x_3^8+x_4^8=0$  is fully resolved by
\be
\ba
&(x_1^{(3)},x_2^{(2)},x_3^{(1)},x_4^{(1)};\delta_1)\ ,\ (x_1,x_2,\delta_1;\delta_2)\,.
\ea
\ee
The resolved equation is completely smooth:
\be
x_1^2+x_2^3\delta_2+x_3^8\delta_1^2+x_4^8\delta_1^2=0\,.
\ee
The triple intersection numbers of $S_1:\delta_1=0$ and $S_2:\delta_2=0$ are computed to be
\be
\includegraphics[height=2cm]{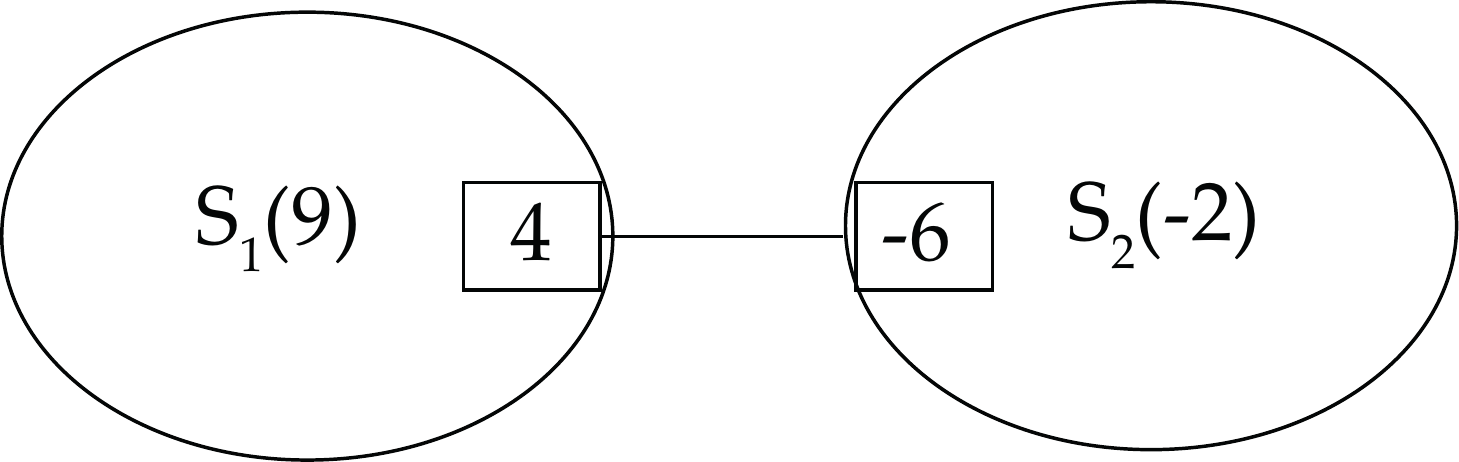}
\ee
These numbers appears to match the rank-2 5d SCFT with an $SU(3)_{1/2}+9\bm{F}$ IR description and $G_F=SO(20)$ flavor symmetry. The intersection curve $S_1\cdot S_2$ is an irreducible curve. 

However, this model has flavor rank $f=0$, which leads to a contradiction. The reason is that the equation of $S_2$ is
\be
x_1^2+x_3^8\delta_1^2+x_4^8\delta_1^2=0\,.
\ee
At a generic locus $\delta_1\neq 0$, $x_1\neq 0$, the equation is a smooth genus-3 curve, because the Newton polytope with vertices $(2,0,0)$, $(0,8,0)$ and $(0,0,8)$ has exactly 3 interior points. On the other hand, $S_2$ is singular at the point
\be
x_1=x_3=x_4=\delta_2=0\,.
\ee
Note that this point is not a singular point in the resolved CY3 $\t \MG$, but it is a singular point on $S_2$.

From the genus-3 curve on $S_2$, we have
\be
b_3(S_2)=b_3(\t \MG)=6\,,
\ee
although $S_1\cdot S_2$ has genus 0. The 5d physical interpretation of this geometry is unclear.

\subsubsection{Rank 3: \{1,1\}\{2,3,9,9\}}
\label{sec:2399}
The 4d SCFT has the Lagrangian description:
\be
  \begin{tikzpicture}[x=.7cm,y=.7cm]
\draw[ligne, black](0,0)--(10,0);
\node[bd] at (0,0) [label=below:{{\scriptsize Spin(8)}}] {};
\node[bd] at (1,0) [label=above:{{\scriptsize$Sp(6)$}}] {};
\node[bd] at (2,0) [label=below:{{\scriptsize Spin(20)}}] {};
\node[bd] at (3,0) [label=above:{{\scriptsize$Sp(8)$}}] {};
\node[bd] at (4,0) [label=below:{{\scriptsize Spin(16)}}] {};
\node[bd] at (5,0) [label=above:{{\scriptsize$Sp(6)$}}] {};
\node[bd] at (6,0) [label=below:{{\scriptsize Spin(12)}}] {};
\node[bd] at (7,0) [label=above:{{\scriptsize$Sp(4)$}}] {};
\node[bd] at (8,0) [label=below:{{\scriptsize Spin(8)}}] {};
\node[bd] at (9,0) [label=above:{{\scriptsize$Sp(2)$}}] {};
\node[bd] at (10,0) [label=below:{{\scriptsize Spin(4)}}] {};
\draw[ligne, black](2,0)--(2,1);
\node[bd] at (2,1) [label=above:{{\scriptsize $Sp(4)$ }}] {};
\end{tikzpicture} 
\ee
It is a simple exercise to check that this quiver has a 1-form symmetry $\mathfrak{f}=\Z_2^7$, in agreement with the geometric determination.  The resolution of the singularity $x_1^2+x_2^3+x_3^9+x_4^9=0$ is achieved by
\be
\ba
&(x_1^{(3)},x_2^{(2)},x_3^{(1)},x_4^{(1)};\delta_1)\ ,\ (x_1,x_2,\delta_1;\delta_2)\ ,\ (x_1,\delta_2;\delta_3)\ ,\ (\delta_2,\delta_3;\delta_4) \,.
\ea
\ee
The resolved equation is
\be
x_1^2\delta_3+x_2^3\delta_2+(x_3^9+x_4^9)\delta_1^3 \delta_2=0\,.
\ee
The set $\delta_2=0$ is empty. The other surface components are irreducible. 
The triple intersection numbers of $S_1:\delta_1=0$, $S_2:\delta_3=0$ and $S_3:\delta_4=0$ are
\be
\includegraphics[height=2cm]{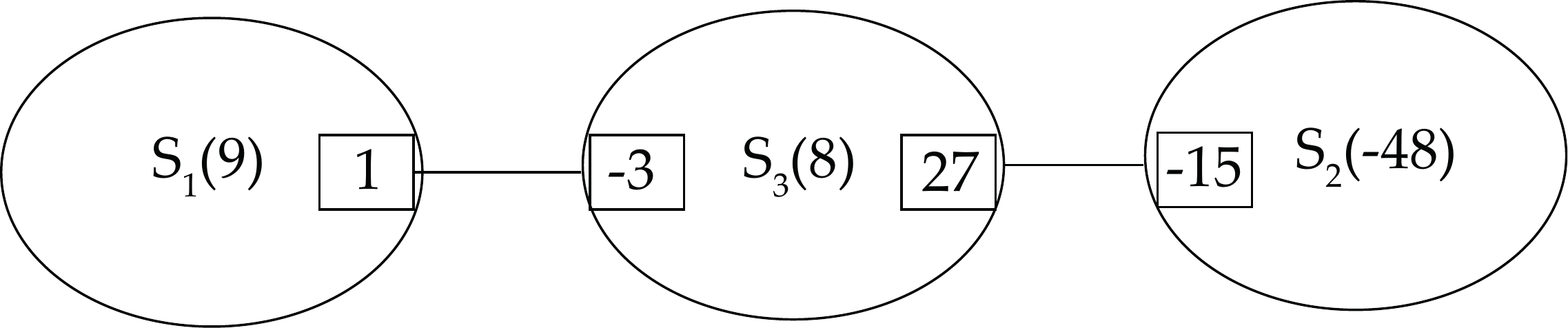}
\ee
The intersection curve $S_2\cdot S_3$ is a genus-7 curve, and the surface $S_2$ is ruled over that curve. $S_1$ is a $\mb{P}^2$ and $S_4$ is a $\mb{F}_3$. There are 14 3-cycles on $S_3$, and $b_3(\MG)=14$. 
After the flop that results from blowing up $S_3$ at seven double points on $S_2\cdot S_3$, the triple intersection numbers become
\be
\includegraphics[height=2cm]{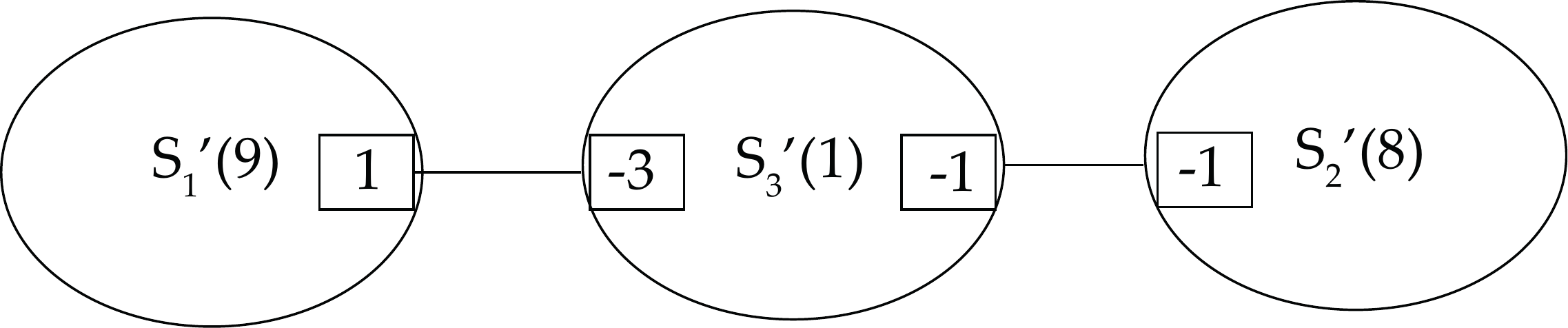}
\ee
The model does not appear to have a 5d gauge-theory description.

\subsubsection{Rank 3: \{1,1\}\{3,4,4,4\}}
\label{sec:3444}

A non-Lagrangian description of $\FTfour$ was found in  \cite{Closset:2020afy}:
\be
 \begin{tikzpicture}[x=.6cm,y=.6cm]
\node at (-4,0) {$\FTfour=$};
\node at (-2.3,0) {$D_2^{12}$};
\node at (0,0) {$E_6$};
\node at (2.3,0) {$D_{4}^{12}$};
\node at (0,2) {$D_{4}^{12}$};
\draw[ligne, black](-1.6,0)--(-0.6,0);
\draw[ligne, black](0.6,0)--(1.6,0);
\draw[ligne, black](0,0.5)--(0,1.3);
\end{tikzpicture} 
\ee
The singularity $x_1^3+x_2^4+x_3^4+x_4^4=0$ leads to the resolution sequence
\be
\ba
&(x_1,x_2,x_3,x_4;\delta_1)\ ,\ (x_1,\delta_1;\delta_2)\ ,\ (\delta_1,\delta_2;\delta_3)\ ,\ (\delta_1,\delta_3;\delta_4)\,.
\ea
\ee
The resolved equation is
\be
x_1^3\delta_2^2\delta_3+(x_2^4+x_3^4+x_4^4)\delta_1=0\,.
\ee
The set $\delta_1=0$ is empty, hence there are only three exceptional divisors $S_1:\ \delta_2=0$, $S_2:\ \delta_3=0$ and $S_3:\ \delta_4=0$.  The triple intersection numbers are
\be
\includegraphics[height=2cm]{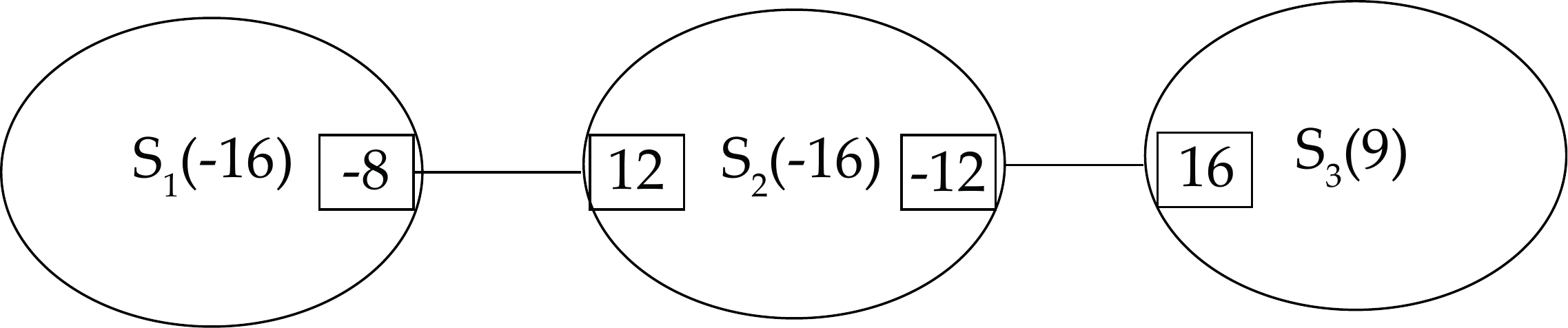}
\ee
Hence the intersection curves $S_1\cdot S_2$ and $S_2\cdot S_3$ both have genus $g=3$, and  both $S_1$ and $S_2$ are  ruled over genus-three curves. The divisor $S_3$ is a $\mb{P}^2$. Since $S_1$ and $S_2$ both have 6 3-cycles, we have $b_3(\t \MG)=12$. 
If we perform flops by blowing up $S_3$ at three double points on $S_2\cdot S_3$,  we end up with
\be
\includegraphics[height=2cm]{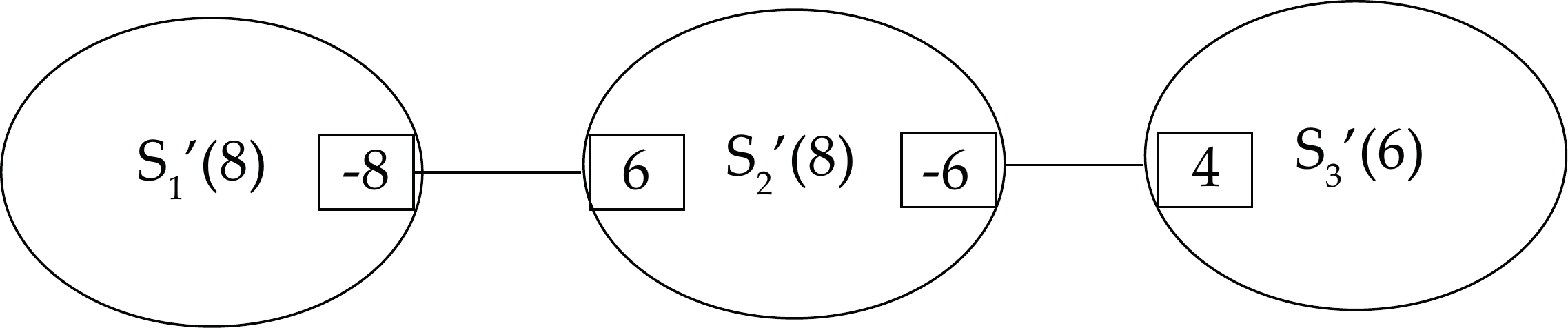}
\ee

\subsubsection{Rank 3: \{2,1\}\{2,7,7,3\}}
\label{sec:2773}

The 4d SCFT for this model has the Lagrangian description:
\be
  \begin{tikzpicture}[x=.7cm,y=.7cm]
\draw[ligne, black](0,0)--(9,0);
\node[bd] at (0,0) [label=above:{{\scriptsize $Sp(1)$}}] {};
\node[bd] at (1,0) [label=below:{{\scriptsize Spin(8)}}] {};
\node[bd] at (2,0) [label=above:{{\scriptsize $Sp(5)$}}] {};
\node[bd] at (3,0) [label=below:{{\scriptsize Spin(16)}}] {};
\node[bd] at (4,0) [label=above:{{\scriptsize $Sp(6)$}}] {};
\node[bd] at (5,0) [label=below:{{\scriptsize Spin(12)}}] {};
\node[bd] at (6,0) [label=above:{{\scriptsize $Sp(4)$}}] {};
\node[bd] at (7,0) [label=below:{{\scriptsize Spin(8)}}] {};
\node[bd] at (8,0) [label=above:{{\scriptsize $Sp(2)$}}] {};
\node[bd] at (9,0) [label=below:{{\scriptsize Spin(4)}}] {};
\draw[ligne, black](3,0)--(3,1);
\node[bd] at (3,1) [label=above:{{\scriptsize $Sp(3)$ }}] {};
\end{tikzpicture} 
\ee
This quiver has a 1-form symmetry $\mathfrak{f}=\Z_2^6$. The singularity $x_1^2+x_2^7+x_3^7+x_4^3 x_3=0$ has a  resolution sequence
\be
\ba
&(x_1^{(2)},x_2^{(1)},x_3^{(1)},x_4^{(1)};\delta_1)\ ,\ (x_1,x_4,\delta_1;\delta_2)\ ,\ (x_1,\delta_2;\delta_3)\ ,\ (\delta_2,\delta_3;\delta_4)
\ea
\ee
The resolved equation is
\be
x_1^2\delta_3+x_2^7\delta_1^3\delta_2+x_3^7\delta_1^3\delta_2+x_4^3 x_3\delta_2=0\,.
\ee
The triple intersection numbers of $S_1:\delta_1=0$, $S_2:\delta_3=0$, $S_3:\delta_4=0$ are
\be
\includegraphics[height=4cm]{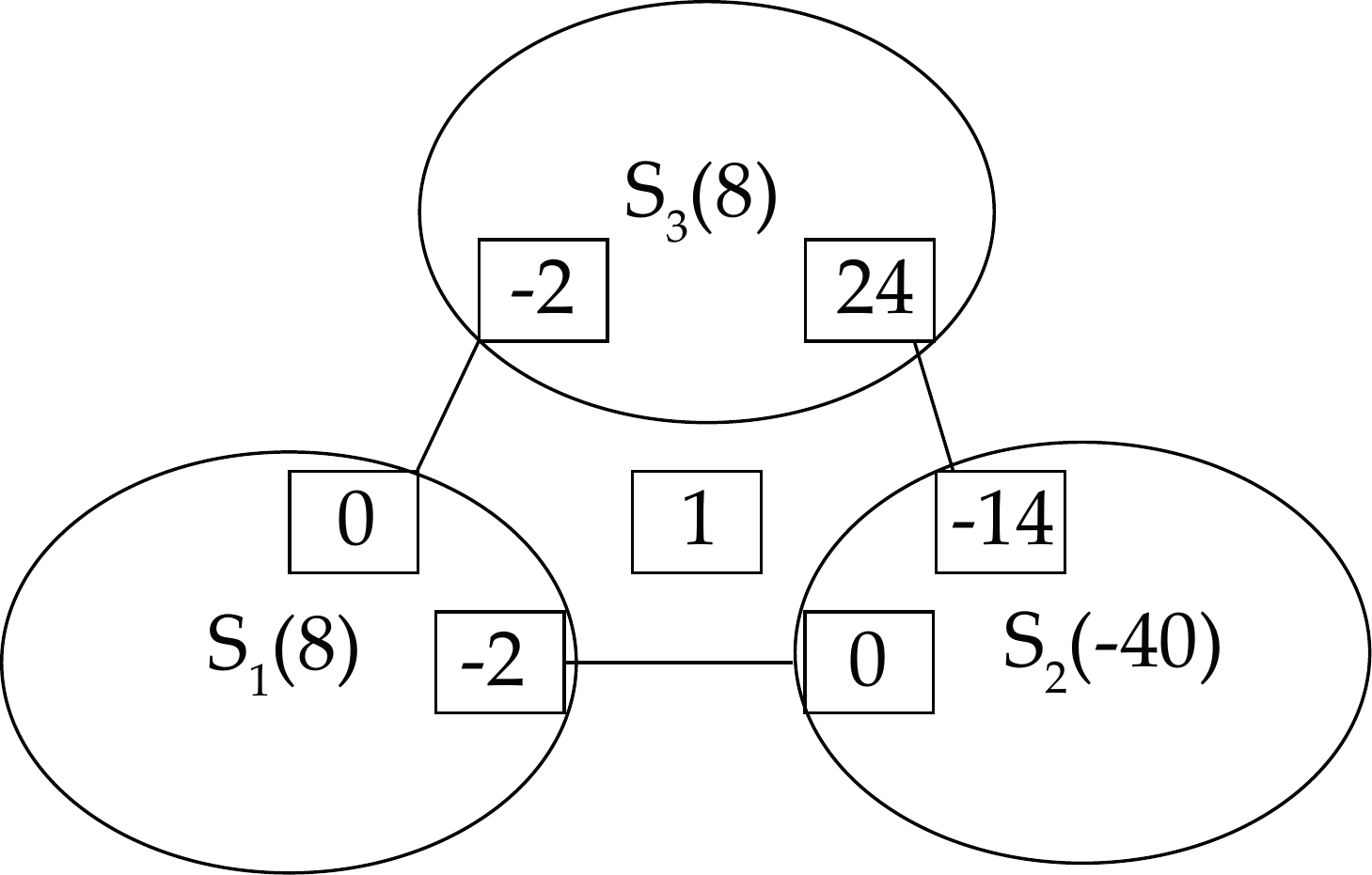}
\ee
$S_3$ is ruled over a genus-6 curve $S_3\cdot S_4$.
After the flop by blowing up $S_3\cdot S_4$ at the six double points on $S_4$, the new intersection numbers are
\be
\includegraphics[height=4cm]{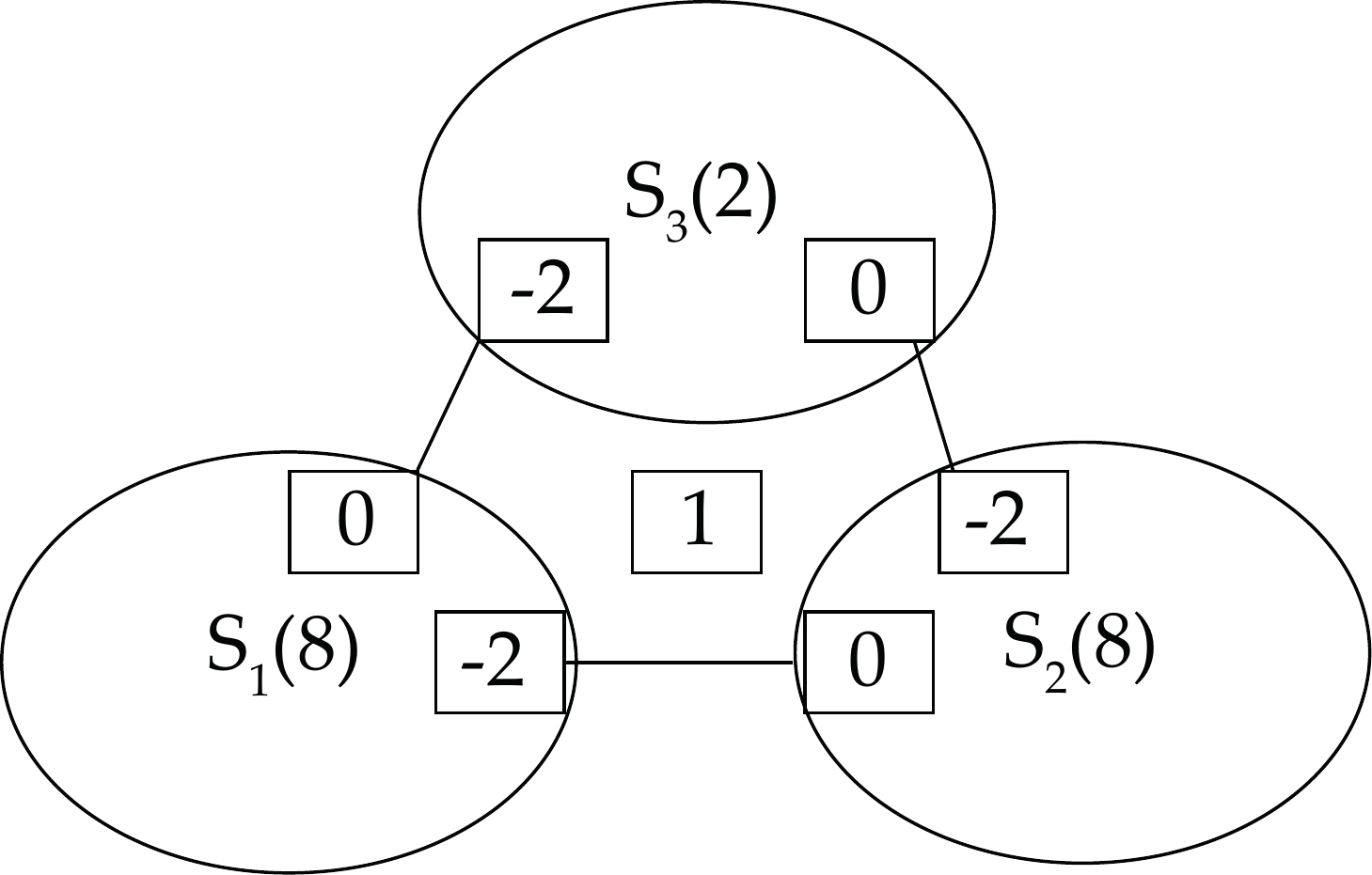}
\ee
Since the assignment of base and ruling curves on $S_i$ cannot match, this geometry does not have a gauge theory description.

\subsubsection{Rank 5: \{1,1\}\{2,3,10,10\}}
\label{sec:231010}

As we noted earlier this model has a trinion represetation in 4d 
\be
  \begin{tikzpicture}[x=.6cm,y=.6cm]
\node at (-2.3,0) {$D_{6}^{24}$};
\node at (0,0) {$SO(26)$};
\node at (2.3,0) {$D_{3}^{24}$};
\node at (0,2) {$D_{2}^{24}$};
\draw[ligne, black](-1.6,0)--(-1,0);
\draw[ligne, black](1,0)--(1.6,0);
\draw[ligne, black](0,0.5)--(0,1.3);
\end{tikzpicture} 
\ee
The singularity $x_1^2+x_2^3+x_3^{10}+x_4^{10}=0$ has a resolution sequence
\be
\ba
&(x_1^{(3)},x_2^{(2)},x_3^{(1)},x_4^{(1)};\delta_1)\ ,\ (x_1,x_2,\delta_1;\delta_2)\ ,\ (x_1,\delta_2;\delta_3)\ ,\ (x_2,\delta_3;\delta_4)\cr
&(\delta_2,\delta_3;\delta_5)\ ,\ (\delta_3,\delta_4;\delta_6)\ ,\ (\delta_3,\delta_6;\delta_7)
\ea
\ee
The resolved equation is
\be
x_1^2\delta_3+x_2^3\delta_2 \delta_4^2\delta_6+(x_3^{10}+x_4^{10})\delta_1^4 \delta_2^2\delta_3\delta_5^2=0\,.
\ee
The sets $\delta_2=0$ and $\delta_3=0$ is empty. The other surface components are irreducible. The Coulomb branch rank is $r=5$.
The triple intersection numbers of $S_1:\delta_1=0$, $S_2:\delta_4=0$, $S_3:\delta_5=0$, $S_4:\delta_6=0$ and $S_5:\delta_7=0$ are
\be
\includegraphics[height=2cm]{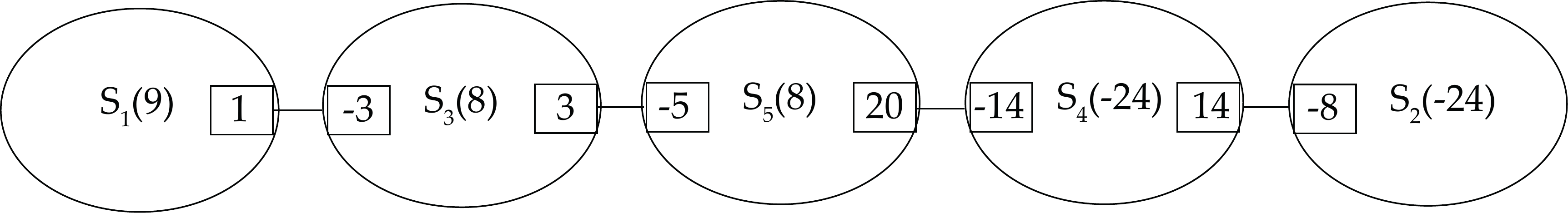}
\ee
The surface components $S_2$ and $S_4$ are ruled over a genus-4 curve, and the other surface components are rational. On each of $S_2$ and $S_4$, there are eight 3-cycles. In total we have $b_3(\t \MG)=16$.
After a flop performed by blowing up four double points on $S_5\cdot S_4$, we obtain the geometry
\be
\includegraphics[height=2cm]{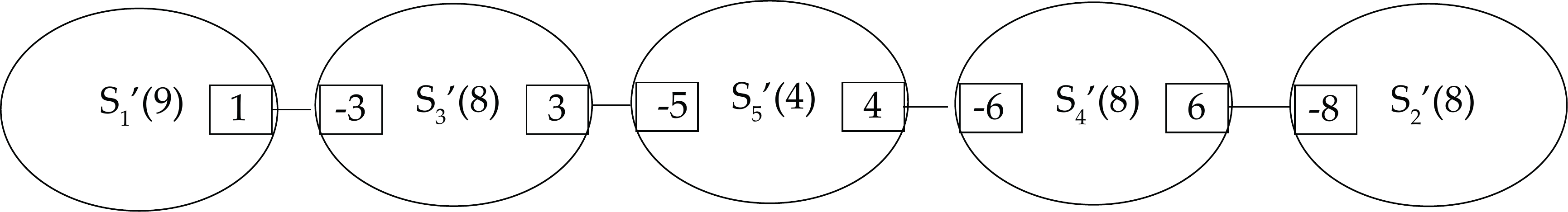}
\ee

\subsubsection{Rank 5: \{2,1\}\{2, 3, 15, 7\}}
\label{sec:23157}

For this model, we did not find a Lagrangian nor a generalised quiver description. The equation $x_1^2+x_2^3+x_3^{15}+x_4^7 x_3=0$ leads to the resolution sequence
\be
\ba
&(x_1^{(3)},x_2^{(2)},x_3^{(1)},x_4^{(1)};\delta_1)\ ,\ (x_1,x_2,\delta_1;\delta_2)\ ,\ (x_1^{(2)},x_2^{(1)},x_4^{(1)},\delta_2^{(1)};\delta_3)\ ,\ (x_1,x_2,\delta_3;\delta_4)\cr
&(x_1,\delta_4;\delta_5)\ ,\ (\delta_4,\delta_5;\delta_6)
\ea
\ee
The resolved equation is
\be
x_1^2\delta_5+x_2^7\delta_1^8\delta_2^{13}\delta_3^{16}\delta_4^{21}\delta_5^{20}\delta_6^{19}+x_3^{15}\delta_1^9\delta_2^7\delta_3^3\delta_4+x_4^7 x_3\delta_1^2\delta_3^3\delta_4=0\,.
\ee
The set $\delta_4=0$ is empty, and the other divisors are irreducible, hence the 5d rank is $r=5$. The triple intersection numbers of $S_1:\delta_1=0$, $S_2:\delta_2=0$, $S_3:\delta_3=0$, $S_4:\delta_5=0$ and $S_5:\delta_6=0$ are
\be
\includegraphics[height=6cm]{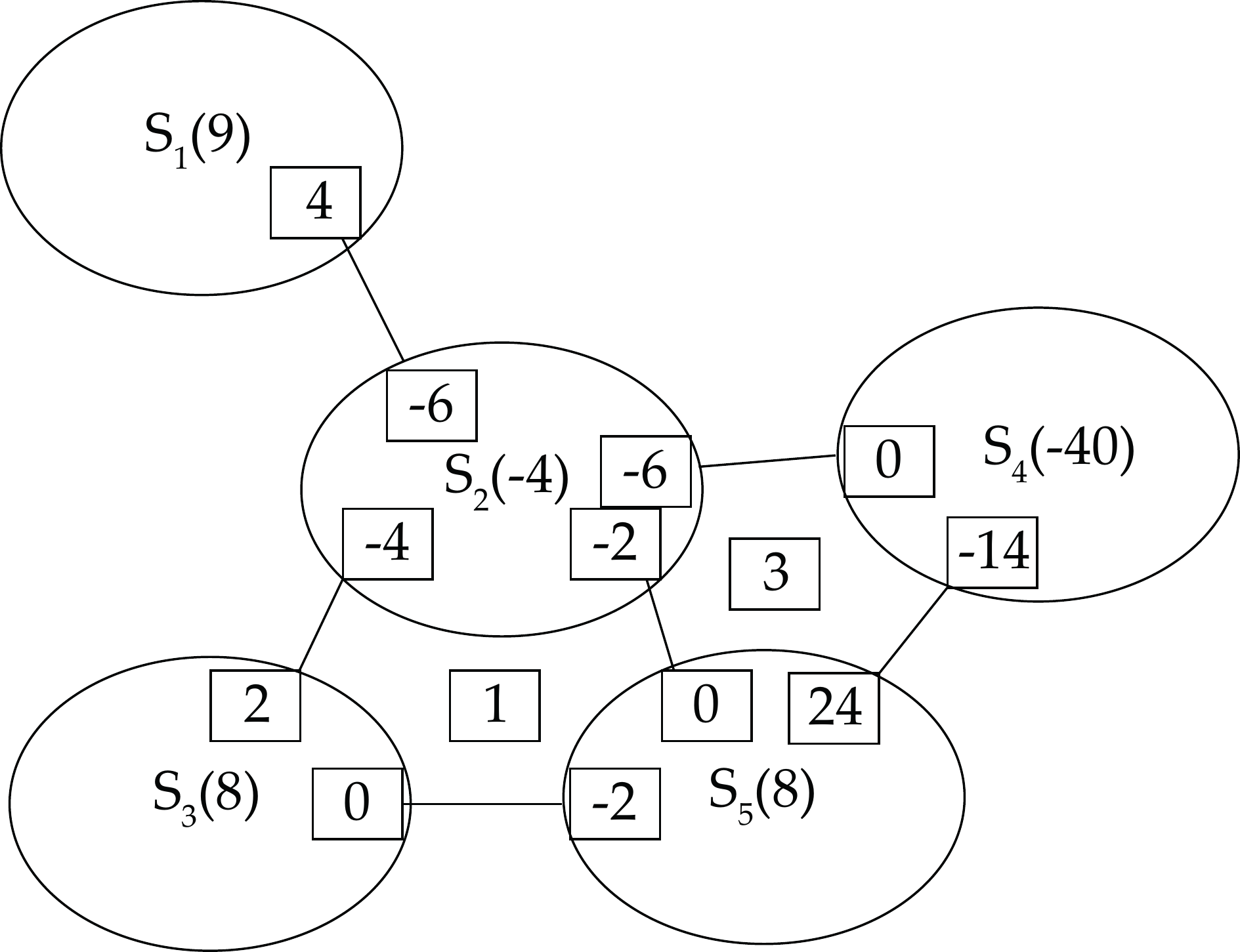}
\ee
$S_4$ is ruled over a genus-6 curve, the other surfaces are rational. There are 12 3-cycles on $S_4$, and $b_3(\t \MG)=12$.

\subsubsection{Rank 6: \{1,1\}\{2,5,6,6\}}
\label{sec:2566}

This 4d SCFT for this model  has a trinion description \cite{Closset:2020afy}:
\be
 \begin{tikzpicture}[x=.6cm,y=.6cm]
 \node at (-4,0) {$\FTfour=$};
\node at (-2.3,0) {$D_{6}^{30}$};
\node at (0,0) {$E_8$};
\node at (2.3,0) {$D_{6}^{30}$};
\node at (0,2) {$D_{1}^{20}$};
\draw[ligne, black](-1.6,0)--(-0.6,0);
\draw[ligne, black](0.6,0)--(1.6,0);
\draw[ligne, black](0,0.5)--(0,1.3);
\end{tikzpicture} 
\ee
The singularity $x_1^2+x_2^5+x_3^6+x_4^6=0$ leads to the  resolution sequence
\be
\ba
&(x_1^{(2)},x_2^{(1)},x_3^{(1)},x_4^{(1)};\delta_1)\ ,\ (x_1,\delta_1;\delta_2)\ ,\ (x_2,\delta_2;\delta_3)\ ,\ (\delta_1,\delta_2;\delta_4)\cr
&(\delta_2,\delta_3;\delta_5)\ ,\ (\delta_2,\delta_5;\delta_6)\ ,\ (\delta_2,\delta_6;\delta_7)\ ,\ (\delta_2,\delta_7;\delta_8)
\ea
\ee
The resolved equation is
\be
x_1^2\delta_2+x_2^5\delta_1\delta_3^4\delta_5^3\delta_6^2\delta_7+(x_3^6+x_4^6)\delta_1^2\delta_2\delta_4^2=0\,.
\ee
The sets $\delta_1=0$ and $\delta_2=0$ are empty, and the other divisors are irreducible, so that $r=6$. The triple intersection numbers of $S_1:\delta_3=0$, $S_2:\delta_4=0$, $S_3:\delta_5=0$, $S_4:\delta_6=0$, $S_5:\delta_7=0$ and $S_6:\delta_8=0$ are
\be
\includegraphics[height=1.8cm]{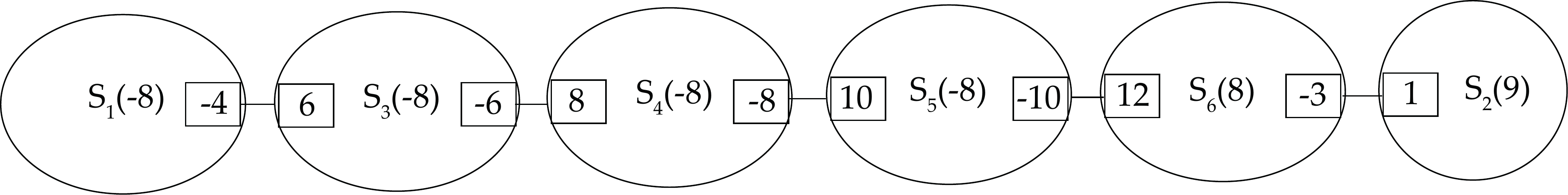}
\ee
The divisors $S_1$, $S_3$, $S_4$ and $S_5$ are all  ruled over a genus-2 curve. $S_6$ is $\mb{F}_3$ and $S_2$ is $\mb{P}^2$. $S_1$, $S_3$, $S_4$ and $S_5$ each has four 3-cycles, and thus $b_3(\t \MG)=16$. After the flop performed by blowing up two double points on $S_6\cdot S_5$, we obtain the geometry
\be
\includegraphics[height=1.8cm]{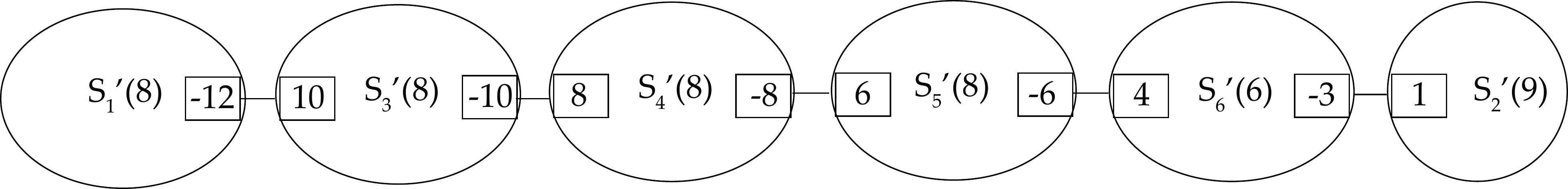}
\ee

\subsection{4d SCFTs with $a=c$}
Another noteworthy class of models consists of the smoothable models such that $a=c$. They all satisfy
\be
f=0~,\qquad  r=\frac{1}{2}b_3=\h d_H~,\qquad  \h r=d_H~.
\ee
Therefore, any such model with $r>0$ is either smoothable or it has residual terminal singularities that correspond to some $a=c$ model.%
\footnote{When there are several terminal singularities, the residual IR theory on the 4d HB can be a direct sum of irreducible theories so that the total central charges satisfy $a=c$.} 
They also all have higher-form symmetries ($\mathfrak{t_2}$ non-trivial). 
These cases include the following infinite series, which are called $E_k^{(l)}$  models in \cite{Closset:2020afy}. We have $E_6^{(3n-2)}$ and $E_6^{(3n-1)}$ models:
\be\nn
\begin{array}{|c|c|c|c|c|c|c|c|c|}\hline
\text{Type} (a,b,c,d) & E_k^{(l)} & F & r & f & d_H & a=c & b_3 & \mathfrak{t}_2\cr
\hline\hline
\{1,1\}  \{3,3,3,3n+1\} & E_6^{(3n-2)} & x_1^3+x_2^3+x_3^3+x_4^{3n+1} & n & 0 & 12n & 2n(3n+2) & 2n & \Z_{3n+1}^2\cr
\{1,1\}  \{3,3,3,3n+2\} & E_6^{(3n-1)} & x_1^3+x_2^3+x_3^3+x_4^{3n+2} & n & 0 & 12n+4 & (2n+2)(3n+1) & 2n & \Z_{3n+2}^2\cr
\hline
\end{array}
\ee
The model $E_6^{(l)}$ have a 4d $G=SU(l+3)$ trinion description \cite{Closset:2020afy}:
\be
\begin{tikzpicture}[x=.6cm,y=.6cm]
\node at (-3.3,0) {$D_3$};
\node at (0,0) {$SU(l+3)$};
\node at (3.3,0) {$D_3$};
\node at (0,2) {$D_3$};
\draw[ligne, black](-2.6,0)--(-1.6,0);
\draw[ligne, black](1.6,0)--(2.6,0);
\draw[ligne, black](0,0.5)--(0,1.3);
\end{tikzpicture}
\ee
We similarly have the $E_7^{(4n-3)}$ and $E_7^{(4n-1)}$ models:
\be\nn
\begin{array}{|c|c|c|c|c|c|c|c|c|}\hline
\text{Type}  (a,b,c,d)  & E_k^{(l)} & F & r & f & d_H & a=c & b_3 & \mathfrak{t}_2\cr
\hline\hline
\{1,1\}  \{2,4,4,4n+1\} & E_7^{(4n-3)}& x_1^2+x_2^4+x_3^4+x_4^{4n+1} & n & 0 & 18n & 6n(2n+1) & 2n & \Z_{4n+1}^2\cr
\{1,1\}  \{2,4,4,4n+3\} & E_7^{(4n-1)}& x_1^2+x_2^4+x_3^4+x_4^{4n+3} & n & 0 & 18n+9 & 6(n+1)(2n+1) & 2n & \Z_{4n+3}^2\cr
\hline
\end{array}
\ee
as well as the $E_8^{(6n-5)}$ and $E_8^{(6n-1)}$ models:
\be\nn
\begin{array}{|c|c|c|c|c|c|c|c|c|}\hline
\text{Type}  (a,b,c,d)  & E_k^{(l)} & F & r & f & d_H & a=c & b_3 & \mathfrak{t}_2\cr
\hline\hline
\{1,1\}  \{2,3,6,6n+1\} & E_8^{(6n-5)}& x_1^2+x_2^3+x_3^6+x_4^{6n+1} & n & 0 & 30n & 10n(3n+1) & 2n & \Z_{6n+1}^2\cr
\{1,1\}  \{2,3,6,6n+5\} & E_8^{(6n-1)}& x_1^2+x_2^3+x_3^6+x_4^{6n+5} & n & 0 &  30n+20 & 10(n+1)(3n+2) & 2n & \Z_{6n+5}^2\cr
\hline
\end{array}
\ee
The $E_7^{(l)}$ and $E_8^{(l)}$ models have a 4d $SU(l+4)$ and $SU(l+6)$ trinion description, respectively:
\be
\begin{tikzpicture}[x=.6cm,y=.6cm]
\node at (-3.3,0) {$D_2$};
\node at (0,0) {$SU(l+4)$};
\node at (3.3,0) {$D_4$};
\node at (0,2) {$D_4$};
\draw[ligne, black](-2.6,0)--(-1.6,0);
\draw[ligne, black](1.6,0)--(2.6,0);
\draw[ligne, black](0,0.5)--(0,1.3);
\end{tikzpicture}\qquad\qquad
\begin{tikzpicture}[x=.6cm,y=.6cm]
\node at (-3.3,0) {$D_2$};
\node at (0,0) {$SU(l+6)$};
\node at (3.3,0) {$D_3$};
\node at (0,2) {$D_6$};
\draw[ligne, black](-2.6,0)--(-1.6,0);
\draw[ligne, black](1.6,0)--(2.6,0);
\draw[ligne, black](0,0.5)--(0,1.3);
\end{tikzpicture}
\ee
These trinion models were also recently discussed in \cite{Kang:2021lic}.
 We also find an interesting infinite family of $a=c$ models with $r=0$:
\be\nn
\begin{array}{|c|c|c|c|c|c|c|c|}\hline
\text{Type}  (a,b,c,d)  & F & r & f & d_H & a=c & b_3 & \mathfrak{t}_2\cr
\hline\hline
\{2,1\}\{2,2k+1,2k+1,2\} & x_1^2+x_2^{2k+1}+x_3^{2k+1}+x_4^2 x_3 & 0 & 0 & 2k(k+1) & \frac{1}{9}k(k+1)(4k+5) & 0 & \mb{Z}_2^{2k}\\
\hline
\end{array}
\ee
Finally, we find six `sporadic' models with $r>0$:
\be
\begin{array}{|c|c|c|c|c|c|c|c|}\hline
\text{Type}  (a,b,c,d)  & F & r & f & d_H & a=c & b_3 & \mathfrak{t}_2\cr
\hline\hline
\{1,1\}\{2,3,7,14\} & x_1^2+x_2^3+x_3^7+x_4^{14}  & 3 & 0 & 78 & 286 & 6 &  \Z_3^6 \cr
\{1,1\}\{2,3,7,21\} & x_1^2+x_2^3+x_3^7+x_4^{21}  & 6 & 0 & 120 & 860 & 12 &   \Z_2^{12} \cr
\{2,1\}\{3,5,5,2\}  & x_1^3+x_2^5+x_3^5+x_3 x_4^2 & 2 & 0 & 24 & 34 & 4  & \Z_3^4 \cr
\{2,1\}\{3,7,7,2\} &  x_1^3 + x_2^7 +x_3^7+x_3 x_4^2   & 6 & 0 & 48 & 176 & 12 &   \Z_3^6 \cr
\{2,1\}\{4,5,5,2\} & x_1^4 + x_2^5 +x_3^5+x_3 x_4^2   &  6 & 0 & 36 & 126 & 6 &   \Z_4^4 \cr
\{3,1\}\{2,11,3,4\} & x_1^2 + x_2^{11} + x_4^4 x_3 + x_4 x_3^3    & 5 & 0 & 60 & 230 & 10 &    \Z_2^{10}  \cr\hline
\end{array}
\ee
All of the six models have a crepant resolution with no terminal singularities. Among these models, the type $\{2,1\}\{4,5,5,2\}$ has a crepant resolution with no singular divisor, see the entry in table~\ref{tab:SmoothSmoothb33}. The other five models have singular divisors in their crepant resolution. We will now discuss the `fully smooth' model in more detail. 

\subsubsection{Rank 6: \{2,1\}\{4,5,5,2\} }

The theory does not have known quivernines $\EQfour$ and $\MQfive$. The resolution sequence of the singularity $x_1^4 + x_2^5 +x_3^5+x_3 x_4^2=0$ reads
\be
\ba
&(x_1,x_2,x_3,x_4;\delta_1)\ ,\ (x_4,\delta_1;\delta_2)\ ,\ (x_3,\delta_1;\delta_3)\ ,\ (x_1,\delta_2;\delta_4)\ ,\ (\delta_1,\delta_2;\delta_5)\cr
&(\delta_2,\delta_4;\delta_6)\ ,\ (\delta_2,\delta_6;\delta_7)\ ,\ (\delta_2,\delta_7;\delta_8) \,.
\ea
\ee
The resolved equation is
\be
x_1^4\delta_1\delta_4^3\delta_6^2\delta_7 + x_2^5\delta_1^2\delta_2\delta_3\delta_5^2 +x_3^5\delta_1^2\delta_2\delta_3\delta_5^2+x_3 x_4^2\delta_2=0 \,.
\ee
The sets $\delta_1=0$ and $\delta_2=0$ are empty, while the other exceptional divisors $\delta_i=0$ are irreducible. The triple intersection numbers of $S_1:\delta_3=0$, $S_2:\delta_4=0$, $S_3:\delta_5=0$, $S_4:\delta_6=0$, $S_5:\delta_7=0$ and $S_6:\delta_8=0$ are
\be
\includegraphics[height=6cm]{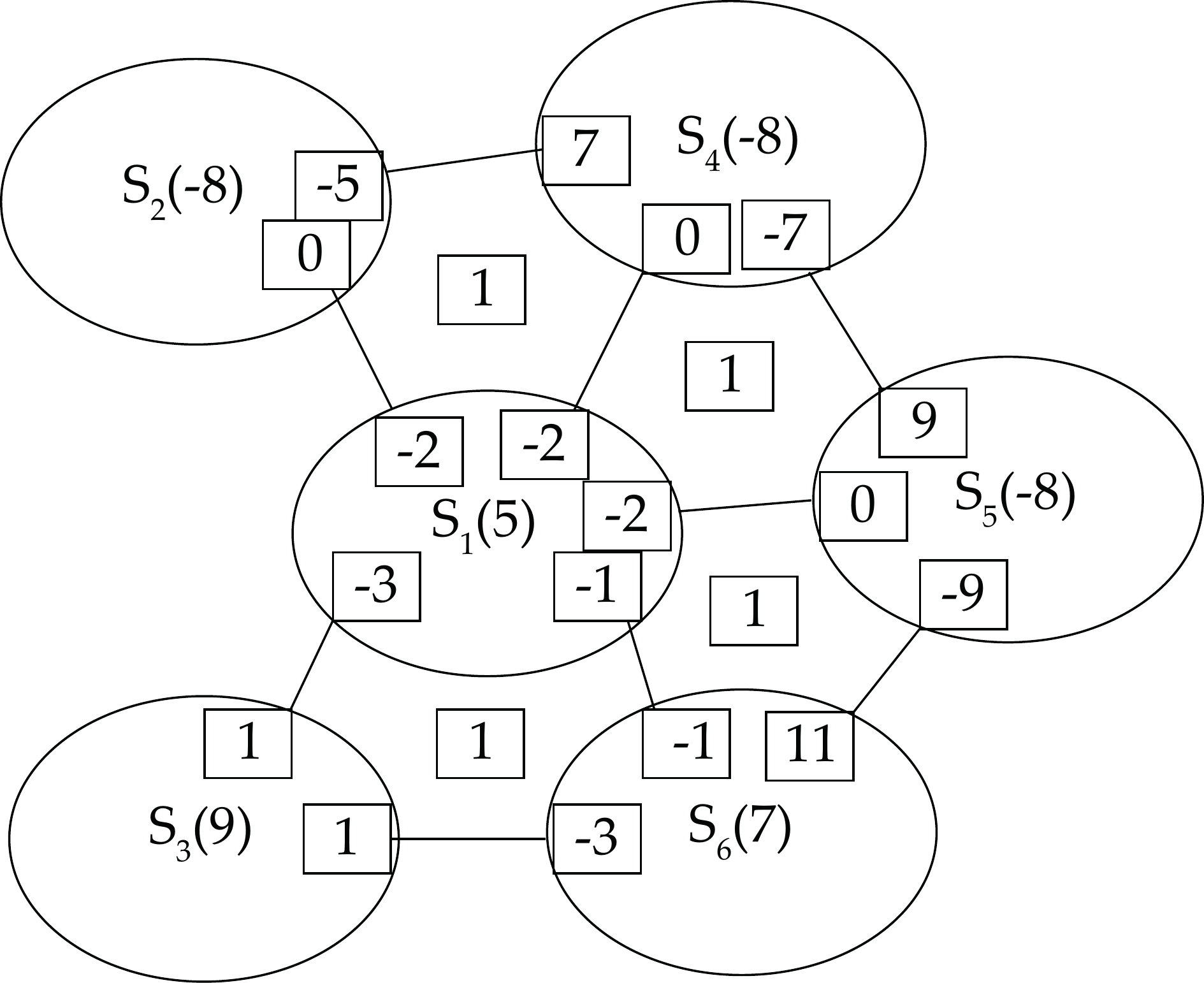}
\ee
The compact divisors $S_2$, $S_4$ and $S_5$ are rules over genus-two curves. After the flop performed by blowing up two double points on $S_6\cdot S_5$, the resulting geometry is
\be
\includegraphics[height=6cm]{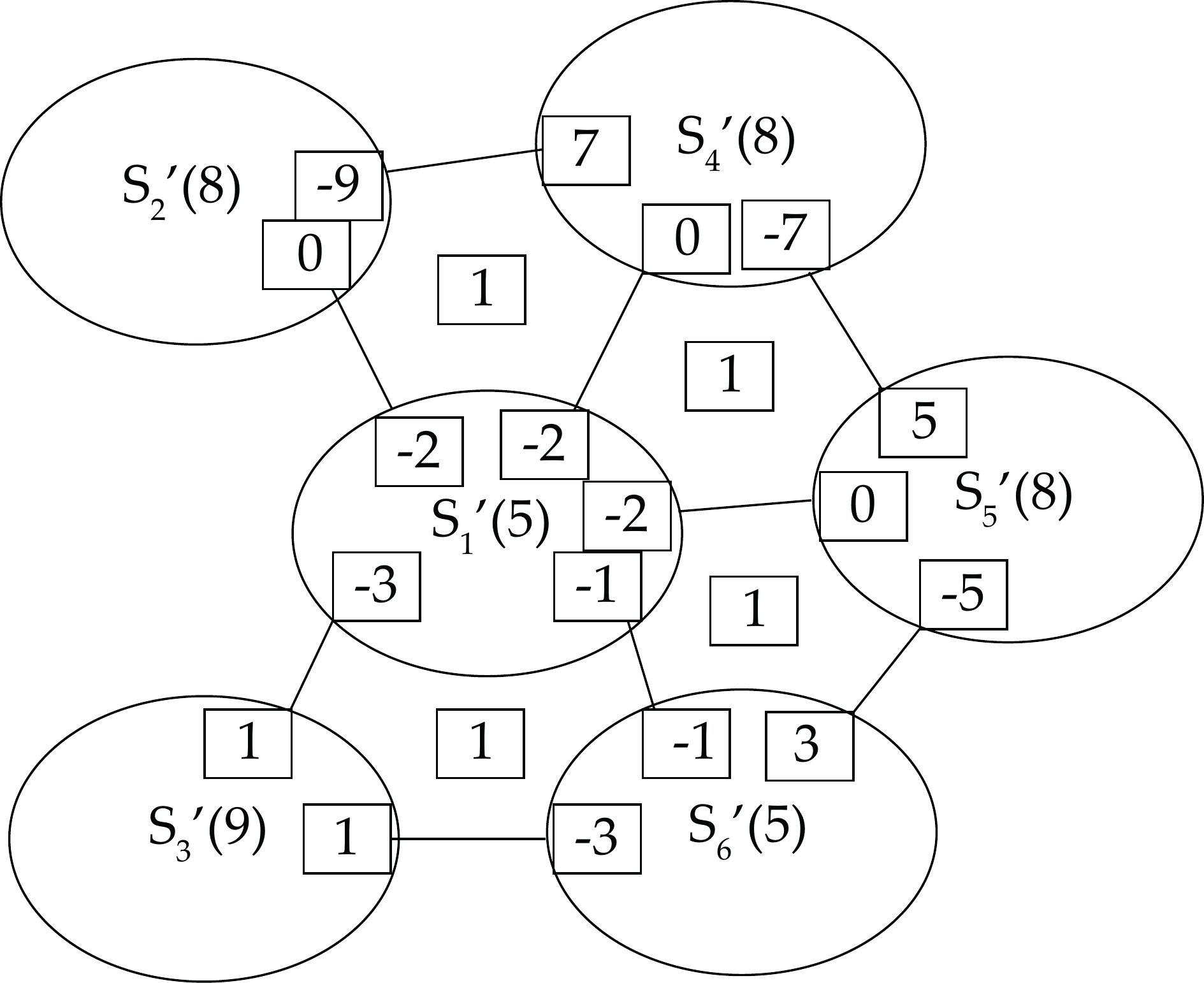}
\ee

\subsection*{Acknowledgements}

We thank Simone Giacomelli,  Ilarion Melnikov and Dave Morrison for discussions. 
The work of CC is supported by a University Research Fellowship 2017, “Supersymmetric gauge theories across dimensions”, of the Royal Society. CC is also a Birmingham Fellow.
The work of SSN and in part of Y-NW is supported by the European Union's Horizon 2020 Framework: ERC grant 682608. SSN is also supported in part by the ``Simons Collaboration on Special Holonomy in Geometry, Analysis and Physics''. Y-NW is also supported by National Science Foundation of China under Grant No. 12175004 and by Peking University under startup Grant No. 7100603534.

\appendix

\section{Crepantly Resolvable Geometries with smooth divisors and $b_3=0$}
\label{app:b3Zero}


We already discussed one example of a completely smoothable model with $b_3=0$ in the main text, section \ref{sec:ADE7E7}. Here we will discuss the remaining theories in this class up until rank $r=10$. 


\subsection{AD$[D_{2n},D_{2n}]$ theories}
\label{sec:D2nD2n}

There is an interesting infinite series of IHS which are `fully smooth' with $b_3=0$. 
In Type IIB, they gives rise to the 4d generalised Argyres-Douglas $[D_{2n},D_{2n}]$ $(n\geq 3)$ theories. 
Note that for $n=2$, the theory is the rank 1 $E_6$ SCFT. Consider the type $IV(2, 2n-1, 2, 2n-1)$ canonical singularity
\be
\label{AD-D2nD2n-sing}
x_1^{2n-1}+ x_1 x_2^2 + x_3^{2n-1} + x_3  x_4^2=0\,.
\ee
M-theory on this canonical singularity AD$[D_{2n},D_{2n}]$ results in a class of 5d SCFTs with  the IR gauge theory description:
 \be\label{gthdesciptDD}
 SU(n)_{\pm 3/2}+(2n+1)\bm{F}\,.
 \ee 
 The UV flavor symmetry is $G_F=SO(4n+2)\times U(1)$.
 Let us first summarize the properties of these singularities: 
 \be
\begin{tabular}{ |c |c | c|c|| c|c| c|| c| c|c|}\hline
$r$ &$f$ & $G_F^{\rm 5d}$  &$d_H$ &$\h r$& $\h d_H$  & $\Delta \CA_r$  & $b_3$ & $\mathfrak{t}_2$  \\ [0.5ex] 
\hline 
 $n-1$ & $2n+2$ & $SO(4n+2)\times U(1)$ & $2n^2+n+1$ & $2n^2-n-1$ & $3n+1$ & 0 & 0 & -- \\\hline
\end{tabular}
 \ee
Now we check these properties from both deformation and resolution of $\MG$.

\subsubsection{Deformation and Magnetic Quiver}

\begin{figure}
\begin{center}
\begin{tikzpicture}
\draw[ligne, black](0,2)--(0,1.8);
\draw[ligne, black](0,1)--(0,1.2);
\node (1) [hasse] at (0,5) {};
\node (2) [hasse] at (0,4) {};
\node (3) [hasse] at (0,3) {};
\node (4) [hasse] at (0,2) {};
\node (5) [hasse] at (0,1) {};
\node (6) [hasse] at (0,0) {};
\node (7) [] at (0,1.6) [label=center:{{$\vdots$}}] {};
\draw (1) edge [] node[label=right:$\mathfrak{e}_6$] {} (2);
\draw (2) edge [] node[label=right:$\mathfrak{d}_7$] {} (3);
\draw (3) edge [] node[label=right:$\mathfrak{d}_9$] {} (4);
\draw (5) edge [] node[label=right:$\mathfrak{d}_{2n+1}$] {} (6);
\end{tikzpicture}
\caption{Hasse diagram for the AD$[D_{2n}, D_{2n}]$ singularity, which has the IR gauge theory description $SU(n)_{\pm 3/2}+(2n+1)\bm{F}$. \label{fig:D2nD2nHasse}}
\end{center}
\end{figure}
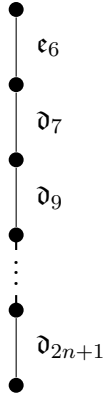

From the spectrum of the singularity, we can check that it is a Lagrangian theory with
\be
f= 2n+2~, \qquad n_h= {4\ov 3} n(1+2n^2)~, \quad n_v= {1\ov 3}(n-1)(3+8n + 8n^2)~, 
\ee
so that we find the virtual dimension of the 4d HB, $n_h-n_v= 3n+1$.  This equals the actual dimension as computed from the geometry, $\h d_H=r+f= 3n+1$. The 4d Lagrangian reads \cite{Closset:2020afy}:
\be\label{EQ4 DD series}
   \begin{tikzpicture}[x=1cm,y=1cm]
\draw[ligne, black](0,0)--(2.5,0);
\draw[ligne, black](3.5,0)--(7,0);
\draw[ligne, black](5,0)--(5,2);
\node[] at (-2,0) {$\EQfour_{\mathrm{AD}[D_{2n}, D_{2n}]}=$};
\node[flavor] at (0,0) [label=below:{{\scriptsize$1$}}] {};
\node[SUd] at (1,0) [label=below:{{\scriptsize$2$}}] {};
\node[SUd] at (2,0) [label=below:{{\scriptsize$3$}}] {};
\node[] at (3,0) [label=center:{{$\cdots$}}] {};
\node[SUd] at (4,0) [label=below:{{\scriptsize$2n-2$}}] {};
\node[SUd] at (5,0) [label=below:{{\scriptsize$2n-1$}}] {};
\node[SUd] at (6,0) [label=below:{{\scriptsize$n$}}] {};
\node[flavor] at (7,0) [label=below:{{\scriptsize$1$}}] {};
\node[SUd] at (5,1) [label=right:{{\scriptsize$n$}}] {};
\node[flavor] at (5,2) [label=right:{{\scriptsize$1$}}] {};
\end{tikzpicture}
\ee
where the nodes are special unitary gauge groups. The 5d magnetic quiver is obtained by gauging all the special-unitary and flavor nodes in \eqref{EQ4 DD series}, and then modding out a common $U(1)$:
\be
   \begin{tikzpicture}[x=1cm,y=1cm]
\draw[ligne, black](0,0)--(2.5,0);
\draw[ligne, black](3.5,0)--(7,0);
\node[] at (-2,0) {$\MQfive_{\mathrm{AD}[D_{2n}, D_{2n}]}=$};
\node[bd] at (0,0) [label=below:{{\scriptsize$1$}}] {};
\node[bd] at (1,0) [label=below:{{\scriptsize$2$}}] {};
\node[bd] at (2,0) [label=below:{{\scriptsize$3$}}] {};
\node[] at (3,0) [label=center:{{$\cdots$}}] {};
\node[bd] at (4,0) [label=below:{{\scriptsize$2n-2$}}] {};
\node[bd] at (5,0) [label=below:{{\scriptsize$2n-1$}}] {};
\node[bd] at (6,0) [label=below:{{\scriptsize$n$}}] {};
\node[bd] at (7,0) [label=below:{{\scriptsize$1$}}] {};
\draw[ligne, black](5,0)--(5,2);
\node[bd] at (5,1) [label=right:{{\scriptsize$n$}}] {};
\node[bd] at (5,2) [label=right:{{\scriptsize$1$}}] {};
\end{tikzpicture}
\ee
The Hasse diagram is shown in Figure~\ref{fig:D2nD2nHasse} (it was already computed in \cite{Bourget:2019aer}). Note that the AD$[D_4, D_4]$ theory is also equivalent to the rank-1 Seiberg $E_6$ theory.  The 3d mirror $\MQfour$ to \eqref{EQ4 DD series} was recently discussed in \cite{Carta:2021dyx}, together with the magnetic quiver for any of the AD$[D_n, D_m]$ 4d SCFTs.

\subsubsection{Resolution and 5d SCFT Interpretation}
The singularity (\ref{AD-D2nD2n-sing})  allows for a complete resolution, with the following sequence:
\be
\ba
\label{D2nD2n-res}
&(x_1,x_2,x_3,x_4;\delta_1)\cr
&(x_2,x_4,\delta_i;\delta_{i+1})\quad (i=1,\dots,n-2)\,.
\ea
\ee
The resulting equation
\be\label{D2nD2n-resEq}
x_1^{2n-1}\prod_{i=1}^{n-2}\delta_i^{2n-2i-2}+x_1 x_2^2+x_3^{2n-1}\prod_{i=1}^{n-2}\delta_i^{2n-2i-2}+x_3 x_4^2=0\,
\ee
is completely smooth, and all the exceptional divisors $S_i:\delta_i=0$ are smooth as well. We can compute the non-vanishing triple intersection numbers among the different divisors, which are represented as
\be
\includegraphics[height=2cm]{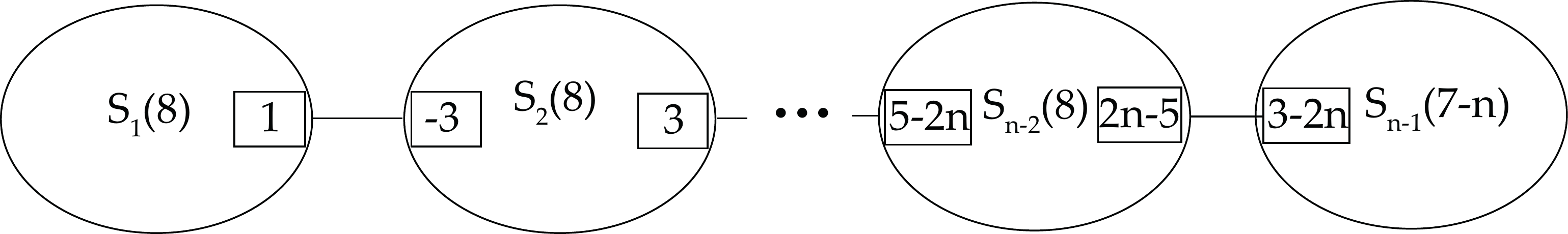}
\ee
The intersection numbers are consistent with the 5d IR gauge theory description \eqref{gthdesciptDD} as well.

\subsection{Rank 1}
\label{sec:rank1}

There are only three rank $r=1$  5d SCFTs in this cateory:  the Seiberg rank 1 $E_6$, $E_7$ and $E_8$ SCFTs, 
and we discussed them in \cite{Closset:2020scj}. They are described by the following type $\{1,1\}$ isolated hypersurface singularities:
\bea\label{En-sings-rank1}
&\MG_{E_6} \; : \; &&x_1^3+ x_2^3+ x_3^3+ x_4^3=0 \cr 
&\MG_{E_7} \; : \;&&x_1^2+ x_2^4+ x_3^4+ x_4^4=0 \cr
&\MG_{E_8} \; : \;&&x_1^2+ x_2^3+ x_3^6+ x_4^6=0\,.
\eea
The resolution of these singularities consists of a single weighted blow-up in the ambient space:
\bea\label{En-rank1-res}
&\MG_{E_6} \; : \; && (x_1,x_2,x_3,x_4;\delta_1) \cr 
&\MG_{E_7} \; : \;&& (x_1^{(2)},x_2^{(1)},x_3^{(1)},x_4^{(1)};\delta_1) \cr 
&\MG_{E_8} \; : \;&& (x_1^{(3)},x_2^{(2)},x_3^{(1)},x_4^{(1)};\delta_1) \,.
\eea
The exceptional divisors $S_1:\delta_1=0$ are del Pezzo surfaces $dP_6$, $dP_7$ and $dP_8$ respectively. We also list the full data of these theories in table  \ref{ref:rank1smoothsmooth}. 
\begin{table}
\centering 
\begin{tabular}{ | c ||c |c | c|c|| c|c| c|| c| c|c|} 
\hline
 $F$& $r$ &$f$ & $G_F^{\rm 5d}$  &$d_H$ &$\h r$& $\h d_H$  & $\Delta \CA_r$  & $b_3$ & $\frak{t}_2$  \\ [0.5ex] 
\hline 
$x_1^2+x_2^3+x_3^6+x_4^6$ & $1$ & $6$ & $E_6$ & $11$ & $5$ & $N+6$ &$0$&  $0$& -- \\
$x_1^2+x_2^4+x_3^4+x_4^4$ & $1$ & $7$ & $E_7$ & $17$ & $10$ & $N+7$ &$0$&  $0$& -- \\
$x_1^2+x_2^3+x_3^6+x_4^6$ & $1$ & $8$ & $E_8$ & $29$ & $21$ & $N+8$ &$0$&  $0$& -- \\
\hline 
\end{tabular}
\caption{All smooth rank 1 models with smooth divisors and $b_3=0$. \label{ref:rank1smoothsmooth}}
\end{table}
The $\EQfour$ for the rank 1 $E_k$ theory are $SU(n)$ quivers in the shape of affine $E_k$ Dynkin diagram, while the $\MQfive$ are $U(n)$ quivers in the same shape. The Hasse diagram is simply a single $\mathfrak{e}_k$, which corresponds to the minimal nilpotent orbit of $E_k$.

\subsection{Rank 3: $\{4,1\}\{3,3,9,2\}$}

The model is characterized by the following data: 
 \be
\small
\begin{tabular}{ | c ||c |c | c|c|| c|c| c|| c| c|c|}\hline
$F$& $r$ &$f$ & $G_F^{\rm 5d}$  &$d_H$ &$\h r$& $\h d_H$  & $\Delta \CA_r$  & $b_3$ & $\mathfrak{t}_2$ \\ [0.5ex] 
\hline 
$x_1^3+x_1 x_2^3+x_3^9+x_3 x_4^2$ & $3$ & $8$ & $E_7\times U(1)$ & $39$ & $31$ & 11 & 0 & 0 & --\\ \hline
\end{tabular}
 \ee
\subsubsection{Deformations and Magnetic Quiver}
 
From the CB spectrum we can infer the $\EQfour$ and $\MQfive$. The magnetic quiver for the 5d theory is 
\be
   \begin{tikzpicture}[x=1cm,y=1cm]
\draw[ligne, black](0,0)--(6,0);
\node[] at (-1.5,0) {$\MQfive=$};
\node[bd] at (0,0) [label=below:{{\scriptsize$3$}}] {};
\node[bd] at (1,0) [label=below:{{\scriptsize$6$}}] {};
\node[bd] at (2,0) [label=below:{{\scriptsize$9$}}] {};
\node[bd] at (3,0) [label=below:{{\scriptsize$7$}}] {};
\node[bd] at (4,0) [label=below:{{\scriptsize$5$}}] {};
\node[bd] at (5,0) [label=below:{{\scriptsize$3$}}] {};
\node[bd] at (6,0) [label=below:{{\scriptsize$1$}}] {};
\draw[ligne, black](2,0)--(2,2);
\node[bd] at (2,1) [label=right:{{\scriptsize$5$}}] {};
\node[bd] at (2,2) [label=right:{{\scriptsize$1$}}] {};
\end{tikzpicture}
\ee
and the Hasse diagram is
\be
\begin{tikzpicture}
\node (1) [hasse] at (0,3) {};
\node (2) [hasse] at (0,2) {};
\node (3) [hasse] at (0,1) {};
\node (4) [hasse] at (0,0) {};
\draw (1) edge [] node[label=right:$\mathfrak{e}_6$] {} (2);
\draw (2) edge [] node[label=right:$\mathfrak{d}_7$] {} (3);
\draw (3) edge [] node[label=right:$\mathfrak{e}_7$] {} (4);
\end{tikzpicture}
\ee
From the Hasse diagram, we can read off the flavor symmetry $G_F=E_7\times U(1)$.

\subsubsection{Resolution and 5d SCFT}

The resolution sequence that completely resolves this model is 
\be
\ba
\label{rank3-2-res}
&(x_1,x_2,x_3,x_4;\delta_1)\ ,\ (x_1^{(2)},x_2^{(1)},x_4^{(3)},\delta_1^{(1)};\delta_3)\ ,\ (x_1,x_4,\delta_1;\delta_2)
\ea
\ee
with the intersection numbers
\be
\includegraphics[height=4cm]{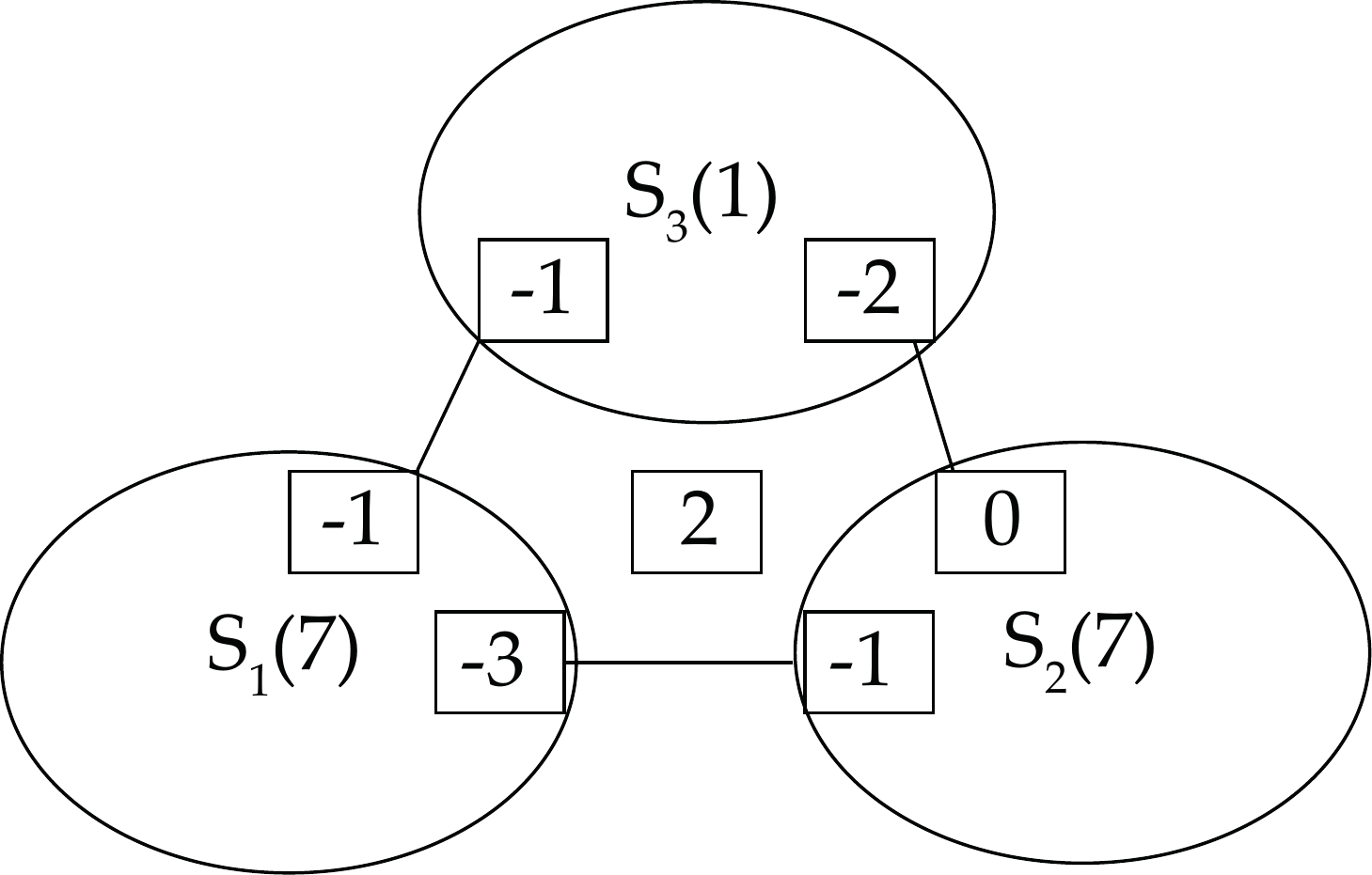}
\ee
This theory does not have a 5d IR gauge theory description. Nonetheless, the theory is a descendant of 6d $(E_7,SO(7))$ conformal matter \cite{Apruzzi:2019enx} with the following CFD transitions
\be
\includegraphics[height=5cm]{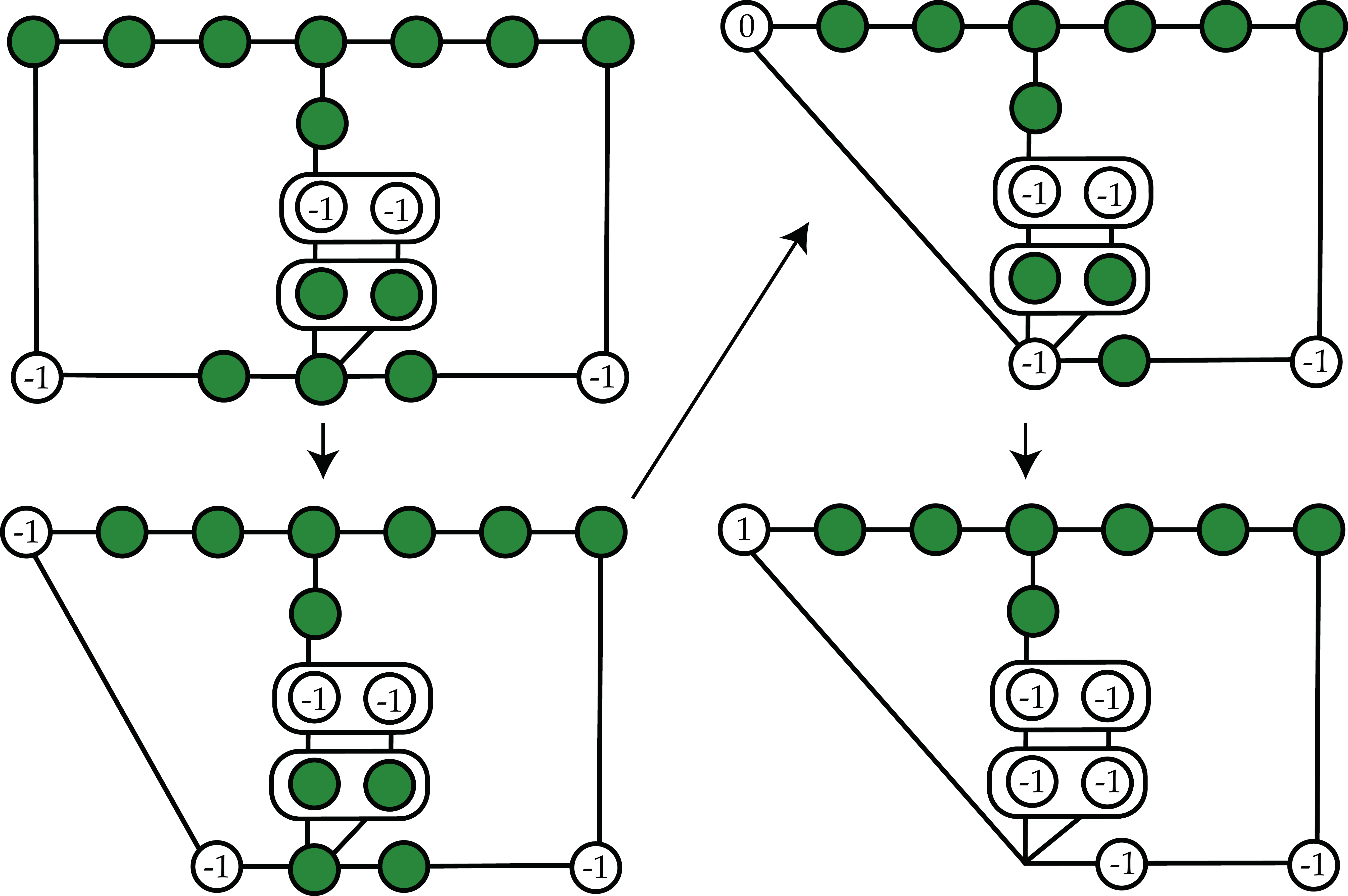}
\ee
The CFD of this 5d SCFT with $G_F=E_7\times U(1)$ is at the bottom-right corner.

\subsection{Rank 3: \{12,1\}\{7,2,3,2\}}

The data of this equation is
 \be
\small
\begin{tabular}{ | c ||c |c | c|c|| c|c| c|| c| c|c|}\hline
$F$& $r$ &$f$ & $G_F^{\rm 5d}$  &$d_H$ &$\h r$& $\h d_H$  & $\Delta \CA_r$  & $b_3$ & $\mathfrak{t}_2$  \\ [0.5ex] 
\hline 
$x_1^7+x_1 x_2^2+x_1 x_3^3+x_2 x_4^2+x_2 x_3^2$ & $3$ & $8$ & $E_6\times SU(2)\times U(1)$ & $29$ & $21$ & 11 & 0 & 0 & --
\\\hline
\end{tabular}
 \ee

\subsubsection{Deformations and Magnetic Quiver}

The magnetic quiver again follows from the CB spectrum of $\FTfour$:
\be
   \begin{tikzpicture}[x=1cm,y=1cm]
\draw[ligne, black](0,0)--(6,0);
\node[] at (-1.5,0) {$\MQfive=$};
\node[bd] at (0,0) [label=below:{{\scriptsize$1$}}] {};
\node[bd] at (1,0) [label=below:{{\scriptsize$3$}}] {};
\node[bd] at (2,0) [label=below:{{\scriptsize$5$}}] {};
\node[bd] at (3,0) [label=below:{{\scriptsize$7$}}] {};
\node[bd] at (4,0) [label=below:{{\scriptsize$5$}}] {};
\node[bd] at (5,0) [label=below:{{\scriptsize$3$}}] {};
\node[bd] at (6,0) [label=below:{{\scriptsize$1$}}] {};
\draw[ligne, black](3,0)--(3,2);
\node[bd] at (3,1) [label=right:{{\scriptsize$4$}}] {};
\node[bd] at (3,2) [label=right:{{\scriptsize$1$}}] {};
\end{tikzpicture}
\ee
and the Hasse diagram is
\be
\begin{tikzpicture}
\node (1) [hasse] at (0,3) {};
\node (2) [hasse] at (-0.5,2) {};
\node (3) [hasse] at (0.5,2) {};
\node (4) [hasse] at (-0.5,1) {};
\node (5) [hasse] at (0.5,1) {};
\node (6) [hasse] at (0,0) {};
\draw (1) edge [] node[label=left:$\mathfrak{e}_7$] {} (2);
\draw (1) edge [] node[label=right:$\mathfrak{e}_6$] {} (3);
\draw (2) edge [] node[label=left:$\mathfrak{e}_6$] {} (4);
\draw (3) edge [] node[label=right:$\mathfrak{a}_7$] {} (5);
\draw (4) edge [] node[label=left:$\mathfrak{a}_1$] {} (6);
\draw (5) edge [] node[label=right:$\mathfrak{e}_6$] {} (6);
\draw (2) edge [] node[label=left:$\mathfrak{a}_1$] {} (5);
\draw (3) edge [] node[label=right:$\mathfrak{e}_7$] {} (4);
\end{tikzpicture}
\ee
From the Hasse diagram, we can read off the flavor symmetry is $G_F=E_6\times SU(2)\times U(1)$, which is different from the subgraph of the balanced nodes in the magnetic quiver.

\subsubsection{Resolution and 5d SCFT}

The resolution sequence is
\be
(x_1,x_2,x_3,x_4;\delta_1)\,,\qquad 
(x_2^{(2)},x_3^{(1)},x_4^{(1)},\delta_1^{(1)};\delta_3)\,,\qquad 
(x_2,\delta_1;\delta_2)\,.
\ee
The non-vanishing triple intersection numbers are
\be
\includegraphics[height=4cm]{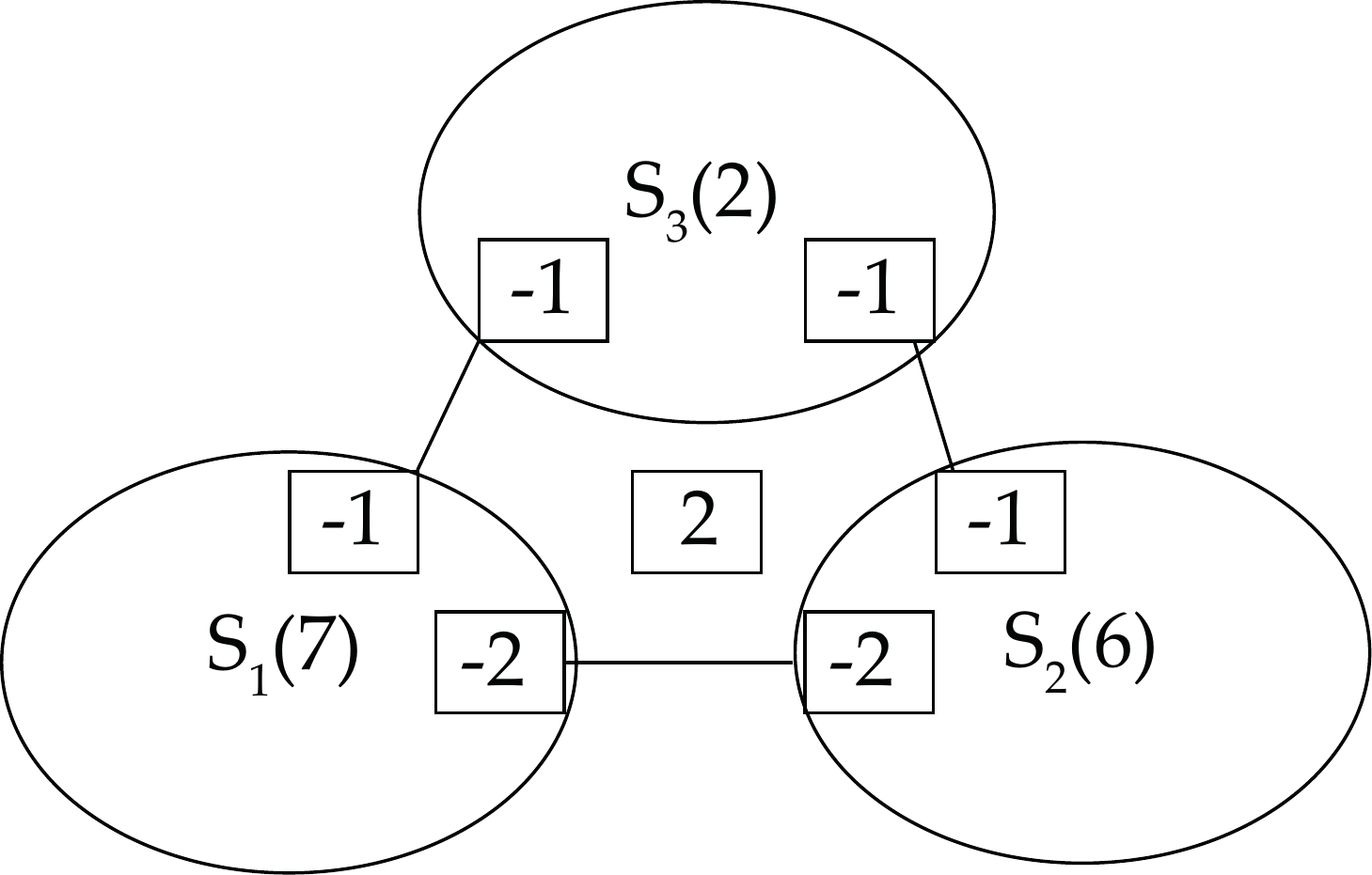}
\ee
This theory does not have a 5d IR gauge theory description. Nonetheless, the theory is a descendant of 6d $(E_7,SO(7))$ conformal matter as well, with the following CFD transitions
\be
\includegraphics[height=6cm]{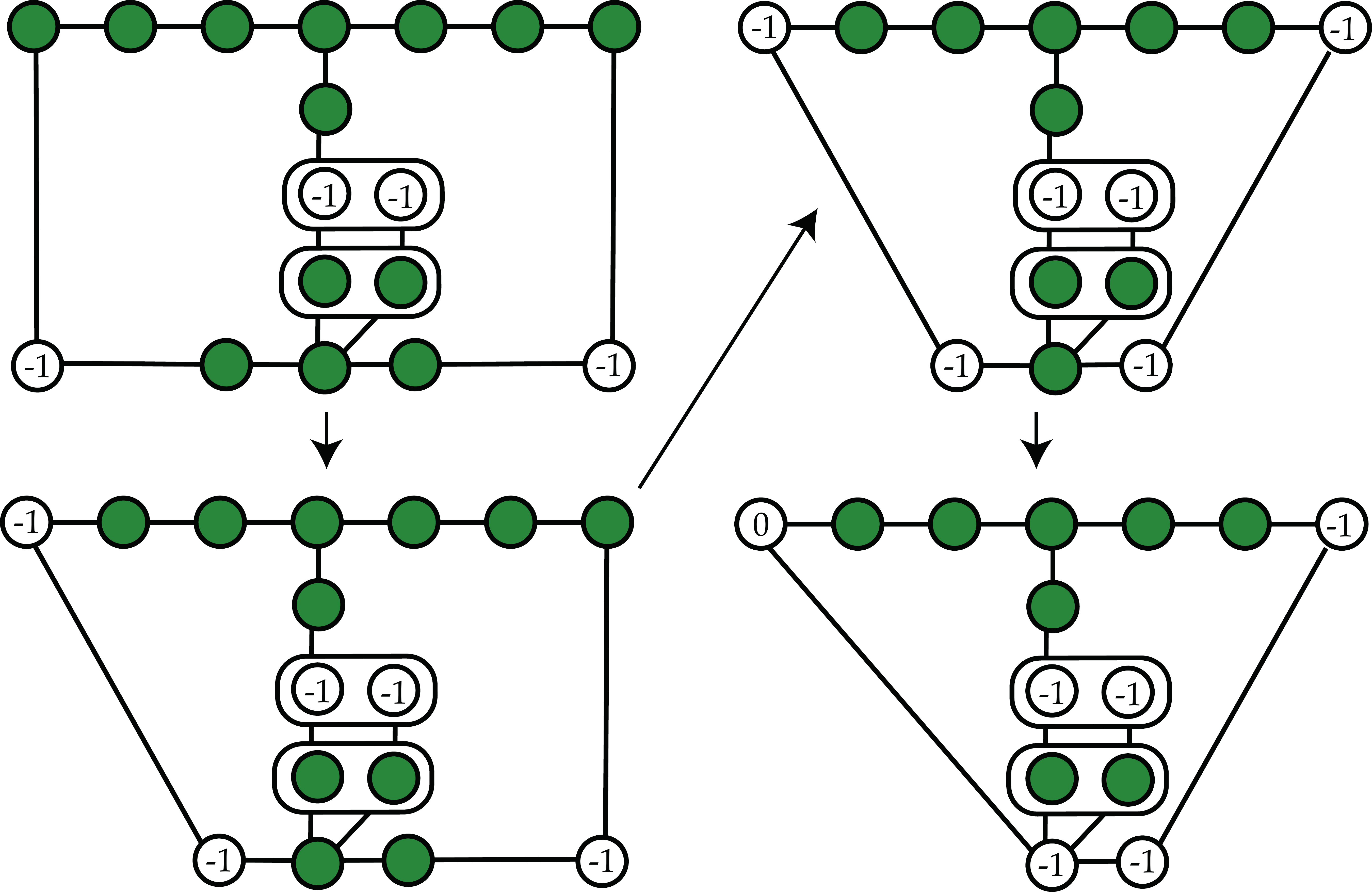}
\ee
The theory on the bottom-right has the correct $r_F$ and $G_F$. There is no valid BG-CFD embedding.

\subsection{Rank 6: \{18,1\}\{2,3,2,5\}}

The model with integral spectrum at this rank is:
 \be\label{rank-6-smooth-model1}
\small
\begin{tabular}{ | c ||c |c | c|c|| c|c| c|| c| c|c|}\hline
$F$& $r$ &$f$ & $G_F^{\rm 5d}$  &$d_H$ &$\h r$& $\h d_H$  & $\Delta \CA_r$  & $b_3$ & $\mathfrak{t}_2$ \\ [0.5ex] 
\hline 
$x_2 x_4^5 + x_3 x_4^4 + x_1 x_2^3 + x_2 x_3^2 + x_1^2 x_3$ & $6$ & $6$ & $SU(5)\times U(1)^2$ & $36$ & $30$ & 12 & 0 & 0 & --
\\ \hline \end{tabular}
 \ee

\subsubsection{Deformations and 4d SCFT}

In this case the CB spectrum of $\FTfour$  is
\be
\Delta= \{2^4,3^5,4^4,5^4,6^3,7^3,8^2,9^2,10,11,13\} \,,
\ee
which contains operators with $\Delta=11$ and $\Delta=13$, but there is no operator with $\Delta=12$. The CB spectrum cannot match any set of Casimir operators of Lie groups, and  there is no Lagrangian description of $\FTfour$. Again there is a conjectural generalised quiver that matches all the data of this model (flavor rank and CB spectrum)
\be
[1]- SU(5) - SU(9)- D_2(SO(28)) - SO(12) \times {\rm Spin}(6) -D_2(SO(16)) - [4]
\ee
Note that  ${\rm Spin}(6)$ is gauged as a subgroup of the $SO(18) \times U(1)$ flavor of the $D_2 (SO(16))$.
Similarly, we gauge an $SO(12) \times SU(9)$ subgroup of the $SO(30)\times U(1)$ flavor symmetry of $D_2(SO(28))$.

\subsubsection{Resolution and 5d SCFT}

Resolution sequence that fully resolves this model is
\be
\ba
&(x_1,x_2,x_3,x_4;\delta_1)\ ,\ (x_1,x_2,x_3,\delta_1;\delta_4)\ ,\ (x_1^{(1)},x_3^{(2)},\delta_1^{(1)},\delta_4^{(1)};\delta_6)\cr
&(x_3,\delta_1;\delta_2)\ ,\ (x_1,\delta_2;\delta_3)\ ,\ (x_3,\delta_4;\delta_5)\,.
\ea
\ee
The intersection relations among the compact surfaces $S_i:\delta_i=0$ are
\be
\label{fig:rank-6-smooth}
\includegraphics[height=7cm]{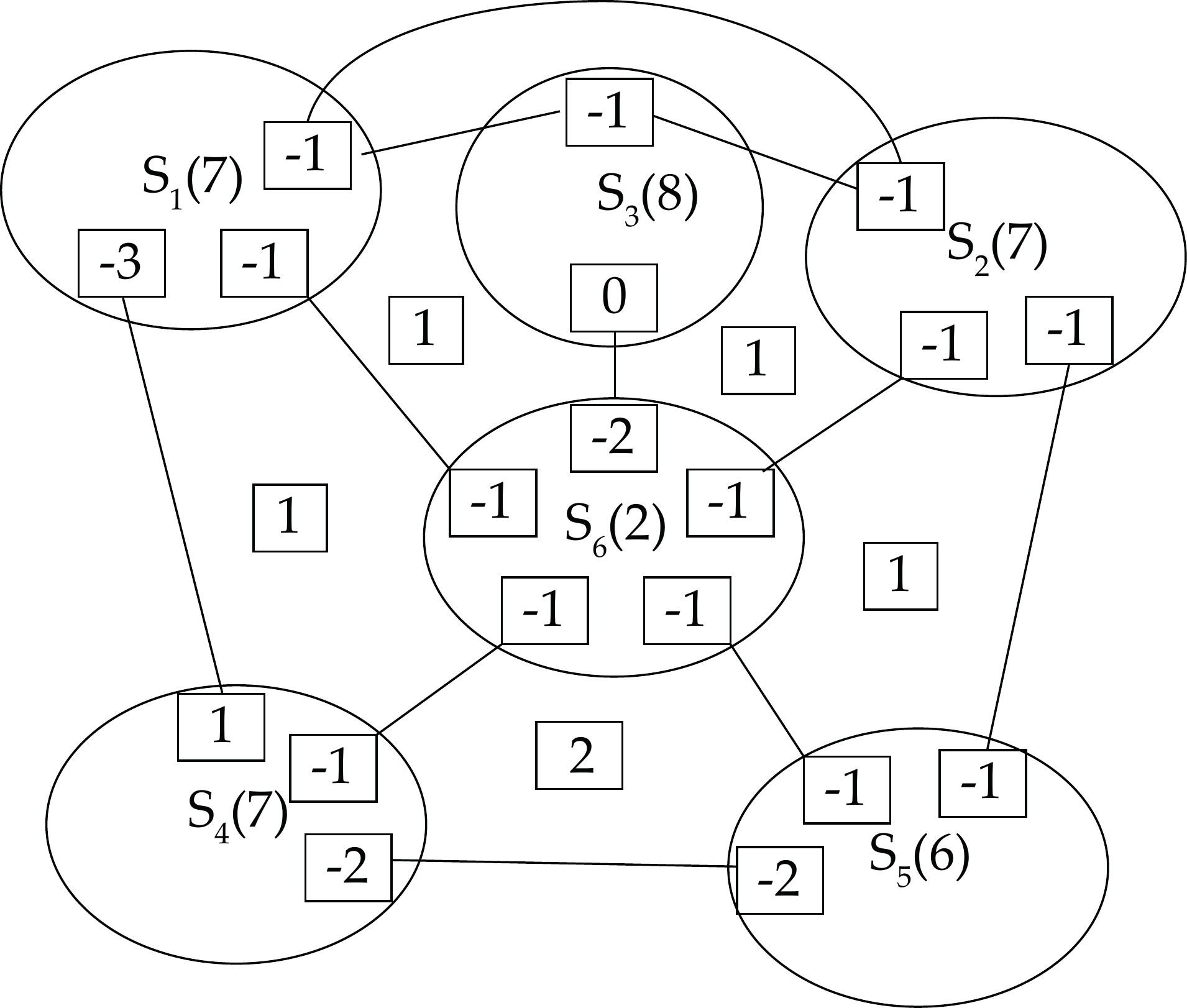}
\ee
Note that $S_1\cdot S_2\cdot S_6=S_1\cdot S_3\cdot S_6=S_2\cdot S_3\cdot S_6=1$, the curve $S_1\cdot S_2=S_2\cdot S_3=S_1\cdot S_3$ coincides, $S_1\cdot S_2\cdot S_3=-1$. 
Since the curve $S_3\cdot S_1$ is a section curve on $S_3$ and a ruling curve on $S_1$, the geometry does not have a consistent ruling structure.

For the non-abelian part of flavor symmetry, it comes from the extra $(-2)$-curves on $S_6$, which has the topology of gdP$_7$. The curves on $S_6$ shown in (\ref{fig:rank-6-smooth}) are written in the Picard group generators $h,e_1,\dots,e_7$:
\be
\ba
S_6\cdot S_3&=e_6-e_7\cr
S_6\cdot S_2&=e_7\cr
S_6\cdot S_5&=h-e_6-e_7\cr
S_6\cdot S_4&=2h-e_1-e_2-e_3-e_4-e_5\cr
S_6\cdot S_1&=3h-e_1-e_2-e_3-e_4-e_5-2e_6-e_7\,.
\ea
\ee
The maximal set of extra $(-2)$-curves with non-negative intersection numbers with the above curves are
\be
\ba
C_1&=e_1-e_2\cr
C_2&=e_2-e_3\cr
C_3&=e_3-e_4\cr
C_4&=e_4-e_5\,,
\ea
\ee
which generate an $SU(5)$ flavor symmetry factor. Hence we have at least
\be
G_F=SU(5)\times U(1)^2\,.
\ee

\subsection{Rank 6: $\{12,2\}\{13,3,2,3\}$}

The model has the following basic characteristics
 \be
\small
\begin{tabular}{ | c ||c |c | c|c|| c|c| c|| c| c|c|}\hline
$F$& $r$ &$f$ & $G_F^{\rm 5d}$  &$d_H$ &$\h r$& $\h d_H$  & $\Delta \CA_r$  & $b_3$ & $\frak{f}$  \\ [0.5ex] 
\hline 
$x_1^{13}+x_1 x_2^3+x_1 x_3^2+x_2 x_3 x_4+x_2 x_4^3$ & $6$ & $9$ & $E_7\times U(1)^2$ & $57$ & $48$ & 15 & 0 & 0 & --\\\hline
\end{tabular}
 \ee
\subsubsection{Deformations and Magnetic Quiver}

From the CB spectrum we can identify the magnetic quiver as
\be
   \begin{tikzpicture}[x=1cm,y=1cm]
\draw[ligne, black](0,0)--(7,0);
\node[] at (-1.5,0) {$\MQfive=$};
\node[bd] at (0,0) [label=below:{{\scriptsize$1$}}] {};
\node[bd] at (1,0) [label=below:{{\scriptsize$5$}}] {};
\node[bd] at (2,0) [label=below:{{\scriptsize$9$}}] {};
\node[bd] at (3,0) [label=below:{{\scriptsize$13$}}] {};
\node[bd] at (4,0) [label=below:{{\scriptsize$10$}}] {};
\node[bd] at (5,0) [label=below:{{\scriptsize$7$}}] {};
\node[bd] at (6,0) [label=below:{{\scriptsize$4$}}] {};
\node[bd] at (7,0) [label=below:{{\scriptsize$1$}}] {};
\draw[ligne, black](3,0)--(3,2);
\node[bd] at (3,1) [label=right:{{\scriptsize$7$}}] {};
\node[bd] at (3,2) [label=right:{{\scriptsize$1$}}] {};
\end{tikzpicture}
\ee
From the balanced nodes and the Hasse diagram, we can read off the non-abelian part of flavor symmetry $G_{F,nA}=E_7$. Hence the full UV flavor symmetry is $G_F=E_7\times U(1)^2$.

\subsubsection{Resolution and 5d SCFT}

The following sequence of resolutions smoothes this model
\be
\ba
&(x_1,x_2,x_3,x_4;\delta_1)\ ,\ (x_2^{(1)},x_3^{(2)},x_4^{(1)},\delta_1^{(1)};\delta_4)\ ,\ (x_2^{(2)},x_3^{(3)},x_4^{(1)},\delta_4^{(1)};\delta_6)\cr
&(x_3,\delta_1;\delta_2)\ ,\ (x_2,\delta_2;\delta_3)\ ,\ (x_2,x_3,\delta_4;\delta_5)\,.
\ea
\ee
The triple intersection numbers of $S_i:\ \delta_i=0$ are given in the following diagram:
\be
\includegraphics[height=7cm]{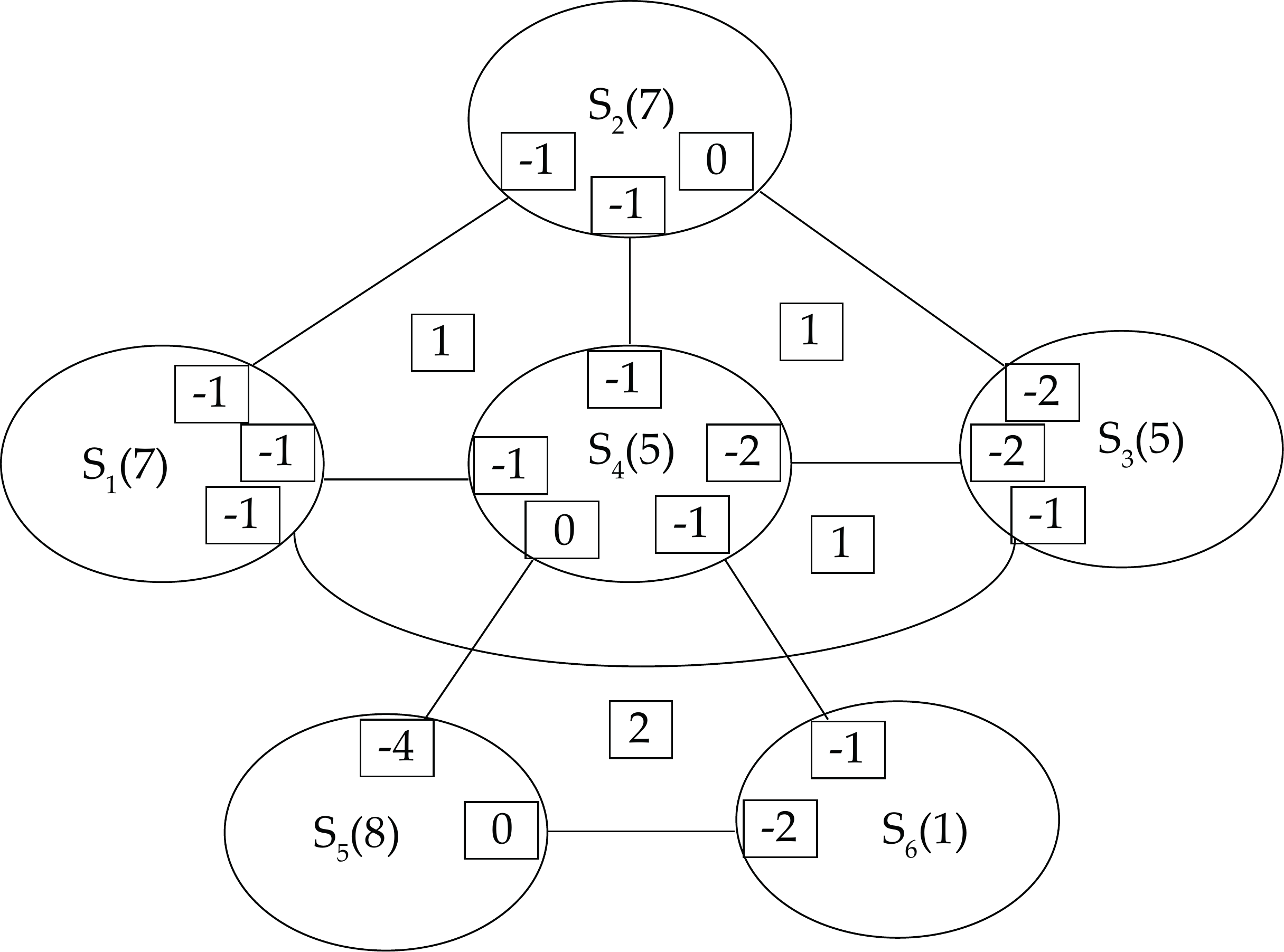}
\ee
For the non-Abelian part of flavor symmetry $G_F$, it comes from the extra $(-2)$-curves on $S_6$. The topology of $S_6$ can be chosen as gdP$_8$ of type $\mathbf{E_7+A_1}$ \cite{derenthal2014singular}, and the $(-2)$-curves arranged in an $E_7$ Dynkin diagram exactly gives rise to $G_{F,nA}=E_7$.

\subsection{Rank 9: \{4,1\}\{4,2,4,4\}}

We consider the type $\{4,1\}\{4,2,4,4\}$ singularity, with the data 
 \be
\small
\begin{tabular}{ | c ||c |c | c|c|| c|c| c|| c| c|c|}\hline
$F$& $r$ &$f$ & $G_F^{\rm 5d}$  &$d_H$ &$\h r$& $\h d_H$  & $\Delta \CA_r$  & $b_3$ & $\mathfrak{t}_2$ \\ [0.5ex] 
\hline 
$x_1^4+x_2^2 x_1+x_3^4+x_3 x_4^4$ & $9$ & $5$ & $SU(2)^4\times U(1)$ & $35$ & $30$ & 14 & 0 & 0 & --\\ \hline
\end{tabular}
 \ee

In this case the CB spectrum of $\FTfour$ contains operators with $\Delta=13$ and $\Delta=16$, but there is no operator with $\Delta=14$ or $15$. The CB spectrum cannot match any set of Casimir operators of Lie groups, and there is no Lagrangian description of $\FTfour$. It would be interesting to identify similarly to the other models a generalised quiver. 

The resolution sequence for this model is 
\be
\ba
&(x_1,x_2,x_3,x_4;\delta_1)\ ,\ (x_1,x_2,x_3,\delta_1;\delta_4)\ ,\ (x_2,\delta_1;\delta_2)\ ,\ (x_1,\delta_1;\delta_3)\cr
&(\delta_1,\delta_2;\delta_5)\ ,\ (\delta_2,\delta_4;\delta_6)\ ,\ (\delta_2,\delta_6;\delta_7)
\ea
\ee
The equation $\delta_1=0$ is empty, and the equation $\delta_2=0$ has four irreducible components, which correspond to compact divisors $S_2^{(1)},\dots,S_2^{(4)}$.
The triple intersection numbers of $S_i:\delta_i=0$ are given in the following diagram:
\be
\label{fig:rank-9-smooth}
\includegraphics[height=9cm]{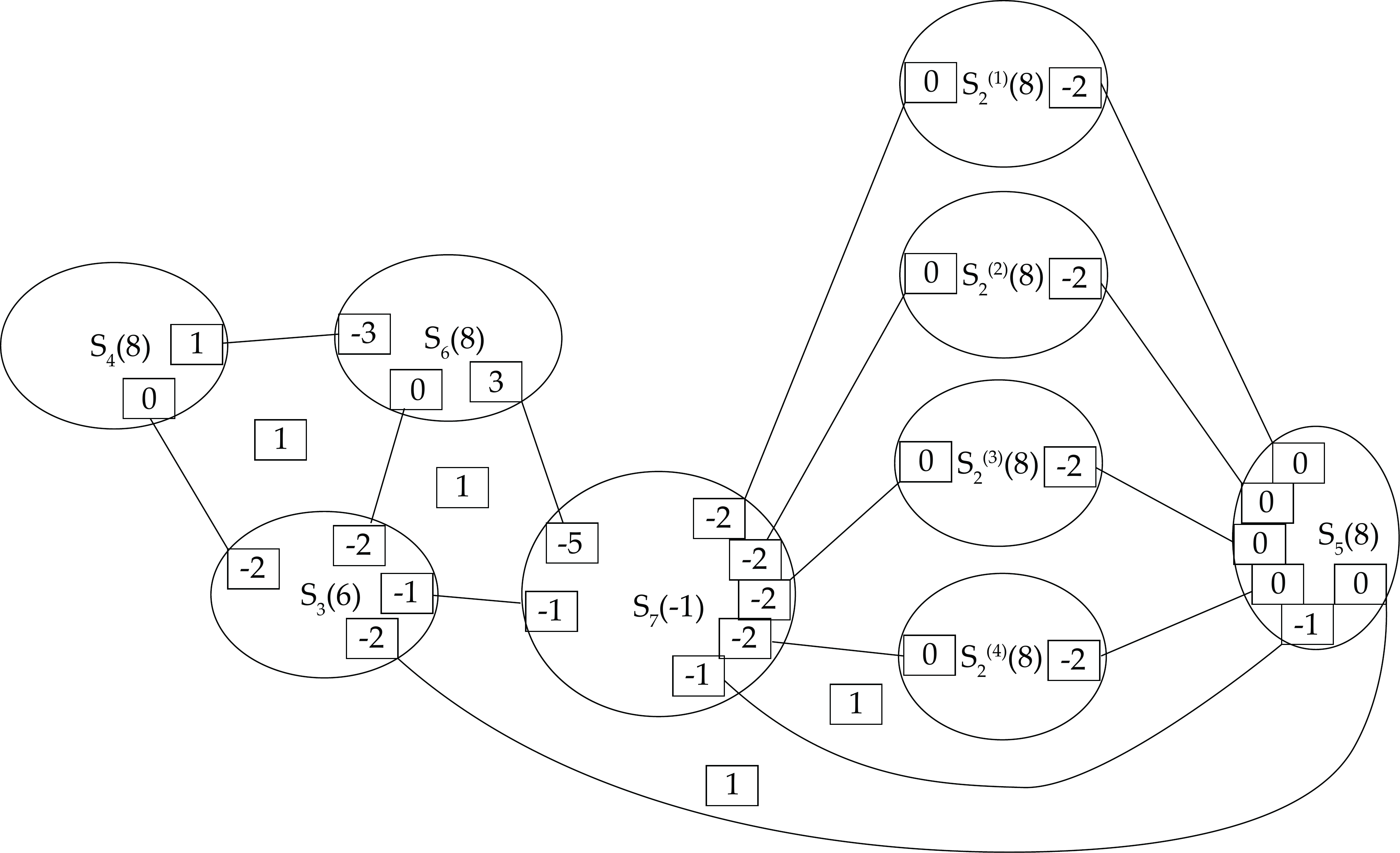}
\ee
Note that $S_5\cdot S_7\cdot S_2^{(i)}=1$ for $i=1,2,3,4$. Since $S_2^{(i)}$ $(i=1,2,3,4)$ are Hirzebruch surface $\mb{F}_2$, and $S_5$ is $\mb{F}_1$, $S_2^{(i)}\cdot S_5$ is a section curve on $S_2^{(i)}\cdot S_5$ but a ruling curve on $S_5$. Hence the geometry does not have a ruling structure and IR gauge theory description.

For the non-abelian part of flavor symmetry $G_{F,nA}$, it comes from additional $(-2)$-curves on $S_7$. Note that the curves on $S_7$ appeared in (\ref{fig:rank-9-smooth}) can be written in terms of Picard group generators $h,e_1,\dots,e_{10}$ as
\be
\ba
S_7\cdot S_6&=2h-e_1-e_2-e_3-e_4-e_5-e_6-e_7-e_8-e_9\cr
S_7\cdot S_3&=h-e_9-e_{10}\cr
S_7\cdot S_5&=e_{10}\cr
S_7\cdot S_2^{(1)}&=h-e_1-e_2-e_{10}\cr
S_7\cdot S_2^{(2)}&=h-e_3-e_4-e_{10}\cr
S_7\cdot S_2^{(3)}&=h-e_5-e_6-e_{10}\cr
S_7\cdot S_2^{(4)}&=h-e_7-e_8-e_{10}\cr
\ea
\ee
The maximal set of additional $(-2)$-curves on $S_7$ with non-negative intersection numbers with the above curves are
\be
\ba
C_1&=e_1-e_2\cr
C_2&=e_3-e_4\cr
C_3&=e_5-e_6\cr
C_4&=e_7-e_8\,.
\ea
\ee
These curves generate non-abelian flavor symmetry $G_{F,nA}=SU(2)^4$, hence we have (at least)
\be
G_F=SU(2)^4\times U(1)\,.
\ee

\subsection{Rank 10: \{10,1\}\{5,2,2,6\}}

The model has the following data 
 \be
\small
\begin{tabular}{ | c ||c |c | c|c|| c|c| c|| c| c|c|}\hline
$F$& $r$ &$f$ & $G_F^{\rm 5d}$  &$d_H$ &$\h r$& $\h d_H$  & $\Delta \CA_r$  & $b_3$ & $\mathfrak{t}_2$  \\ [0.5ex] 
\hline 
$x_2 x_4^6+x_3^2 x_4+x_1^5+x_2^2 x_3$ & $10$ & $4$ & $SU(5)$ & $50$ & $46$ & 14 & 0 & 0 & -- \\\hline
\end{tabular}
 \ee
In this case the CB spectrum of $\FTfour$ contains operators with $\Delta=22$ and $\Delta=25$, but there is no operator with $\Delta=23$ or $24$. The CB spectrum cannot match any set of Casimir operators of Lie groups, and there is no Lagrangian description of $\FTfour$.

Resolution sequence is 
\be
\ba
&(x_1,x_2,x_3,x_4;\delta_1)\ ,\ (x_1^{(1)},x_2^{(1)},x_3^{(2)},\delta_1^{(1)};\delta_5)\ ,\ (x_2^{(1)},x_3^{(2)},\delta_1^{(1)},\delta_5^{(1)};\delta_9)\cr
&(x_2^{(1)},x_3^{(2)},\delta_5^{(1)},\delta_9^{(1)};\delta_{11})\ ,\ (x_3,\delta_1;\delta_2)\ ,\ (x_3,\delta_2;\delta_3)\ ,\ (x_2,\delta_2;\delta_4)\ ,\ (x_3,\delta_5;\delta_6)\cr
&(x_2,\delta_6;\delta_7)\ ,\ (\delta_2,\delta_4;\delta_8)\,.
\ea
\ee
The equation $\delta_9=0$ is empty, and the triple intersection numbers among the remaining 10 compact divisors $S_i:\ \delta_i=0$ are given in the following diagram:
\be
\includegraphics[height=9cm]{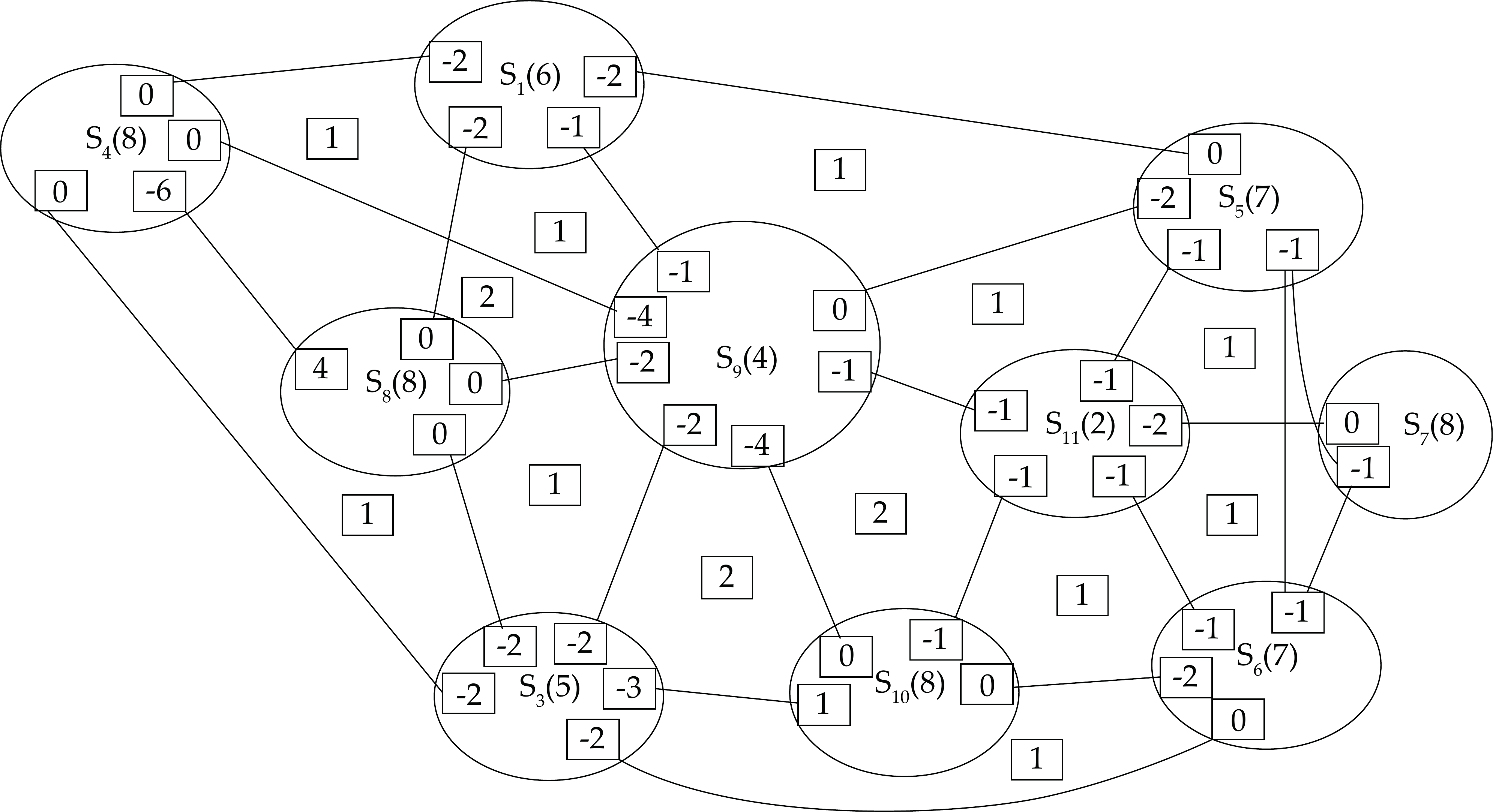}
\ee
Note that $S_1\cdot S_4\cdot S_8=S_1\cdot S_8\cdot S_9=1$, $S_4\cdot S_8\cdot S_9=2$, $S_{11}\cdot S_5\cdot S_6=S_{11}\cdot S_5\cdot S_7=S_{11}\cdot S_6\cdot S_7=1$, $S_5\cdot S_6\cdot S_7=-1$. The curve $S_5\cdot S_6=S_5\cdot S_7=S_6\cdot S_7$ coincides.  
Since $S_6\cdot S_7$ is a section curve on $S_7$ but a ruling curve on $S_5$ and $S_6$, the whole geometry does not have a ruling structure.

For the flavor symmetry $G_F$, note that the topology of $S_{11}$ is exactly the same as the topology of $S_6$ in the rank-6 model (\ref{rank-6-smooth-model1}). Following the same argument based on extra $(-2)$-curves on $S_{11}$, we get
\be
G_F=SU(5)\,.
\ee

\section{Tables: Smoothable Models}


{\tiny{
\begin{sidewaystable}[htp]
$$
\begin{array}{|c|c|c|c|c|c||c|c|c|c|c|c|}
\hline
\text{Type} & (a,b,c,d) & F & r &  f&d_H & \widehat{r} & \widehat{d_H} & \Delta \CA_r  & b_3 & \mathfrak{t}_2& \text{Magnetic Quiver}\, (\MQfive). \cr \hline\hline
 \{4,1\} & \{5,2,5,2\} & x_1^5+x_2^2 x_1+x_3^5+x_3 x_4^2 & 2 & 8 & 22 & 14 & 10 & 0 & 0 & 0  & 
  \begin{tikzpicture}[x=.5cm,y=.5cm]
\draw[ligne, black](0,0)--(6,0);
\node[bd] at (0,0) [label=below:{{\scriptsize$1$}}] {};
\node[bd] at (1,0) [label=below:{{\scriptsize$2$}}] {};
\node[bd] at (2,0) [label=below:{{\scriptsize$3$}}] {};
\node[bd] at (3,0) [label=below:{{\scriptsize$4$}}] {};
\node[bd] at (4,0) [label=below:{{\scriptsize$5$}}] {};
\node[bd] at (5,0) [label=below:{{\scriptsize$3$}}] {};
\node[bd] at (6,0) [label=below:{{\scriptsize$1$}}] {};
\draw[ligne, black](4,0)--(4,2);
\node[bd] at (4,1) [label=right:{{\scriptsize$3$}}] {};
\node[bd] at (4,2) [label=right:{{\scriptsize$1$}}] {};
\end{tikzpicture} 
 \\ \hline\hline 
 \{4,1\} & \{3,3,9,2\} & x_3^9+x_4^2 x_3+x_1^3+x_1 x_2^3 & 3 & 8 & 39 & 31 & 11 & 0 & 0 & 0 & 
  \begin{tikzpicture}[x=.5cm,y=.5cm]
\draw[ligne, black](0,0)--(6,0);
\node[bd] at (0,0) [label=below:{{\scriptsize$3$}}] {};
\node[bd] at (1,0) [label=below:{{\scriptsize$6$}}] {};
\node[bd] at (2,0) [label=below:{{\scriptsize$9$}}] {};
\node[bd] at (3,0) [label=below:{{\scriptsize$7$}}] {};
\node[bd] at (4,0) [label=below:{{\scriptsize$5$}}] {};
\node[bd] at (5,0) [label=below:{{\scriptsize$3$}}] {};
\node[bd] at (6,0) [label=below:{{\scriptsize$1$}}] {};
\draw[ligne, black](2,0)--(2,2);
\node[bd] at (2,1) [label=right:{{\scriptsize$5$}}] {};
\node[bd] at (2,2) [label=right:{{\scriptsize$1$}}] {};
\end{tikzpicture} 
 \\
\hline\hline
\end{array}
$$
\caption{Rank $r>1$ Model with smooth divisors and $b_3=0$. The red colored models have a generalised $D_pG$ quiver representation.\label{tab:b3zero1}  }
\end{sidewaystable}
}}


\begin{sidewaystable}[!htbp]
$$
\begin{array}{|c|c|c|c|c|c||c|c|c|c|c|c|}
\hline
\text{Type} & (a,b,c,d) & F & r &  f&d_H & \widehat{r} & \widehat{d_H} & \Delta \CA_r  & b_3 & \mathfrak{t}_2& \text{Magnetic Quiver}\, (\MQfive) \cr \hline\hline
 \{12,2\} & \{13,3,2,3\} & x_1^{13}+x_2^3 x_1+x_3^2 x_1+x_2 x_4^3+x_2 x_3 x_4 & 6 & 9 & 57 & 48 & 15 & 0 & 0 & 0 &
   \begin{tikzpicture}[x=.5cm,y=.5cm]
\draw[ligne, black](0,0)--(7,0);
\node[bd] at (0,0) [label=below:{{\scriptsize$1$}}] {};
\node[bd] at (1,0) [label=below:{{\scriptsize$5$}}] {};
\node[bd] at (2,0) [label=below:{{\scriptsize$9$}}] {};
\node[bd] at (3,0) [label=below:{{\scriptsize$13$}}] {};
\node[bd] at (4,0) [label=below:{{\scriptsize$10$}}] {};
\node[bd] at (5,0) [label=below:{{\scriptsize$7$}}] {};
\node[bd] at (6,0) [label=below:{{\scriptsize$4$}}] {};
\node[bd] at (7,0) [label=below:{{\scriptsize$1$}}] {};
\draw[ligne, black](3,0)--(3,2);
\node[bd] at (3,1) [label=right:{{\scriptsize$7$}}] {};
\node[bd] at (3,2) [label=right:{{\scriptsize$1$}}] {};
\end{tikzpicture}
\\ \hline\hline
\textcolor{red}{ \{18,1\} }& \textcolor{red}{\{2,3,2,5\}} & x_2 x_4^5+x_3 x_4^4+x_1 x_2^3+x_2 x_3^2+x_1^2 x_3 & 6 & 6 & 36 & 30 & 12 & 0 & 0 & 0 & 2^4,3^5,4^4,5^4,6^3,7^3,8^2,9^2,10,11,13
\\
\hline
\end{array}
$$
\caption{Rank $r>1$ Model with smooth divisors and $b_3=0$ (cont.); Orange shows models which do not have a $D_p^b(G)$ trinion representation in 4d.\label{tab:b3zero2} }
\end{sidewaystable}


\begin{table}
$$
\begin{array}{|c|c|c|c|c|c||c|c|c|c|c|c|}
\hline
\text{Type} & (a,b,c,d) & F & r &  f&d_H & \widehat{r} & \widehat{d_H} & \Delta \CA_r  & b_3 & \mathfrak{t}_2& \text{Magnetic Quiver}\,  (\MQfive) \cr \hline \hline
 \{1,1\} & \{2,3,6,12\} & x_4^{12}+x_3^6+x_2^3+x_1^2 & 2 & 8 & 59 & 51 & 10 & -1 & 2 & \Z_2^2 & 
  \begin{tikzpicture}[x=.5cm,y=.5cm]
\draw[ligne, black](0,0)--(7,0);
\node[bd] at (0,0) [label=below:{{\scriptsize$2$}}] {};
\node[bd] at (1,0) [label=below:{{\scriptsize$4$}}] {};
\node[bd] at (2,0) [label=below:{{\scriptsize$6$}}] {};
\node[bd] at (3,0) [label=below:{{\scriptsize$8$}}] {};
\node[bd] at (4,0) [label=below:{{\scriptsize$10$}}] {};
\node[bd] at (5,0) [label=below:{{\scriptsize$12$}}] {};
\node[bd] at (6,0) [label=below:{{\scriptsize$8$}}] {};
\node[bd] at (7,0) [label=below:{{\scriptsize$4$}}] {};
\draw[ligne, black](5,0)--(5,1);
\node[bd] at (5,1) [label=right:{{\scriptsize$6$}}] {};
\end{tikzpicture} 

\\\hline
 \{1,1\} & \{2,4,4,8\} & x_4^8+x_2^4+x_3^4+x_1^2 & 2 & 7 & 35 & 28 & 9 & -1 & 2 & \Z_2^2 &
  \begin{tikzpicture}[x=.5cm,y=.5cm]
\draw[ligne, black](0,0)--(6,0);
\node[bd] at (0,0) [label=below:{{\scriptsize$2$}}] {};
\node[bd] at (1,0) [label=below:{{\scriptsize$4$}}] {};
\node[bd] at (2,0) [label=below:{{\scriptsize$6$}}] {};
\node[bd] at (3,0) [label=below:{{\scriptsize$8$}}] {};
\node[bd] at (4,0) [label=below:{{\scriptsize$6$}}] {};
\node[bd] at (5,0) [label=below:{{\scriptsize$4$}}] {};
\node[bd] at (6,0) [label=below:{{\scriptsize$2$}}] {};
\draw[ligne, black](3,0)--(3,1);
\node[bd] at (3,1) [label=right:{{\scriptsize$4$}}] {};
\end{tikzpicture} 

\\
\hline\hline
   \end{array}
$$
\caption{Rank $r=2,3$ Models with smooth divisors and $b_3>0$. Blue indicates models with trinion representations in 4d. \label{tab:SmoothSmoothb31}}
\end{table}


\begin{table}
$$
\begin{array}{|c|c|c|c|c|c||c|c|c|c|c|c|}
\hline
\text{Type} & (a,b,c,d) & F & r &  f&d_H & \widehat{r} & \widehat{d_H} & \Delta \CA_r  & b_3 & \mathfrak{t}_2& \text{Magnetic Quiver}\, (\MQfive) \cr \hline \hline
\{1,1\} & \{2,3,6,24\} & x_4^{24}+x_3^6+x_2^3+x_1^2 & 4 & 8 & 119 & 111 & 12 & -3 & 6 & \Z_4^2 &
  \begin{tikzpicture}[x=.5cm,y=.5cm]
\draw[ligne, black](0,0)--(7,0);
\node[bd] at (0,0) [label=below:{{\scriptsize$4$}}] {};
\node[bd] at (1,0) [label=below:{{\scriptsize$8$}}] {};
\node[bd] at (2,0) [label=below:{{\scriptsize$12$}}] {};
\node[bd] at (3,0) [label=below:{{\scriptsize$16$}}] {};
\node[bd] at (4,0) [label=below:{{\scriptsize$20$}}] {};
\node[bd] at (5,0) [label=below:{{\scriptsize$24$}}] {};
\node[bd] at (6,0) [label=below:{{\scriptsize$16$}}] {};
\node[bd] at (7,0) [label=below:{{\scriptsize$8$}}] {};
\draw[ligne, black](5,0)--(5,1);
\node[bd] at (5,1) [label=right:{{\scriptsize$12$}}] {};
\end{tikzpicture} 

\\\hline
 \{1,1\} & \{2,4,4,16\} & x_4^{16}+x_2^4+x_3^4+x_1^2 & 4 & 7 & 71 & 64 & 11 & -3 & 6 & \Z_4^2 &
  \begin{tikzpicture}[x=.5cm,y=.5cm]
\draw[ligne, black](0,0)--(6,0);
\node[bd] at (0,0) [label=below:{{\scriptsize$4$}}] {};
\node[bd] at (1,0) [label=below:{{\scriptsize$8$}}] {};
\node[bd] at (2,0) [label=below:{{\scriptsize$12$}}] {};
\node[bd] at (3,0) [label=below:{{\scriptsize$16$}}] {};
\node[bd] at (4,0) [label=below:{{\scriptsize$12$}}] {};
\node[bd] at (5,0) [label=below:{{\scriptsize$8$}}] {};
\node[bd] at (6,0) [label=below:{{\scriptsize$4$}}] {};
\draw[ligne, black](3,0)--(3,1);
\node[bd] at (3,1) [label=right:{{\scriptsize$8$}}] {};
\end{tikzpicture} 
\\
\hline
 \hline
 \end{array}
$$
\caption{Rank $r=4,5$ Models with smooth divisors and $b_3>0$. 
Blue indicates models that have a $D_p^b (G)$ trinion representation. \label{tab:SmoothSmoothb32}}
\end{table}

\begin{sidewaystable}
$$
\begin{array}{|c|c|c|c|c|c||c|c|c|c|c|c|}
\hline
\text{Type} & (a,b,c,d) & F & r &  f&d_H & \widehat{r} & \widehat{d_H} & \Delta \CA_r  & b_3 & \mathfrak{t}_2& \text{Magnetic Quiver}\, (\MQfive) \cr \hline \hline
  \{1,1\} & \{2,3,6,36\} & x_4^{36}+x_3^6+x_2^3+x_1^2 & 6 & 8 & 179 & 171 & 14 & -5 & 10 & \Z_6^2 &
  \begin{tikzpicture}[x=.5cm,y=.5cm]
\draw[ligne, black](0,0)--(7,0);
\node[bd] at (0,0) [label=below:{{\scriptsize$6$}}] {};
\node[bd] at (1,0) [label=below:{{\scriptsize$12$}}] {};
\node[bd] at (2,0) [label=below:{{\scriptsize$18$}}] {};
\node[bd] at (3,0) [label=below:{{\scriptsize$24$}}] {};
\node[bd] at (4,0) [label=below:{{\scriptsize$30$}}] {};
\node[bd] at (5,0) [label=below:{{\scriptsize$36$}}] {};
\node[bd] at (6,0) [label=below:{{\scriptsize$24$}}] {};
\node[bd] at (7,0) [label=below:{{\scriptsize$12$}}] {};
\draw[ligne, black](5,0)--(5,1);
\node[bd] at (5,1) [label=right:{{\scriptsize$18$}}] {};
\end{tikzpicture} 
 \\ \hline
 \{1,1\} & \{2,4,4,24\} & x_4^{24}+x_2^4+x_3^4+x_1^2 & 6 & 7 & 107 & 100 & 13 & -5 & 10 & \Z_6^2 &
  \begin{tikzpicture}[x=.5cm,y=.5cm]
\draw[ligne, black](0,0)--(6,0);
\node[bd] at (0,0) [label=below:{{\scriptsize$6$}}] {};
\node[bd] at (1,0) [label=below:{{\scriptsize$12$}}] {};
\node[bd] at (2,0) [label=below:{{\scriptsize$18$}}] {};
\node[bd] at (3,0) [label=below:{{\scriptsize$24$}}] {};
\node[bd] at (4,0) [label=below:{{\scriptsize$18$}}] {};
\node[bd] at (5,0) [label=below:{{\scriptsize$12$}}] {};
\node[bd] at (6,0) [label=below:{{\scriptsize$6$}}] {};
\draw[ligne, black](3,0)--(3,1);
\node[bd] at (3,1) [label=right:{{\scriptsize$12$}}] {};
\end{tikzpicture} 
\\
\hline
\hline
\end{array}
$$
\caption{Rank $r=6,7$ Models with smooth divisors and $b_3>0$.  Orange labels models with integral spectrum, but no known trinion representation in 4d. \label{tab:SmoothSmoothb33}}
\end{sidewaystable}

\begin{sidewaystable}
$$
\begin{array}{|c|c|c|c|c|c||c|c|c|c|c|c|}
\hline
\text{Type} & (a,b,c,d) & F & r &  f&d_H & \widehat{r} & \widehat{d_H} & \Delta \CA_r  & b_3 & \mathfrak{t}_2& \text{Magnetic Quiver}\, (\MQfive) \cr \hline \hline
 \{1,1\} & \{2,3,6,48\} & x_4^{48}+x_3^6+x_2^3+x_1^2 & 8 & 8 & 239 & 231 & 16 & -7 & 14 & \Z_8^2 &
  \begin{tikzpicture}[x=.5cm,y=.5cm]
\draw[ligne, black](0,0)--(7,0);
\node[bd] at (0,0) [label=below:{{\scriptsize$8$}}] {};
\node[bd] at (1,0) [label=below:{{\scriptsize$16$}}] {};
\node[bd] at (2,0) [label=below:{{\scriptsize$24$}}] {};
\node[bd] at (3,0) [label=below:{{\scriptsize$32$}}] {};
\node[bd] at (4,0) [label=below:{{\scriptsize$40$}}] {};
\node[bd] at (5,0) [label=below:{{\scriptsize$48$}}] {};
\node[bd] at (6,0) [label=below:{{\scriptsize$32$}}] {};
\node[bd] at (7,0) [label=below:{{\scriptsize$16$}}] {};
\draw[ligne, black](5,0)--(5,1);
\node[bd] at (5,1) [label=right:{{\scriptsize$24$}}] {};
\end{tikzpicture} 
\\ \hline
\textcolor{orange}{\multirow{4}{*}{ \{1,1\} } }& \textcolor{orange}{\multirow{4}{*}{\{2,3,9,12\} }} & \multirow{4}{*}{$x_4^{12}+x_3^9+x_2^3+x_1^2$} & \multirow{4}{*}{8} & \multirow{4}{*}{4} & \multirow{4}{*}{90} & \multirow{4}{*}{86} & \multirow{4}{*}{12} & \multirow{4}{*}{$-2$} & \multirow{4}{*}{4} & \multirow{4}{*}{$\Z_3^2$} &
2^5,3^4,4^4,5^5,6^5,7^3,8^5,9^5,\\
& & & & & & & & & & & 10^3,11^3,12^5,13^3,14^3,15^3,16^3,\\
& & & & & & & & & & & 17^3,18^3,19,20^3,21^3,22,23,24^3,\\
& & & & & & & & & & & 25,26,27,28,30,32,33,36
 \\
\hline
\hline 
\end{array}
$$
\caption{Rank $r=8,9$ Models with smooth divisors and $b_3>0$. \label{tab:SmoothSmoothb34}}
\end{sidewaystable}

\begin{sidewaystable}
$$
\begin{array}{|c|c|c|c|c|c||c|c|c|c|c|c|}
\hline
\text{Type} & (a,b,c,d) & F & r &  f&d_H & \widehat{r} & \widehat{d_H} & \Delta \CA_r  & b_3 & \mathfrak{t}_2& \text{Magnetic Quiver}\, (\MQfive) \cr \hline \hline
 \{1,1\} & \{2,4,4,40\} & x_4^{40}+x_2^4+x_3^4+x_1^2 & 10 & 7 & 179 & 172 & 17 & -9 & 18 & \Z_{10}^2 &
  \begin{tikzpicture}[x=.5cm,y=.5cm]
\draw[ligne, black](0,0)--(6,0);
\node[bd] at (0,0) [label=below:{{\scriptsize$10$}}] {};
\node[bd] at (1,0) [label=below:{{\scriptsize$20$}}] {};
\node[bd] at (2,0) [label=below:{{\scriptsize$30$}}] {};
\node[bd] at (3,0) [label=below:{{\scriptsize$40$}}] {};
\node[bd] at (4,0) [label=below:{{\scriptsize$30$}}] {};
\node[bd] at (5,0) [label=below:{{\scriptsize$20$}}] {};
\node[bd] at (6,0) [label=below:{{\scriptsize$10$}}] {};
\draw[ligne, black](3,0)--(3,1);
\node[bd] at (3,1) [label=right:{{\scriptsize$20$}}] {};
\end{tikzpicture} 
 \\ \hline
 \{1,1\} & \{3,3,3,30\} & x_4^{30}+x_1^3+x_2^3+x_3^3 & 10 & 6 & 119 & 113 & 16 & -9 & 18 & \Z_{10}^2 &
  \begin{tikzpicture}[x=.5cm,y=.5cm]
\draw[ligne, black](0,0)--(4,0);
\node[bd] at (0,0) [label=below:{{\scriptsize$10$}}] {};
\node[bd] at (1,0) [label=below:{{\scriptsize$20$}}] {};
\node[bd] at (2,0) [label=below:{{\scriptsize$30$}}] {};
\node[bd] at (3,0) [label=below:{{\scriptsize$20$}}] {};
\node[bd] at (4,0) [label=below:{{\scriptsize$10$}}] {};
\draw[ligne, black](2,0)--(2,2);
\node[bd] at (2,1) [label=right:{{\scriptsize$20$}}] {};
\node[bd] at (2,2) [label=right:{{\scriptsize$10$}}] {};
\end{tikzpicture} 
\\
\hline
\hline 
\end{array}
$$
\caption{Rank $r=10$ Models with smooth divisors and $b_3>0$. \label{tab:SmoothSmoothb35}}
\end{sidewaystable}



\begin{sidewaystable}
\centering
$$
\begin{array}{|c|c|c|c|c|c|c|c|c|c|c|c|c|c|}
\hline
\text{Type} & (a,b,c,d) & F & \text{Divs}& r &  f&d_H & \widehat{r} & \widehat{d_H} & \Delta \CA_r  & b_3 & \mathfrak{t}_2 & \text{CB Spectrum}\cr \hline \hline
  \{1,1\} & \{2,3,6,6\} & x_3^6+x_4^6+x_2^3+x_1^2 & \text{Smt} & 1 & 8 & 29 & 21 & 9 & 0 & 0&0 & \{2^8,3^6,4^4,5^2, 6\} \\\hline
\textcolor{blue}{\multirow{2}{*}{ \{1,1\} }} & \textcolor{blue}{\multirow{2}{*}{\{2,3,6,7\}}} & \multirow{2}{*}{$x_4^7+x_3^6+x_2^3+x_1^2 $}& \multirow{2}{*}{\text{Sgl}} & \multirow{2}{*}{1} & \multirow{2}{*}{0} & \multirow{2}{*}{30} & \multirow{2}{*}{30 }& \multirow{2}{*}{1} & \multirow{2}{*}{-1} & \multirow{2}{*}{2} & \multirow{2}{*}{$\mathbb{Z}_7^2 $}&
 \left\{\frac{7}{6},\frac{4}{3},\frac{4}{3},\frac{3}{2},\frac{3}{2},\frac{5}{3},\frac{5}{3},\frac{11}{6},2,\frac{7}{3},\frac{7}{3},\frac{5}{2},\frac{5}{2},\frac{8}{3},\frac{8}{3},\frac{17}{6}, \right. 
\\ 
&&&&&&&&&&& &\left.3,\frac{7}{2},\frac{7}{2},\frac{11}{3},\frac{11}{3},\frac{23}{6},4,\frac{14}{3},\frac{14}{3},\frac{29}{6},5,\frac{35}{6},6,7\right\}
   \\\hline
 \{1,1\} & \{2,4,4,4\} & x_2^4+x_3^4+x_4^4+x_1^2 & \text{Smt} & 1 & 7 & 17 & 10 & 8 & 0 & 0 & 0 & \{2^6,3^3,4\} \\\hline
\textcolor{blue}{ \{1,1\} } & \textcolor{blue}{\{2,4,4,5\} } & x_4^5+x_2^4+x_3^4+x_1^2 & \text{Sgl} & 1 & 0 & 18 & 18 & 1 & -1 & 2 & \mathbb{Z}_5^2 &
   \left\{\frac{5}{4},\frac{5}{4},\frac{3}{2},\frac{3}{2},\frac{3}{2},\frac{7}{4},\frac{7}{4},2,\frac{5}{2},\frac{5}{2},\frac{5}{2},\frac{11}{4},\frac{11}{4},3,\frac{15}{4},\frac{15}{4},4,5\right\} \\\hline
 \{1,1\} & \{3,3,3,3\} & x_1^3+x_2^3+x_3^3+x_4^3 & \text{Smt} & 1 & 6 & 11 & 5 & 7 & 0 & 0 & 0 & \{2,2,2,2,3\} \\\hline
\textcolor{blue}{ \{1,1\} } & \textcolor{blue}{\{3,3,3,4\} } & x_4^4+x_1^3+x_2^3+x_3^3 & \text{Sgl} & 1 & 0 & 12 & 12 & 1 & -1 & 2 & \mathbb{Z}_4^2 &
   \left\{\frac{4}{3},\frac{4}{3},\frac{4}{3},\frac{5}{3},\frac{5}{3},\frac{5}{3},2,\frac{8}{3},\frac{8}{3},\frac{8}{3},3,4\right\} \\
\hline
   \end{array}
$$
\caption{Rank $r=1$: Smoothable models. `Divs' indicates whether all exceptional divisors of the resolution are smooth (Smt) or some are singular (Sgl). We use black color for the ones with a known Lagrangian $\FTfour$, blue color for trinion models, red color for generalised $D_p^b(G)$ models and orange color for the unknown ones. \label{smoothableModelsI}}
\end{sidewaystable}


\begin{sidewaystable}
\centering
$$
\begin{array}{|c|c|c|c|c|c|c|c|c|c|c|c|c|c|}
\hline
\text{Type} & (a,b,c,d) & F & \text{Divs}& r &  f&d_H & \widehat{r} & \widehat{d_H} & \Delta \CA_r  & b_3 & \mathfrak{t}_2 & \text{CB Spectrum}\cr \hline \hline
\{1,1\} & \{2,3,6,12\} & x_4^{12}+x_3^6+x_2^3+x_1^2 & \text{Smt} & 2 & 8 & 59 & 51 & 10 & -1 & 2 & \mathbb{Z}_2^2 &
   \{2^9,3^8,4^8,5^6,6^6,7^4,8^4,9^2,10^2,11,12\} \\\hline
 \textcolor{blue}{\{1,1\} } & \textcolor{blue}{\{2,3,6,13\} } & x_4^{13}+x_3^6+x_2^3+x_1^2 & \text{Sgl} & 2 & 0 & 60 & 60 & 2 & -2 & 4 & \mathbb{Z}_{13}^2 &
  \left\{\frac{7}{6},\frac{4}{3}^2,\frac{3}{2}^2,\frac{5}{3}^2,\frac{11}{6},2,\frac{13}{6},\frac{7}{3}^2,
   \frac{5}{2}^2,\frac{8}{3}^2,\frac{17}{6},3,\frac{10}{3}^2,\frac{7}{2}^2,\frac{11}{3}^2,\frac{23}{6},4,
   \right. 
\\ 
&&&&&&&&&&& &\left.  
\frac{13}{3}^2,\frac{9}{2}^2,\frac{14}{3}^2,\frac{29}{6},5, \frac{11}{2}^2,\frac{17}{3}^2,\frac{35}{6},
 6,\frac{13}{2}^2,\frac{20}{3}^2,\frac{41}{6},7,
 \right. 
\\ 
&&&&&&&&&&& &\left.  
 \frac{23}{3}^2,\frac{47}{6},8,\frac{26}{3}^2,\frac{53}{6},9,
 \frac{59}{6},10,\frac{65}{6},11,12,13\right\}
 \\\hline
\textcolor{blue}{ \{1,1\} } & \textcolor{blue}{\{2,3,8,8\} } & x_3^8+x_4^8+x_2^3+x_1^2 & \text{Sgl} & 2 & 0 & 49 & 49 & 2 & -3 & 6 & \mathbb{Z}_3^6 &
   \left\{\frac{3}{2}^6,2^5,3^7,\frac{7}{2}^4,\frac{9}{2}^6,5^3,6^5,\frac{13}{2}^2,\frac{15}{2}^4,8,9^3,\frac{21}{2}^2,12\right\} \\\hline
 \{1,1\} & \{2,4,4,8\} & x_4^8+x_2^4+x_3^4+x_1^2 & \text{Smt} & 2 & 7 & 35 & 28 & 9 & -1 & 2 & \mathbb{Z}_2^2 &
   \{2^8,3^6,4^6,5^3,6^3,7,8\} \\\hline
\textcolor{blue}{ \{1,1\} } & \textcolor{blue}{\{2,4,4,9\} } & x_4^9+x_2^4+x_3^4+x_1^2 & \text{Sgl} & 2 & 0 & 36 & 36 & 2 & -2 & 4 & \mathbb{Z}_9^2 &
   \left\{\frac{5}{4}^2, \frac{3}{2}^3,\frac{7}{4}^2,2,\frac{9}{4}^2,\frac{5}{2}^3,\frac{11}{4}^2,3,\frac{7}{2}^3,\frac{15}{4}^2,4,\frac{9}{2}^3,
   \frac{19}{4}^2,5,\frac{23}{4}^2,6,\frac{27}{4}^2,7,8,9\right\} \\
\end{array}
$$
\caption{Rank $r=2$: Smoothable models. `Divs' indicates whether all exceptional divisors of the resolution are smooth (Smt) or some are singular (Sgl). \label{smoothableModelsII}}
\end{sidewaystable}


\begin{sidewaystable}
\centering
$$
\begin{array}{|c|c|c|c|c|c|c|c|c|c|c|c|c|c|}
\hline
\text{Type} & (a,b,c,d) & F & \text{Divs}& r &  f&d_H & \widehat{r} & \widehat{d_H} & \Delta \CA_r  & b_3 & \mathfrak{t}_2 & \text{CB Spectrum}\cr \hline \hline
\textcolor{orange}{ \{7,1\} }& \textcolor{orange}{\{2,3,5,7\} } & x_3 x_4^7+x_2 x_3^5+x_2^3+x_1^2 & \text{Sgl} & 2 & 0 & 46 & 46 & 2 & 0 & 0 & 0 &
   \left\{\frac{20}{19},\frac{22}{19},\frac{24}{19},\frac{28}{19},\frac{30}{19},\frac{32}{19},\frac{34}{19},\frac{36}{19},\frac{42}{19},\frac{44}{19
   },\frac{46}{19},\frac{48}{19},\frac{50}{19},
      \right. 
\\ 
&&&&&&&&&&& &\left.  
\frac{52}{19},   \frac{54}{19},\frac{56}{19},\frac{58}{19},
      \frac{60}{19},\frac{62}{19},\frac{70}{19},\frac{72}{19},\frac{74}{19},4,\frac{78}{19},\frac{80}{19},\frac{84}{19},\frac{86}{19},\frac{88}{19},\frac{98}{19},
    \frac{100}{19},    \frac{102}{19},
       \right. 
\\ 
&&&&&&&&&&& &\left.  
\frac{104}{19},\frac{106}{19},\frac{112}{19},6,\frac{126}{19},
    \frac{128}{19},\frac{130}{19},\frac{132}{19},\frac{140}{19},
   \frac{154}{19},\frac{156}{1   9},\frac{158}{19},\frac{182}{19},\frac{184}{19},\frac{210}{19}\right\} \\\hline
\textcolor{orange}{ \{7,1\} } & \textcolor{orange}{\{2,6,3,4\} } & x_2^6+x_3^3 x_2+x_3 x_4^4+x_1^2 & \text{Sgl} & 2 & 1 & 30 & 29 & 3 & 0 & 0 & 0 &
   \left\{\frac{10}{9},\frac{11}{9},\frac{4}{3},\frac{14}{9},\frac{5}{3},\frac{16}{9},\frac{19}{9},\frac{20}{9},\frac{7}{3},\frac{22}{9},\frac{23}{9
   },\frac{8}{3},\frac{26}{9},3,\frac{28}{9},\frac{32}{9},
      \right. 
\\ 
&&&&&&&&&&& &\left.  
   \frac{11}{3},\frac{34}{9},\frac{35}{9},4,\frac{13}{3},\frac{40}{9},\frac{46}{9},\frac{47}{
   9},\frac{16}{3},\frac{52}{9},\frac{59}{9},\frac{20}{3},8\right\} \\
\hline
   \end{array}
$$
\caption{Rank $r=2$: Smoothable models. `Divs' indicates whether all exceptional divisors of the resolution are smooth (Smt) or some are singular (Sgl).  (Continued) \label{smoothableModelsIII}}
\end{sidewaystable}


\begin{sidewaystable}
\centering
$$
\begin{array}{|c|c|c|c|c|c|c|c|c|c|c|c|c|c|}
\hline
\text{Type} & (a,b,c,d) & F & \text{Divs}& r &  f&d_H & \widehat{r} & \widehat{d_H} & \Delta \CA_r  & b_3 & \mathfrak{t}_2 & \text{CB Spectrum}\cr \hline \hline
 \{1,1\} & \{2,3,6,18\} & x_4^{18}+x_3^6+x_2^3+x_1^2 & \text{Smt} & 3 & 8 & 89 & 81 & 11 & -2 & 4 & \mathbb{Z}_3^2 &
   \{2^9,3^9,4^8,5^8,6^8,7^6,8^6,9^6,10^4,11^4
   ,12^4,13^2,14^2,15^2,16,17,18\} \\\hline 
\textcolor{blue}{ \{1,1\} } & \textcolor{blue}{ \{2,3,6,19\} } & x_4^{19}+x_3^6+x_2^3+x_1^2 & \text{Sgl} & 3 & 0 & 90 & 90 & 3 & -3 & 6 & \mathbb{Z}_{19}^2 &
   \left\{\frac{7}{6},\frac{4}{3}^2,\frac{3}{2}^2,\frac{5}{3}^2,\frac{11}{6},2,\frac{13}{6},\frac{7}{3}^2,\frac{5}{2}^2,\frac{8}{3}^2,\frac{17}{6},3,\frac{19}{6},\frac{10}{3}^2,\frac{7}{2}^2,   \frac{11}{3}^2,\frac{23}{6},
  \right. 
\\ 
&&&&&&&&&&& &\left.  
4,\frac{13}{3}^2,\frac{9}{2}^2,\frac{14}{3}^2,\frac{29}{6},5,\frac{16}{3}^2,
   \frac{11}{2}^2,\frac{17}{3}^2,\frac{35}{6},6,\frac{19}{3}^2,\frac{13}{2}^2,\frac{20}{3}^2,\frac{41   }{6},
     \right. 
\\ 
&&&&&&&&&&& &\left.  
  7, \frac{15}{2}^2,\frac{23}{3}^2,\frac{47}{6},8,\frac{17}{2}^2,\frac{26}{3}^2,\frac{53}{6},9,\frac{19}{2}^2,\frac{29}{3}^2,\frac{59}{6},10,\frac{32}{3}^2,\frac{65}{6},   \right. 
\\ 
&&&&&&&&&&& &\left.  
11,   \frac{35}{3}^2,  \frac{71}{6},
   12,\frac{38}{3}^2,\frac{77}{6},13,\frac{83}{6},14,\frac{89}{6},15,\frac{95}{6},16,17,18,19\right\} \\\hline 
\textcolor{orange}{ \{1,1\} } & \textcolor{orange}{\{2,3,7,14\} } & x_4^{14}+x_3^7+x_2^3+x_1^2 & \text{Sgl} & 3 & 0 & 78 & 78 & 3 & -3 & 6 & \mathbb{Z}_3^6 &
   \left\{\frac{3}{2}^5,2^5, 3^6,\frac{7}{2}^4,\frac{9}{2}^6,5^4,6^6,\frac{13}{2}^3,\frac{15}{2}^5,8^3,9^5,   \frac{19}{2}^2,\frac{21}{2}^4,
    \right. 
\\ 
&&&&&&&&&&& &\left.  
11^2,
   12^4,\frac{25}{2},\frac{27}{2}^3,14,15^3,\frac{33}{2}^2,18^2,\frac{39}{2},21\right\} \\
\hline 
    \end{array}
$$
\caption{Rank $r=3$: Smoothable models. `Divs' indicates whether all exceptional divisors of the resolution are smooth (Smt) or some are singular (Sgl). \label{smoothableModelsIV}}
\end{sidewaystable}


\begin{sidewaystable}
\centering
$$
\begin{array}{|c|c|c|c|c|c|c|c|c|c|c|c|c|c|}
\hline
\text{Type} & (a,b,c,d) & F & \text{Divs}& r &  f&d_H & \widehat{r} & \widehat{d_H} & \Delta \CA_r  & b_3 & \mathfrak{t}_2 & \text{CB Spectrum}\cr \hline \hline
\textcolor{orange}{ \{2,1\} } & \textcolor{orange}{\{3,5,9,2\} } & x_3^9+x_4^2 x_3+x_2^5+x_1^3 & \text{Sgl} & 3 & 0 & 40 & 40 & 3 & -1 & 2 & \mathbb{Z}_5^2 &
   \left\{\frac{5}{4}^2, \frac{3}{2}^2,\frac{7}{4}^3,2,\frac{5}{2}^3,\frac{11}{4}^2,3^2,\frac{13}{4},\frac{15}{4}^2,   4^3,\frac{17}{4},\frac{9}{2},
     \right. 
\\ 
&&&&&&&&&&& &\left.  
5^2,\frac{21}{4}^2,\frac{11}{2},
   \frac{25}{4}^3,\frac{13}{2},\frac{27}{4},\frac{15}{2}^2,\frac{31}{4},\frac{35}{4},9,10,\frac{45}{4}\right\} \\\hline
\textcolor{blue}{ \{2,1\} } & \textcolor{blue}{\{3,8,4,2\} } & x_2^8+x_3^4+x_1^3+x_3 x_4^2 & \text{Sgl} & 3 & 0 & 35 & 35 & 3 & -2 & 4 & \mathbb{Z}_3^4 &
   \left\{\frac{3}{2}^4,2^4,3^5,\frac{7}{2}^3,\frac{9}{2}^4,5^2,6^4,\frac{13}{2},\frac{15}{2}^3,8,9,9,\frac{21}{2},12\right\} \\
\hline 
 \{2,1\} & \{4,4,4,2\} & x_1^4+x_2^4+x_3^4+x_3 x_4^2 & \text{Smt} & 3 & 3 & 24 & 21 & 6 & -3 & 6 & \mathbb{Z}_2^6 &
   \{2^8,3^2,4^6,5,6^3,8\} \\
\hline 
 \{4,1\} & \{3,3,9,2\} & x_3^9+x_4^2 x_3+x_1^3+x_1 x_2^3 & \text{Smt} & 3 & 8 & 39 & 31 & 11 & 0 & 0 & 0 &
   \{2^7,3^7,4^5,5^5,6^3,7^2,8,9\} \\
\hline 
\textcolor{orange}{ \{4,1\} } & \textcolor{orange}{\{3,3,10,2\} } & x_3^{10}+x_4^2 x_3+x_1^3+x_1 x_2^3 & \text{Sgl} & 3 & 1 & 39 & 38 & 4 & 0 & 0 & 0 &
   \left\{\frac{20}{19},\frac{24}{19},\frac{26}{19},\frac{28}{19},\frac{30}{19},\frac{32}{19},\frac{36}{19},\frac{39}{19},\frac{42}{19},\frac{44}{19
   },\frac{46}{19},   \frac{48}{19},\frac{50}{19},\frac{54}{19},
     \right. 
\\ 
&&&&&&&&&&& &\left.  
\frac{59}{19},\frac{60}{19},\frac{62}{19},\frac{64}{19},\frac{66}{19},\frac{68}{19},\frac{72}{19},\frac{80}{19},\frac{82}{19},\frac{84}{19},\frac{86}{19},\frac{90}{19},\frac{99}{19},\frac{100}{19},       \right. 
\\
\hline 
   \end{array}
$$
\caption{Rank $r=3$: Smoothable models. `Divs' indicates whether all exceptional divisors of the resolution are smooth (Smt) or some are singular (Sgl). \label{smoothableModelsV}}
\end{sidewaystable}


\begin{sidewaystable}
\centering
$$
\begin{array}{|c|c|c|c|c|c|c|c|c|c|c|c|c|c|}
\hline
\text{Type} & (a,b,c,d) & F & \text{Divs}& r &  f&d_H & \widehat{r} & \widehat{d_H} & \Delta \CA_r  & b_3 & \mathfrak{t}_2 & \text{CB Spectrum}\cr \hline \hline
\textcolor{orange}{ \{10,1\} } & \textcolor{orange}{\{3,2,5,3\} } & x_4 x_3^5+x_2^2 x_3+x_1^3+x_2 x_4^3 & \text{Sgl} & 3 & 0 & 30 & 30 & 3 & 0 & 0 & 0 &
   \left\{\frac{11}{10},\frac{6}{5},\frac{7}{5},\frac{3}{2},\frac{17}{10},\frac{9}{5},\frac{21}{10},\frac{23}{10},\frac{12}{5},\frac{13}{5},\frac{27
   }{10},\frac{29}{10},   3,\frac{16}{5},\frac{33}{10},
       \right. 
\\ 
&&&&&&&&&&& &\left.  
\frac{18}{5},\frac{39}{10},\frac{21}{5},\frac{22}{5},\frac{9}{2},\frac{47}{10},\frac{24}{5},\frac{27}{5},\frac{57}{10},6,\frac{31}{5},\frac{63}{10},\frac{15}{2},\frac{39}{5},\frac{93}{10}\right\} \\
\hline 
 \textcolor{orange}{\{10,1\} }& \textcolor{orange}{\{5,2,2,3\} } & x_1^5+x_2 x_4^3+x_2^2 x_3+x_3^2 x_4 & \text{Sgl} & 3 & 0 & 24 & 24 & 3 & 0 & 0 & 0 &
   \left\{\frac{9}{8},\frac{5}{4},\frac{11}{8},\frac{3}{2},\frac{7}{4},\frac{15}{8},\frac{17}{8},\frac{19}{8},\frac{5}{2},\frac{11}{4},   3,\frac{25}{8 },\frac{13}{4},
     \right. 
\\ 
&&&&&&&&&&& &\left.  
\frac{27}{8},\frac{15}{4},4,\frac{35}{8},\frac{37}{8},\frac{39}{8},5,\frac{45}{8},\frac{25}{4},\frac{13}{2},\frac{65}{8}\right\}   \\
   \hline 
\textcolor{orange}{ \{11,1\} } & \textcolor{orange}{\{4,5,2,2\} } & x_1 x_2^5+x_3^2 x_2+x_1^4+x_3 x_4^2 & \text{Sgl} & 3 & 1 & 29 & 28 & 4 & 0 & 0 & 0 &
   \left\{\frac{10}{9},\frac{4}{3},\frac{13}{9},\frac{14}{9},\frac{16}{9},\frac{17}{9},\frac{20}{9},\frac{7}{3},\frac{22}{9},\frac{8}{3},\frac{25}{9
   },\frac{26}{9},\frac{28}{9},\frac{32}{9},\frac{11}{3},\frac{34}{9},
     \right. 
\\ 
&&&&&&&&&&& &\left. 
4,\frac{37}{9}, \frac{40}{9},\frac{44}{9},5,\frac{46}{9},\frac{16}{3},\frac{56}
   {9},\frac{19}{3},\frac{20}{3},\frac{68}{9},\frac{80}{9}\right\} \\
\hline
   \end{array}
$$
\caption{Rank $r=3$: Smoothable models. `Divs' indicates whether all exceptional divisors of the resolution are smooth (Smt) or some are singular (Sgl).  (Continued)\label{smoothableModelsVI}}
\end{sidewaystable}

\begin{table}
$$
\begin{array}{|c|c|c|c|c|c|c|c|c|c|c|c|c|}
\hline
\text{Type} & (a,b,c,d) & F & \text{Divs}& r &  f&d_H & \widehat{r} & \widehat{d_H} & \Delta \CA_r  & b_3 & \mathfrak{t}_2 \\ \hline \hline 
 \{1,1\} & \{2,3,6,24\} & x_4^{24}+x_3^6+x_2^3+x_1^2 & \text{Smt} & 4 & 8 & 119 & 111 & 12 & -3 & 6 & \mathbb{Z}_4^2 \\
 \textcolor{blue}{\{1,1\}} &  \textcolor{blue}{\{2,3,6,25\}} & x_4^{25}+x_3^6+x_2^3+x_1^2 & \text{Sgl} & 4 & 0 & 120 & 120 & 4 & -4 & 8 & \mathbb{Z}_{25}^2 \\
 \{1,1\} & \{2,4,4,16\} & x_4^{16}+x_2^4+x_3^4+x_1^2 & \text{Smt} & 4 & 7 & 71 & 64 & 11 & -3 & 6 & \mathbb{Z}_4^2 \\
 \textcolor{blue}{ \{1,1\}} &  \textcolor{blue}{\{2,4,4,17\}} & x_4^{17}+x_2^4+x_3^4+x_1^2 & \text{Sgl} & 4 & 0 & 72 & 72 & 4 & -4 & 8 & \mathbb{Z}_{17}^2 \\
 \{1,1\} & \{2,4,6,6\} & x_3^6+x_4^6+x_2^4+x_1^2 & \text{Smt} & 4 & 5 & 40 & 35 & 9 & -2 & 4 & \mathbb{Z}_2^4 \\
 \{1,1\} & \{3,3,3,12\} & x_4^{12}+x_1^3+x_2^3+x_3^3 & \text{Smt} & 4 & 6 & 47 & 41 & 10 & -3 & 6 & \mathbb{Z}_4^2 \\
  \textcolor{blue}{\{1,1\}} &  \textcolor{blue}{\{3,3,3,13\}} & x_4^{13}+x_1^3+x_2^3+x_3^3 & \text{Sgl} & 4 & 0 & 48 & 48 & 4 & -4 & 8 & \mathbb{Z}_{13}^2 \\
\hline
\end{array}
$$
\caption{Rank $r=4$: Smoothable models. `Divs' indicates whether all exceptional divisors of the resolution are smooth (Smt) or some are singular (Sgl).  
The CB spectra are rather expansive and we include them instead in the ancillary mathematica file. \label{smoothableModelsVII}}
\end{table}


\bibliography{FM}
\bibliographystyle{JHEP}


\end{document}